\documentclass[12pt,preprint,twoside]{article} 
 \usepackage{ams}
 \usepackage{dsfont}
 \usepackage{amsmath}
 \usepackage{amssymb}
 \usepackage{amsthm}
 \usepackage{rotating}
 \usepackage{fancyhdr}
 \usepackage{fancyheadings}
 \RequirePackage[OT1]{fontenc}
 \RequirePackage{amsthm,amsmath}
 \RequirePackage[numbers]{natbib}
 \RequirePackage[colorlinks,citecolor=red,linkcolor=blue,urlcolor=red]{hyperref}

 \usepackage{titlesec}
 \titleformat{\section}{\large\bfseries}{\thesection}{1em}{}
 \titleformat{\subsection}{\normalsize\bfseries}{\thesubsection}{1em}{}
 \titleformat{\subsubsection}{\normalsize\it}{{\rm \thesubsubsection}}{1em}{\vspace{-.7em}}

 \usepackage{mathptmx,bm}
 \usepackage{times}

 \newtheoremstyle{mytheo}
                {\topsep}
                {\topsep}
                {\rm}
                {}
                {\sc }
                {.}
                {0.5em}
                {}

 \theoremstyle{mytheo}

\newtheorem{theo}{\small\bf Theorem}[section]

 \numberwithin{equation}{section}
 
 \renewcommand{\Pr}{\mathds{P}}

 \newcommand{\be}{\begin{equation}}
 \newcommand{\ee}{\end{equation}}
 \newcommand{\E}{\mathds{E}}
 \newcommand{\Var}{\mbox{\rm \hspace*{.2ex}Var\hspace*{.2ex}}}
 \newcommand{\Cov}{\mbox{\rm \hspace*{.2ex}Cov\hspace*{.2ex}}}

 \renewcommand{\R}{\mathds{R}}

   \font\twelvebm                       = cmmib10 at 12truept
   \font\tenbm                          = cmmib10 at 10truept
   \font\sevenbm                        = cmmib10 at 7truept
\tenbm \scriptscriptfont9              = \sevenbm

  at 10truept
 at 10truept  at 10truept  at 10truept 
at 10truept

\mathchardef \BGamma            = "0900 \mathchardef
\BDelta = "0901 \mathchardef \BTheta            = "0902
\mathchardef \BLambda           = "0903 \mathchardef \BXi
= "0904 \mathchardef \BPi               = "0905
\mathchardef \BSigma = "0906 \mathchardef \BUpsilon
= "0907 \mathchardef \BPhi = "0908 \mathchardef \BPsi
= "0909 \mathchardef \BOmega = "090A \mathchardef \Balpha
= "090B \mathchardef \Bbeta             = "090C
\mathchardef \Bgamma = "090D \mathchardef \Bdelta
= "090E \mathchardef \Bepsilon = "090F \mathchardef \Bzeta
= "0910 \mathchardef \Beta = "0911 \mathchardef \Btheta =
"0912 \mathchardef \Biota = "0913 \mathchardef \Bkappa
= "0914 \mathchardef \Blambda = "0915 \mathchardef \Bmu
= "0916 \mathchardef \Bnu = "0917 \mathchardef \Bxi
= "0918 \mathchardef \Bpi = "0919 \mathchardef \Brho
= "091A \mathchardef \Bsigma            = "091B
\mathchardef \Btau = "091C \mathchardef \Bupsilon
= "091D \mathchardef \Bphi = "091E \mathchardef \Bchi
= "091F \mathchardef \Bpsi = "0920 \mathchardef \Bomega
= "0921 \mathchardef \Bvarepsilon       = "0922
\mathchardef \Bvartheta         = "0923 \mathchardef
\Bvarpi = "0924 \mathchardef \Bvarrho = "0925 \mathchardef
\Bvarsigma = "0926 \mathchardef \Bvarphi           = "0927
\mathchardef \bA        = "0941 \mathchardef \bB        =
"0942 \mathchardef \bC        = "0943 \mathchardef \bD
= "0944 \mathchardef \bE        = "0945 \mathchardef \bF
= "0946 \mathchardef \bG        = "0947 \mathchardef \bH
= "0948 \mathchardef \bI        = "0949 \mathchardef \bJ
= "094A \mathchardef \bK        = "094B \mathchardef \bL
= "094C \mathchardef \bM        = "094D \mathchardef \bN
= "094E \mathchardef \bO        = "094F \mathchardef \bP
= "0950 \mathchardef \bQ        = "0951 \mathchardef \bR
= "0952 \mathchardef \bS        = "0953 \mathchardef \bT
= "0954 \mathchardef \bU        = "0955 \mathchardef \bV
= "0956 \mathchardef \bW        = "0957 \mathchardef \bX
= "0958 \mathchardef \bY        = "0959 \mathchardef \bZ
= "095A \mathchardef \ba        = "0961 \mathchardef \bb
= "0962 \mathchardef \bc        = "0963 \mathchardef \bd
= "0964
\mathchardef \bee       = "0965 
\mathchardef \bff       = "0966 \mathchardef \bg        =
"0967 \mathchardef \bh        = "0968
\mathchardef \bj        = "096A \mathchardef \bk        =
"096B \mathchardef \bl        = "096C \mathchardef \bm
= "096D \mathchardef \bn        = "096E \mathchardef \bo
= "096F \mathchardef \bp        = "0970 \mathchardef \bq
= "0971 \mathchardef \br        = "0972 \mathchardef \bs
= "0973 \mathchardef \bt        = "0974 \mathchardef \bu
= "0975 \mathchardef \bv        = "0976 \mathchardef \bw
= "0977 \mathchardef \bx        = "0978 \mathchardef \by
= "0979 \mathchardef \bz        = "097A

\font\tencb            = cmssbx10 scaled \magstep4
\font\eigcb = cmssbx10 scaled \magstep2 \textfont8
= \tencb \scriptfont8           = \eigcb \scriptscriptfont8
= \eigcb \mathchardef\bAs       = "1841
\def\Asem#1#2{\mathop{\vrule height10.5pt depth5.5pt width0pt\bAs}_{#1}^{#2}}
\def\asem#1#2{
          \ifmmode
         \ifinner
            \raise0.9pt\hbox{$\scriptstyle\bAs$}_{#1}^{#2}
         \else
            \Asem{#1}{#2}
         \fi
          \fi
          }

\renewcommand{\Pr}{\mathbb{P}}

\newtheorem{lem}{\small\bf Lemma}[section]
\newtheorem{prop}{\small\bf Proposition}[section]
\newtheorem{rem}{\small\bf Remark}[section]
\newenvironment{REM}{\begin{rem} \rm}{\end{rem}}
\newtheorem{defi}{\small\bf Definition}[section]
\newenvironment{DEFI}{\begin{defi} \rm}{\end{defi}}
\newtheorem{cor}{\small\bf Corollary}[section]
\newenvironment{THEO}{\begin{theo} \rm}{\end{theo}}
\newenvironment{LEM}{\begin{lem} \rm}{\end{lem}}
\newenvironment{PROP}{\begin{prop} \rm}{\end{prop}}
\newtheorem{notation}{\small\bf Notation}

\newenvironment{COR}{\begin{cor} \rm}{\end{cor}}
\newtheorem{example}{\small\bf Example}[section]
\newenvironment{EXAMPLE}{\begin{example} \rm}{\end{example}}
\newenvironment{Proof}{{\small\bf Proof:}}{}
\newcommand{\Corr}{\mbox{\rm \hspace*{.2ex}Corr\hspace*{.2ex}}}

\newenvironment{PR}[1]{\noindent{{#1}\hspace*{.4ex}}}{}

 \newcommand{\authorrunning}{\textsc{Nickos Papadatos}
 }
 \vspace{-3em}
 \newcommand{\titlerunning}{\textsc{Maximizing the Expected
 Range}}
 \vspace{-3em}

 \newcommand{\bbb}[1]{\mbox{$ #1 $}}
 \newcommand{\bmu}{\mu\hspace{-1.47ex}\mu\hspace{-1.47ex}\mu}
 \newcommand{\bsigma}{\sigma\hspace{-1.52ex}\sigma\hspace{-1.52ex}\sigma}
 \newcommand{\bSigma}{\Sigma\hspace{-1.41ex}\Sigma\hspace{-1.41ex}\Sigma}
 \newcommand{\bXX}{X\hspace{-1.52ex}X\hspace{-1.52ex}X}
 \newcommand{\bxx}{x\hspace{-.95ex}x\hspace{-.95ex}x}
 \newcommand{\byy}{y\hspace{-.93ex}y\hspace{-.93ex}y}
 \newcommand{\balpha}{\alpha\hspace{-1.53ex}\alpha\hspace{-1.53ex}\alpha}
 \newcommand{\bbalpha}{\alpha\hspace{-1.16ex}\alpha\hspace{-1.16ex}\alpha}
 \newcommand{\bVV}{V\hspace{-1.5ex}V\hspace{-1.5ex}V}
 \newcommand{\bbee}{e\hspace{-.95ex}e\hspace{-.95ex}e}
 \newcommand{\bpp}{p\hspace{-1.23ex}p\hspace{-1.23ex}p}
 \newcommand{\bqq}{q\hspace{-1.16ex}q\hspace{-1.16ex}q}
 \newcommand{\bYY}{Y\hspace{-1.44ex}Y\hspace{-1.44ex}Y}
 \newcommand{\bone}{1\hspace{-1.03ex}1\hspace{-1.03ex}1}

\begin{document}

 \pagestyle{fancy} \fancyhf{}
 \fancyhead[RO,LE]{\thepage}
 \renewcommand{\headrulewidth}{.5pt}
 \fancyhead[LO]{\footnotesize\itshape\titlerunning}
 \fancyhead[RE]{\footnotesize\itshape\authorrunning}

 \begin{center}
 {\Large\bf Maximizing the expected range from dependent
 observations under
 mean-variance information\textcolor[rgb]{0,0,1}{\footnote{Work partially
 supported by the University of Athens Research Grant
 70/4/5637}}
 }
 \end{center}

 \begin{center}
 {\large
 Nickos
 Papadatos\textcolor[rgb]{0,0,1}{\footnote{{\it
 }{e-mail:}\
 \textcolor[rgb]{0.98,0.00,0.00}{npapadat@math.uoa.gr}, {url:}\
 \textcolor[rgb]{0.98,0.00,0.00}{users.uoa.gr/$\sim$npapadat/}}}}
 \end{center}

 \begin{center}
 {\small\it
 Department of Mathematics, Section of Statistics and
 O.R., University of Athens,
 \\
 Panepistemiopolis, 157 84 Athens, Greece.
 }
 \end{center}

 \begin{center}
 \begin{minipage}[1]{34em}
 {\small
 {\bf Abstract:}
 In this article we derive the best possible upper bound
 for $\E[\max_i\{X_i\}-\min_i\{X_i\}]$ under given means and
 variances on $n$ random variables $X_i$.
 The random vector $(X_1,\ldots,X_n)$ is allowed to have any
 dependence structure, provided $\E X_i=\mu_i$ and
 $\Var X_i=\sigma_i^2$, $0<\sigma_i<\infty$.
 We provide an explicit characterization of the
 $n$-variate distributions that attain the equality
 (extremal random vectors), and the
 tight bound is compared to other existing results.
 \medskip

 \medskip

 }
 \end{minipage}
 \vspace{-.5em}
 \end{center}

 \thispagestyle{empty}

 \section{Introduction}
 \setcounter{equation}{0}
 \label{sec.intro}
 The problem of determining best possible expectation bounds
 on linear functions of order statistics in terms of means and variances of
 the
 observations has a long history. Especially
 for the sample range based on $n\geq 2$ independent identically
 distributed  (i.i.d.) random variables, the problem goes back to
 Plackett (1947), Gumbel (1954) and Hartley and David (1954)
 who derived
 the inequality
 \be
 \label{eq.plackett}
 \mbox{$
 \E\big[\max_{1\leq i \leq n}\{X_i\}-\min_{1\leq i \leq n}\{X_i\}\big]\leq
 n\sigma \sqrt{\frac{2}{2n-1}\Big(1-\frac{1}{{{2n-2 \choose n-1}}}\Big)},
 $}
 \ee
 where $\sigma^2$ is the common variance of $X_i$. This bound
 is best possible in the sense that for any given values of
 $\mu\in\R$ and $\sigma\in(0,\infty)$ there exist $n$ i.i.d.\ random variables
 with mean $\mu$ and variance $\sigma^2$ that attain the equality in
 (\ref{eq.plackett}).

 Since then, a lot of research has been developed in order to
 drop the assumptions of independence and/or
 identical distributions on the observations, and
 also to extend the results to any
 $L$-statistic of the form
 $
 L=\sum_{i=1}^n c_i X_{i:n},
 $
 where $c_i$ are given constants and $X_{1:n}\leq \cdots\leq X_{n:n}$
 are the order statistics corresponding to the random vector
 $(X_1,\ldots,X_n)$.
 When the components $X_i$ are merely assumed to be i.d.\
 (identically distributed but not necessarily independent)
 with mean $\mu$ and variance $\sigma^2$,
 the best possible bounds for $\E L$
 were established by Rychlik (1993b). In particular, setting
 $c_1=-1$, $c_n=1$ and $c_i=0$ for any other $i$
 in Rychlik's result,
 we get the optimal upper bound
 for the expected range:
 \be
 \label{eq.Rychlik}
 \E\left[X_{n:n}-X_{1:n}\right]\leq
 \sigma \sqrt{2n}.
 \ee
 For a comprehensive review of related
 results and extensions, the reader is
 referred to Rychlik's (2001)
 monograph; see also
 David (1981), Rychlik (1998) and David and Nagaraja (2003).
 Dropping both assumptions of independence and i.d.,
 Arnold and Groeneveld (1979)
 obtained the upper bound
 \be
 \label{eq.AG.general}
 \mbox{$
 \E\left(\sum_{i=1}^n c_i X_{i:n}\right)
 \leq \overline{\mu} \sum_{i=1}^n c_i
 +\sqrt{\sum_{i=1}^n (c_i-\overline{c})^2}
 \sqrt{\sum_{i=1}^n \{(\mu_i-\overline{\mu})^2+\sigma_i^2\}},
 $}
 \ee
 which is valid for any random vector with $\E X_i=\mu_i$
 and $\Var X_i=\sigma_i^2$,
 where $\overline{\mu}=\frac{1}{n}\sum_{i=1}^n \mu_i$,
 $\overline{c}=\frac{1}{n}\sum_{i=1}^n c_i$.
 For other inequalities related to (\ref{eq.AG.general})
 the reader is referred to Nagaraja (1981), Aven (1985),
 Lef\`{e}vre (1986), Papadatos (2001a) and
 Kaluszka, Okolewski and Szymanska (2005); see also
 the monograph by Arnold and Balakrishnan (1989).
 Applied to the range, (\ref{eq.AG.general}) yields
 the inequality
 \be
 \label{eq.AG}
 \mbox{$
 \E[ X_{n:n}-X_{1:n} ]
 \leq AG_n:=
 \sqrt{2\sum_{i=1}^n \{(\mu_i-\overline{\mu})^2+\sigma_i^2\}},
 $}
 \ee
 which, in the homogeneous case $\mu_i=\mu$, $\sigma_i^2=\sigma^2$,
 reduces to (\ref{eq.Rychlik}).
 However, the upper bound in (\ref{eq.AG})
 is not tight under general mean-variance information, and the
 purpose of the present work is to replace the RHS of
 (\ref{eq.AG}) by its best possible value.

 Recently, Bertsimas, Natarajan and Teo (2004, 2006) applied
 convex optimization techniques in order to replace
 the RHS of (\ref{eq.AG.general}) by its tight counterpart
 in some particular cases of interest.
 They obtained, among other things, the best possible upper bound for
 the expected maximum under any mean-variance information and any
 dependence structure, namely,
 \be
 \label{eq.BNT}
 \mbox{$
 \E X_{n:n} \leq BNT_n:=-\frac{n-2}{2}y_0
 +\frac{1}{2}\sum_{i=1}^n \mu_i
 +\frac{1}{2} \sum_{i=1}^n \sqrt{(\mu_i-y_0)^2+\sigma_i^2},
 $}
 \ee
 where $y_0$ is the unique solution to the equation
 \be
 \label{eq.equation}
 \mbox{$
 \sum_{i=1}^n {
 \frac{y_0-\mu_i}{\sqrt{(\mu_i-y_0)^2+\sigma_i^2}}}=n-2.
 $}
 \ee
 The equality in (\ref{eq.BNT}) is attained by the extremely dependent
 random vector with
 \[
 \Pr[X_1=y_0-\alpha_1,
 \ldots,X_j=y_0+\alpha_j,\ldots,
 X_n=y_0-\alpha_n]=p_j, \ \ j=1,\ldots,n,
 \]
 where
 \[
 \mbox{$
 \alpha_j=\sqrt{(\mu_j-y_0)^2+\sigma_j^2},
 \ \
 p_j=\frac{1}{2}\Big(1-\frac{y_0-\mu_j}{\sqrt{(\mu_j-y_0)^2+\sigma_j^2}}\Big),
 \ \ j=1,\ldots,n.
 $}
 \]
 Note that $p_j>0$ and, by (\ref{eq.equation}), $\sum_{j=1}^n p_j=1$.

 In the present work we extend the techniques
 of Lai and Robbins (1976) and of
 Bertsimas, Natarajan and Teo (2006), in order to obtain the best
 possible upper bound for the expected range. Also, we characterize
 the {\it extremal random vectors}, i.e.\
 the vectors that attain the equality in the bound, and we
 provide simple conditions (on $\mu_i$ and $\sigma_i$)
 under which the $AG_n$ bound of (\ref{eq.AG}) is already sharp.
 The main result is given in Theorem \ref{theo.6.1}.
 Particular cases of interest are presented as examples.

 \section{An upper bound for the expected range}
 \label{sec.main.inequality}
 \setcounter{equation}{0}
 Let ${\bXX}=(X_1,\ldots,X_n)$
 be an arbitrary random vector
 with $\E {\bXX}=\bmu:=(\mu_1,\ldots,\mu_n)$
 and
 \\
 $(\Var X_1,\ldots,\Var X_n)=(\sigma_1^2,\ldots,\sigma_n^2)$
 where $0<\sigma_i<\infty$ for all $i$.
 For notational simplicity
 we write
 ${\bsigma}=(\sigma_1,\ldots,\sigma_n)$,
 ${\bsigma}^2=(\sigma_1^2,\ldots,\sigma_n^2)$
 and $\Var{\bXX}={\bsigma}^2$; that is,
 $\Var {\bXX}:=\mbox{diag}({\bSigma})$
 where ${\bSigma}$
 is the dispersion matrix of
 ${\bXX}$.
 The class of random vectors satisfying the above moment requirements
 will be denoted by
 \be
 \label{eq.F}
 {\cal F}_n(\bmu,{\bsigma}):=\{{\bXX}:\E{\bXX}=\bmu,
 \Var{\bXX}={\bsigma}^2\}.
 \ee
 In particular, $X\in{\cal F}_1(\mu,\sigma)$ means that $\E X=\mu$ and
 $\Var X=\sigma^2$.

 Let $X_{1:n}\leq \cdots \leq X_{n:n}$ be the order statistics
 corresponding to ${\bXX}$ and set $R_n=X_{n:n}-X_{1:n}$ for the range.
 Our main interest is in calculating
 \be
 \label{eq.problem}
 \inf_{{\footnotesize{\bXX}}
 \in{\cal F}_n({\footnotesize\bmu},
 {\footnotesize{\bsigma}})} \E R_n,
 \ \ \ \ \ \
 \sup_{{\footnotesize{\bXX}}\in{\cal F}_n({\footnotesize\bmu},
 {\footnotesize{\bsigma}})} \E R_n,
 \ee
 for any given $\bmu\in\R^n$ and ${\bsigma}\in\R_+^n$.
 However, the result is known for the infimum:
 \[
 \inf_{{\footnotesize{\bXX}}
 \in{\cal F}_n({\footnotesize\bmu},
 {\footnotesize{\bsigma}})}
 \E R_n
 = \max_i \{\mu_i\} -\min_i \{\mu_i\}.
 \]
 Indeed, since  $R_n=R_n({\bXX})$ is a convex function of ${\bXX}$ we have
 $\E R_n({\bXX})\geq R_n(\bmu)=\max_i\{\mu_i\}-\min_i\{\mu_i\}$
 from Jensen's inequality.
 Bertsimas, Doan, Natarajan and Teo (2010) showed that
 this lower bound is best possible even
 for the narrowed class of random vectors with given mean vector
 $\bmu$ and (any) given non-negative defined
 dispersion matrix \mbox{${\bSigma}$\hspace{.3ex}.}
 For
 clarity of the presentation
 we provide here the construction
 of
 Bertsimas, Doan, Natarajan and Teo (2010).
 Define
 \[
 {\bXX}_{\epsilon}=\bmu+\frac{I_{\epsilon}}{\sqrt{\epsilon}}
 \,
 \bVV 
 \,
 {\bSigma}^{1/2}, \ \ \
 0<\epsilon<1,
 \]
 where
 ${\bVV}=(V_1,\ldots,V_n)$ with $V_i$ being i.i.d.\
 with zero mean and variance one
 and $I_{\epsilon}$ is a
 Bernoulli random
 variable, independent of ${\bVV}$, with probability of
 success equal to $\epsilon$.
 Then it is easy to verify that for all $\epsilon\in(0,1)$,
 ${\bXX}_\epsilon$ has mean $\bmu$
 and dispersion matrix \mbox{${\bSigma}$\hspace{.3ex}.}
 Let
 $A\subseteq \R^n$ be the finite collection of
 vectors of the form
 $\bbee(i)-\bbee(j)$, $i\neq j$,
 $i,j\in\{1,\ldots,n\}$, where ${\bbee}(i)=(0,\ldots,1,\ldots,0)$
 is the unitary vector of the $i$-th axis.
 With $\bx^{\mbox{t}}$ denoting the transpose
 of any $1\times n$ random vector $\bx$ we have
 \[
 \mbox{$
 R_n({\bXX}_{\epsilon})
 =\max_{\bbalpha\in A}\{
 \balpha{\bXX}_{\epsilon}^{\mbox{t}}\}
 \leq
 \max_{\bbalpha\in A}\{ \balpha\bmu^{\mbox{t}}\}
 + \frac{I_{\epsilon}}{\sqrt{\epsilon}}
 \max_{\bbalpha\in A}\{ \balpha\ {\bSigma}^{1/2}{\bVV}^{\mbox{t}}\}.
 $}
 \]
 Clearly, $\max_{\bbalpha\in A}\{ \balpha\bmu^{\mbox{t}}\}
 =\max_i\{\mu_i\}-\min_i\{\mu_i\}$,
 while
 \[
 \mbox{$
 \E\Big(\frac{I_{\epsilon}}{\sqrt{\epsilon}}
 \max_{\bbalpha\in A}\{ \balpha \,
 {\bSigma}^{1/2}{\bVV}^{\mbox{t}}\}\Big)
 =
 \sqrt{\epsilon}\, \E\Big(\max_{\bbalpha\in A}\{
  \balpha
  \,
  {\bSigma}^{1/2}{\bVV}^{\mbox{t}}\}\Big)
 \leq
 \sqrt{\epsilon}\sum_{\bbalpha\in A}
 \E\Big|\balpha
 \,
 {\bSigma}^{1/2}{\bVV}^{\mbox{t}}\Big|
 =\gamma\sqrt{\epsilon},
 $}
 \]
 where $\gamma\geq 0$ is a finite constant independent of $\epsilon$.
 It follows that
 \[
 \E R_{n}({\bXX}_\epsilon)\leq
 \max_i\{\mu_i\}-\min_i\{\mu_i\}+\gamma\sqrt{\epsilon}
 \]
 and thus,
 \[
 \lim_{\epsilon\searrow 0}\E R_n({\bXX}_{\epsilon})
 = \max_i\{\mu_i\}-\min_i\{\mu_i\}.
 \]
 Hence, the best possible lower bound for $\E R_n$
 is $\max_i\{\mu_i\}-\min_i\{\mu_i\}$.

 Regarding the supremum in
 (\ref{eq.problem}), we shall make
 use of the following definition.
 \begin{DEFI}
 \label{def.3.1}
 A random vector ${\bXX}\in {\cal F}_n(\bmu,{\bsigma})$
 of dimension $n\geq 2$
 will be called {\it extremal random vector} (for the range)
 if
 $\E R_n({\bXX})=\sup \E R_n$, where the supremum is
 taken over ${\cal F}_n(\bmu,{\bsigma})$.
 The class of extremal random vectors is denoted
 by ${\cal E}_n(\bmu,{\bsigma})$.
 \end{DEFI}

 To the best of our knowledge,
 the value of the supremum and the nature of the
 set ${\cal E}_n(\bmu,{\bsigma})$
 have not been analysed elsewhere; it is not even
 known whether ${\cal E}_n(\bmu,{\bsigma})$
 in nonempty for general $\bmu$ and ${\bsigma}$.
 In the present article we shall address both issues.

 We start with a deterministic inequality
 which is the range analogue of the inequality
 given by Lai and Robbins (1976):
 \begin{LEM}
 \label{lem.1}
 For any ${\bXX}\in \R^n$, $c\in\R$ and $\lambda>0$,
 \be
 \label{eq.deterministic}
 \mbox{$
  R_n\leq -(n-2)\lambda+\frac{\lambda}{2}\sum_{i=1}^n
  \left\{\left|\frac{X_i-c}{\lambda}-1\right|
  +\left|\frac{X_i-c}{\lambda}+1\right|\right\}.
  $}
 \ee
 The equality in (\ref{eq.deterministic}) is attained
 if and only if
 \be
 \label{eq.equality}
 X_{1:n}\leq c-\lambda\leq X_{2:n}\leq \cdots \leq X_{n-1:n}
 \leq c+\lambda\leq X_{n:n}.
 \ee
 \end{LEM}

 \noindent
 The Lemma entails that the use of two decision variables
 is sufficient for properly handling $R_n$. Also, it
 suggests the investigation of
 $\sup\E\big\{|X-1|+|X+1|\big\}$ when $X$
 is a random variable with given mean and variance:
 \begin{LEM}
 \label{lem.2}
 For any $X\in{\cal F}_1(\mu,\sigma)$
 ($0<\sigma<\infty$),
 \be
 \label{eq.in}
 \E\big\{|X-1|+|X+1|\big\}\leq U(\mu,\sigma),
 \ee
 where
 \be
 \label{eq.u}
 U(\mu,\sigma):=\left\{
 \begin{array}{lll}
 2\sqrt{\mu^2+\sigma^2}, \medskip & \mbox{if}
 & \mu^2+\sigma^2\geq 4,
 \\
 2+\frac{1}{2}(\mu^2+\sigma^2), \medskip
 & \mbox{if} & 2|\mu|<\mu^2+\sigma^2< 4,
 \\
 |\mu|+1+\sqrt{(|\mu|-1)^2+\sigma^2},
 & \mbox{if} & \mu^2+\sigma^2\leq 2|\mu|< 4.
 \end{array}
 \right.
 \ee
 The equality in (\ref{eq.in}) is attained by a unique random
 variable $X^*\in{\cal F}_1(\mu,\sigma)$. Depending on
 $(\mu,\sigma)$, $X^*$ assumes two or three
  supporting values.
 More
 precisely:

 \noindent
 (a) For $\mu^2+\sigma^2\geq 4$,
 \[
 \mbox{$
 \Pr\big[X^*=\sqrt{\mu^2+\sigma^2}\big]=\frac{1}{2}
 \Big(1+\frac{\mu}{\sqrt{\mu^2+\sigma^2}}\Big)
 =1- \Pr\big[X^*=-\sqrt{\mu^2+\sigma^2}\big].
 $}
 \]

 \noindent
 (b) For $2|\mu|<\mu^2+\sigma^2< 4$,
 \[
 \mbox{$
 \Pr[X^*=0]=1-\frac{\mu^2+\sigma^2}{4}, \ \
 \Pr[X^*=-2]=\frac{\mu^2+\sigma^2-2\mu}{8}, \ \
 \Pr[X^*=2]=\frac{\mu^2+\sigma^2+2\mu}{8}.
 $}
 \]

 \noindent
 (c) For $\mu^2+\sigma^2\leq 2\mu$ (and hence, $0<\mu<2$),
 \[
 \mbox{$
 \Pr\big[X^*=1+\sqrt{(\mu-1)^2+\sigma^2}\big]=\frac{1}{2}
 \Big(1+\frac{\mu-1}{\sqrt{(\mu-1)^2+\sigma^2}}\Big)
 =1- \Pr\big[X^*=1-\sqrt{(\mu-1)^2+\sigma^2}\big].
 $}
 \]

 \noindent
 (d) For $\mu^2+\sigma^2\leq -2\mu$ (and hence, $-2<\mu<0$),
 \[
 \mbox{$
 \Pr\big[X^*=-1+\sqrt{(\mu+1)^2+\sigma^2}\big]=\frac{1}{2}
 \Big(1+\frac{\mu+1}{\sqrt{(\mu+1)^2+\sigma^2}}\Big)
 =1- \Pr\big[X^*=-1-\sqrt{(\mu+1)^2+\sigma^2}\big].
 $}
 \]
 \end{LEM}

 \begin{REM}
 Isii (1963) presented general results
 that include
 inequalities of the form of Lemma \ref{lem.2}; see also
 Karlin and Studden (1966).
 The univariate mean-variance
 inequality in Isii's paper can be stated as
 follows: If $h:\R\to\R$ is a Borel function, $\mu\in\R$ and $\sigma>0$ then
 \[
 \sup_{X\in{\cal F}_1(\mu,\sigma)}
 \E h(X) =
 \inf_{\alpha_0,\alpha_1,\alpha_2}
 \big\{\alpha_0+\alpha_1\mu+\alpha_2 (\mu^2+\sigma^2):\alpha_0+\alpha_1 x
 +\alpha_2 x^2\geq h(x)  \mbox{ for all } x\big\}.
 \]
 Isii showed that the above infimum is attained by some
 ${\balpha}^*=(\alpha_0^*,\alpha_1^*,\alpha_2^*)\in A$,
 where
 \[
  A=\{(\alpha_0,\alpha_1,\alpha_2):\alpha_0+\alpha_1 x
 +\alpha_2 x^2\geq h(x)  \mbox{ for all } x\in\R\}\subseteq \R^3,
 \]
 provided that the infimum is finite.
 However, usually it is not an easy task to specify the subset $A$
 and the extremal point(s) ${\balpha}^*$. Lemma \ref{lem.2}
 shows that this is possible for $h(x)=|x-1|+|x+1|$ and,
 more importantly, characterizes the case of equality.
 \end{REM}

 \noindent
 The following corollary is a straightforward consequence of
 Lemma \ref{lem.2}.
 \begin{COR}
 \label{cor.1}
 Let $X\in{\cal F}_1(\mu,\sigma)$
 ($0<\sigma<\infty$).
 Fix $c\in\R$ and $\lambda>0$.
 Then,
 \be
 \label{eq.in22}
 \mbox{$
 \E\big\{|(X-c)-\lambda|+|(X-c)+\lambda|\big\}\leq \lambda
 \,
 U\big(\frac{\mu-c}{\lambda},\frac{\sigma}{\lambda}\big),
 $}
 \ee
 with
 $U(\cdot,\cdot)$ given by (\ref{eq.u}).
 The equality in (\ref{eq.in22}) is attained
 by a unique two or three-valued random variable.
 Setting
 \[
 \mbox{
 $\xi=\mu-c$,
 \ \
 $\theta=\sqrt{(\mu-c)^2+\sigma^2}$,
 \ \
 $\alpha=\sqrt{(\xi-\lambda)^2+\sigma^2}$,
 \ \
 $\beta=\sqrt{(\xi+\lambda)^2+\sigma^2}$,
 }
 \]
 the distribution that attains the equality is described
 by the
 following table:
 \bigskip

 \noindent
 {\small
 \begin{tabular}{l|ccc}
 \relax
 \begin{tabular}{l}
 No: \ \  Condition on $\mu$, $\sigma$, $c$, $\lambda$
 \\
 Tight Upper Bound $\lambda \, U\big(\frac{\mu-c}{\lambda},
 \frac{\sigma}{\lambda}\big)$
 \end{tabular}
 &
 \begin{tabular}{c} value $x^{-}$ \\
 probability $p^{-}$
 \end{tabular}
 &
 \begin{tabular}{c} value $x^{o}$ \\
 probability
 $p^{o}$
 \end{tabular}
 &
 \begin{tabular}{c} value $x^{+}$ \\
 probability
 $p^{+}$
 \end{tabular}
 \relax
 \\
 \hline
 \hline
 \relax
 \begin{tabular}{c}
 {\bf 1}: \ \ $(\mu-c)^2+\sigma^2\geq 4\lambda^2$
 \vspace{.5ex}
 \\
 \hspace{5ex}
 $2\sqrt{(\mu-c)^2+\sigma^2}$
 \end{tabular}
 &
 \begin{tabular}{c}
 $c-\theta$
 \\
 $\frac{1}{2}\big(1-\frac{\xi}{\theta}\big)$
 \end{tabular}
 &
 \begin{tabular}{c}
 \\
 \end{tabular}
 &
 \begin{tabular}{c}
 $c+\theta$
 \\
 $\frac{1}{2}\big(1+\frac{\xi}{\theta}\big)$
 \end{tabular}
 \vspace{1ex}
 \\
 \hline
 \relax
 \begin{tabular}{l}
 {\bf 2}: \ \  $2\lambda|\mu-c|<(\mu-c)^2+\sigma^2< 4\lambda^2$
 \vspace{.5ex}
 \\
 \hspace{5ex}
 $2\lambda+\frac{1}{2\lambda}\left[(\mu-c)^2+\sigma^2\right]$
 \end{tabular}
 &
 \begin{tabular}{c}
 $c-2\lambda$
 \\
 $\frac{1}{8\lambda^2}\left[\theta^2-2\lambda\xi\right]$
 \end{tabular}
 &
 \begin{tabular}{c}
 $c$
 \\
 $1-\frac{1}{4\lambda^2}\theta^2$
 \end{tabular}
 \vspace{1ex}
 &
 \begin{tabular}{c}
 $c+2\lambda$
 \\
 $\frac{1}{8\lambda^2}\left[\theta^2+2\lambda\xi\right]$
 \end{tabular}
 \\
 \hline
 \relax
 \begin{tabular}{l}
 {\bf 3}: \ \  $(\mu-c)^2+\sigma^2\leq2\lambda (\mu-c)$
 \vspace{.5ex}
 \\
 \hspace{3ex}
 $\mu-c+\lambda+\sqrt{(\mu-c-\lambda)^2+\sigma^2}$
 \end{tabular}
 &
 \begin{tabular}{c}
 \\
 \end{tabular}
 &\begin{tabular}{c}
 $c+\lambda-\alpha$
 \\
 $\frac{1}{2}\big(1-\frac{\xi-\lambda}
 {\alpha}\big)$
 \end{tabular}
 \vspace{1ex}
 &
 \begin{tabular}{c}
 $c+\lambda+\alpha$
 \\
 $\frac{1}{2}\big(1+\frac{\xi-\lambda}
 {\alpha}\big)$
 \end{tabular}
 \\
 \hline
 \relax
 \begin{tabular}{l}
 {\bf 4}: \ \ $(\mu-c)^2+\sigma^2\leq2\lambda (c-\mu)$
 \vspace{.5ex}
 \\
 \hspace{3ex}
 $c-\mu+\lambda+\sqrt{(c-\mu-\lambda)^2+\sigma^2}$
 \end{tabular}
 &
 \begin{tabular}{c}
 $c-\lambda-\beta$
 \\
 $\frac{1}{2}\big(1-\frac{\xi+\lambda}
 {\beta}\big)$
 \end{tabular}
 &
 \begin{tabular}{c}
 $c-\lambda+\beta$
 \\
 $\frac{1}{2}\big(1+\frac{\xi+\lambda}
 {\beta}\big)$
 \end{tabular}
 &
 \begin{tabular}{c}
 \\
 \end{tabular}
 \end{tabular}
 }

 \end{COR}
 \noindent
 \begin{Proof} Write $|(X-c)-\lambda|+|(X-c)+\lambda|
 =\lambda\Big\{\big|\frac{X-c}{\lambda}-1\big|
 +\big|\frac{X-c}{\lambda}+1\big|\Big\}$.
 Since $Y=\frac{X-c}{\lambda}\in{\cal F}_1\big(\frac{\mu-c}{\lambda},
 \frac{\sigma}{\lambda}\big)$,
 Lemma \ref{lem.2}
 yields (\ref{eq.in22})
 as follows:
 \[
 \mbox{$
 \E\big\{|(X-c)-\lambda|+|(X-c)+\lambda|\big\}=\lambda
 \E\big\{|Y-1|+|Y+1|\big\}\leq
 \lambda
 \,
 U\big(\frac{\mu-c}{\lambda},\frac{\sigma}{\lambda}\big).
 $}
 \]
 Since $\lambda>0$, Lemma \ref{lem.2}
 asserts that the equality is attained by a unique random variable
 $Y^{*}\in{\cal F}_1\big(\frac{\mu-c}{\lambda}, \frac{\sigma}{\lambda}\big)$.
 Thus, $X^*=c+\lambda Y^*$ is
 the unique random variable in ${\cal F}_1(\mu,\sigma)$
 that attains the equality in (\ref{eq.in22}). Substituting the
 probability function
 of $Y^*$ in the four distinct cases of Lemma \ref{lem.2} we obtain
 the probabilities and supporting points as in the table.
 $\Box$
 \medskip
 \end{Proof}

 \noindent
 It is important to observe that, whatever the values of
 $\mu,\sigma,c,\lambda$ are,
 the supporting points satisfy the relation
 $x^{-}<c-\lambda<x^{o}<c+\lambda<x^{+}$.

 We can now obtain the proposed upper bound for the expected range.
 \begin{THEO}
 \label{th.main}
 If $\E {\bXX}=\bmu$ and $\Var {\bXX}={\bsigma}^2$ then
 \be
 \label{eq.main}
 \E R_n\leq \inf_{c\in\R,\lambda>0}
 \Big\{-(n-2)\lambda
 +\frac{\lambda}{2}\sum_{i=1}^n
 U\Big(\frac{\mu_i-c}{\lambda},\frac{\sigma_i}{\lambda}\Big)\Big\},
 \ee
 where the function $U(\cdot,\cdot):\R\times(0,\infty)\to(2,\infty)$ is given by (\ref{eq.u}).
 \end{THEO}
 \noindent
 \begin{Proof}
 Fix $c\in\R$, $\lambda>0$. We take expectations in
 (\ref{eq.deterministic}) and then use (\ref{eq.in22})
 to get
 \begin{eqnarray*}
 \E R_n
 \hspace{-1ex}
 & \leq &
 \hspace{-1ex}
 \mbox{$
 -(n-2)\lambda + \frac{1}{2}\sum_{i=1}^n
 \E\big\{|(X_i-c)-\lambda|+|(X_i-c)+\lambda|\big\}
 $}
 \\
 \hspace{-1ex}
 &\leq &
 \hspace{-1ex}
 \mbox{$
 -(n-2)\lambda + \frac{\lambda}{2}\sum_{i=1}^n
 U\big(\frac{\mu_i-c}{\lambda},\frac{\sigma_i}{\lambda}\big).
 $}
 \end{eqnarray*}
 Since for all $c\in\R$
 and $\lambda>0$ the last quantity is an upper bound for $\E R_n$,
 its infimum is an upper bound too.
 $\Box$
 \medskip
 \end{Proof}
 \begin{REM}
 \label{rem.2}
 It is not clear at this stage whether
 the upper bound (\ref{eq.main}) is tight, and it is not
 an obvious task to find $c=c_0$ and $\lambda=\lambda_0$ (if exist) that
 realize the infimum in the RHS of (\ref{eq.main}). However,
 the substitution of any (convenient)
 arguments $c$
 and $\lambda$
 in the function
 \be
 \label{eq.phi}
 \mbox{$
 \phi_n(c,\lambda):=-(n-2)\lambda + \frac{\lambda}{2}\sum_{i=1}^n
 U\big(\frac{\mu_i-c}{\lambda},\frac{\sigma_i}{\lambda}\big)
 $}
 \ee
 will produce an upper bound for $\E R_n$. For example,
 one can choose $c=\overline{\mu}$ and $\lambda=\frac{1}{4}AG_n$
 (see (\ref{eq.AG})).
 A simple way to produce a closed-form upper bound is the following:
 First observe that
 \[
 \mbox{$
 \lambda
 \,
 U\Big(\frac{\mu_i-c}{\lambda},\frac{\sigma_i}{\lambda}\Big)
 \leq 2\lambda+\frac{1}{2\lambda}\big[(\mu_i-c)^2+\sigma_i^2\big],
 $}
 \]
 because the RHS is an upper bound for the expectation
 $\E\big\{|(X_i-c)-\lambda|+|(X_i-c)+\lambda|\big\}$ (since
 $|(X_i-c)-\lambda|+|(X_i-c)+\lambda|\leq 2\lambda+\frac{1}{2\lambda}(X_i-c)^2$
 and $\E\big\{2\lambda+\frac{1}{2\lambda}(X_i-c)^2\big\}
 =2\lambda+\frac{1}{2\lambda}[(\mu_i-c)^2+\sigma_i^2]$),
 while the LHS
 is the least upper bound for the same expectation as
 $X_i$ varies in
 ${\cal F}_1(\mu_i,\sigma_i)$. It follows that
 \[
 \mbox{$
 \phi_n(c,\lambda)\leq \overline{\phi}_n(c,\lambda)
 :=2\lambda+\frac{1}{4\lambda}\sum_{i=1}^n\big\{(\mu_i-c)^2+\sigma_i^2\big\}.
 $}
 \]
 Minimizing $\overline{\phi}_n(c,\lambda)$ is a simple fact: it suffices
 to  take $c=\overline{\mu}$ and $\lambda=\frac{1}{4}AG_n$ as before.
 Observing that
 $\sum_{i=i}^n\big\{(\mu_i-\overline{\mu})^2+\sigma_i^2\big\}=\frac{1}{2}AG_n^2$
 we get
 \[
 \E R_n \leq
 \inf_{c\in\R, \lambda>0}\phi_n(c,\lambda)\leq \inf_{c\in\R, \lambda>0}
 \overline{\phi}_n(c,\lambda)=
 \overline{\phi}_n\Big(\overline{\mu},\frac{1}{4}AG_n\Big)
 =\frac{1}{2}AG_n+\frac{1}{2}AG_n=AG_n.
 \]
 Now it became clear that the bound in (\ref{eq.main}) is reasonable, since it
 outperforms the bound in (\ref{eq.AG}) for any given values of $\bmu$ and
 ${\bsigma}$. As a result, the $AG_n$ bound need no be tight;
 e.g., the infimum of $\phi_n(c,\lambda)$ need no be attained at
 $(c,\lambda)=\big(\overline{\mu},\frac{1}{4}AG_n\big)$.
 We shall prove in the sequel that the new bound is always tight,
 and (for $n\geq 3$) the infimum in the RHS of (\ref{eq.main}) is attained
 by a unique value $(c_0,\lambda_0)$.
 \end{REM}
 \begin{REM}
 \label{rem.2.3}
 Fixing $\mu$ in (\ref{eq.u}) and taking limits for $\sigma\searrow 0$ we see that
 \[
 \lim_{\sigma\searrow 0} U(\mu,\sigma)=2\max\{|\mu|,1\}
 =|\mu-1|+|\mu+1|, \ \ \mu\in\R.
 \]
 Let us now set $\sigma_{n:n}=\max\{\sigma_1,\ldots,\sigma_n\}$
 and fix $\bmu=(\mu_1,\ldots,\mu_n)$. Then,
 \[
 \lim_{\sigma_{n:n} \searrow 0} \phi_n(c,\lambda)
 =-(n-2)\lambda +\frac{\lambda}{2}\sum_{i=1}^n
 \Big\{\Big|\frac{\mu_i-c}{\lambda}-1\Big|
 +\Big|\frac{\mu_i-c}{\lambda}+1\Big|\Big\}, \ \
 \bmu\in\R^n, \ c\in\R, \ \lambda>0.
 \]
 Let $\mu_{1:n}\leq \cdots \leq \mu_{n:n}$ be the ordered values of
 $\mu_1,\ldots,\mu_n$, and assume that
 the $\mu$'s are not all equal, that is,
 $\mu_{1:n}<\mu_{n:n}$.
 Substituting in the above limit
 $c=c_0=\frac{\mu_{1:n}+\mu_{n:n}}{2}$,
 $\lambda=\lambda_0=\frac{\mu_{n:n}-\mu_{1:n}}{2}>0$,
 we obtain
 \[
 \lim_{\sigma_{n:n} \searrow 0} \phi_n(c_0,\lambda_0)
 =-(n-2)\lambda_0 +\frac{\lambda_0}{2}\sum_{i=1}^n
 \Big\{\Big|\frac{\mu_i-c_0}{\lambda_0}-1\Big|
 +\Big|\frac{\mu_i-c_0}{\lambda_0}+1\Big|\Big\}
 =\mu_{n:n}-\mu_{1:n}.
 \]
 Note that the last equality follows from (\ref{eq.deterministic}) and
 (\ref{eq.equality}), applied to ${\bXX}=\bmu$
 (with $R_n(\bmu)=\mu_{n:n}-\mu_{1:n}$), observing that
 for the particular choice of $(c_0,\lambda_0)$,
 \[
 \mu_{1:n}\leq c_0-\lambda_0\leq \mu_{2:n}\leq \cdots\leq
 \mu_{n-1:n}\leq c_0+\lambda_0\leq \mu_{n:n}.
 \]
 For any ${\bXX}\in{\cal F}_n(\bmu,{\bsigma})$
 it is true that
 $\mu_{n:n}-\mu_{1:n}\leq \E R_n({\bXX})\leq
 \inf_{c\in\R,\lambda>0}\big\{\phi_n(c,\lambda)\big\}$.
 Therefore,
 \begin{eqnarray*}
 \mu_{n:n}-\mu_{1:n}
 \leq
 \lim_{\sigma_{n:n}\searrow 0} \E R_n({\bXX})
 \leq
 \lim_{\sigma_{n:n}\searrow 0}
 \Big\{\inf_{c\in\R,\lambda>0}\phi_n(c,\lambda) \Big\}
 \leq \lim_{\sigma_{n:n}\searrow 0} \phi_n(c_0,\lambda_0)
 =\mu_{n:n}-\mu_{1:n},
 \end{eqnarray*}
 and we conclude that
 \be
 \label{eq.2.20}
 \lim_{\sigma_{n:n}\searrow 0} \Big\{\inf_{c\in\R,\lambda>0}\phi_n(c,\lambda) \Big\}
 =\mu_{n:n}-\mu_{1:n}.
 \ee
 The limit (\ref{eq.2.20}) continue to hold even if all
 $\mu_i$'s are equal. Then
  $\mu_{1:n}=\mu_{n:n}$ and
 the inequality
 $
 \inf_{c\in\R,\lambda>0}\big\{\phi_n(c,\lambda)\big\}\leq AG_n$
 (see Remark \ref{rem.2}) shows that
 \[
 \mbox{$
 0
 \leq
 \inf_{c\in\R,\lambda>0}
 \big\{\phi_n(c,\lambda)\big\}
 \leq AG_n=\sqrt{2\sum_{i=1}^n\sigma_i^2}\leq
 \sigma_{n:n}\sqrt{2n}\to 0, \ \  \mbox{ as }
 \sigma_{n:n}\searrow 0.
 $}
 \]
 From these considerations it is again clear that the $AG_n$
 bound
 is not tight in general; for example,
 \[
 \lim_{\sigma_{n:n}\searrow 0} AG_n=
 \mbox{$
 \sqrt{2\sum_{i=1}^n(\mu_i-\overline{\mu})^2} >\mu_{n:n}-\mu_{1:n}
 $}
 \]
 whenever ($n\geq 3$ and) $\mu_{1:n}+\mu_{n:n}\neq 2\overline{\mu}$.
 The $AG_n$ bound need no be tight even for equal $\mu_i$'s; see Theorem
 \ref{theo.AG} and Example
 \ref{ex.2}, below.
 \end{REM}


 \section{When is the Arnold-Groeneveld bound tight?}
 \label{sec.AG}
 \setcounter{equation}{0}
 Arnold and Groeneveld (1979), Rychlik (1993b) and Papadatos (2001a)
 showed that if $\mu_i=\mu$ and $\sigma_i=\sigma$
 for all $i$, the $AG_n$ bound of
 (\ref{eq.AG}), which reduces to (\ref{eq.Rychlik}), is
 attainable.
 In the present section we provide
 an exact characterization of the attainability
 of the $AG_n$ bound
 under any mean-variance information.

 The proof of Theorem \ref{theo.AG}, below, is based on the construction of particular
 bivariate probability distributions supported in a subset of
 $\{1,\ldots,n\}^2$.
 A distribution of this kind corresponds to a
 $n\times n$  matrix with nonnegative
 elements having sum $1$; a probability matrix.
 Matrices of this form with integer-valued entries
 have been extensively studied; for a recent review
 see Barvinok (2012).
 The actual question, related to our problem, is
 whether there exist probability
 matrices with given marginals and vanishing trace.

 The following notation and terminology will be used in the sequel.
 \begin{DEFI}
 \label{def.3.2}
 A $n\times m$ matrix $Q=(q_{ij})$ ($n\geq 1$, $m\geq 1$)
 is called {\it a probability matrix} if it has nonnegative elements
 summing to $1$.
 In particular, a $n$-variate {\it probability vector}
 ${\bpp}=(p_1,\ldots,p_n)$ is
 a probability matrix with dimension $1\times n$, and
 $X\sim {\bpp}$ is a convention for
 $\Pr[X=i]=p_i$ for all $i$.
 The {\it marginals} of $Q$, say ${\bpp}$, ${\bqq}$,
 are the probability vectors obtained by summing
 the rows and columns of $Q$, respectively; and
 ${\cal M}({\bpp},{\bqq})$ denotes the class of probability
 matrices with given marginals ${\bpp}$, ${\bqq}$.
 Moreover, $(X,Y)\sim Q$ is a convention for
 $\Pr[X=i,Y=j]=q_{ij}$ for all $i,j$.
 \end{DEFI}

 We now state a characterization for the $AG_n$ bound.
 \begin{THEO}
 \label{theo.AG}
 Assume that $\E {\bXX}=\bmu$ and $\Var{\bXX}={\bsigma}^2$.
 Then the equality in (\ref{eq.AG}) is attainable if and only if
 both conditions (i) and (ii) below are satisfied.
 \be
 \label{eq.3.1}
 \begin{array}{ll}
 \mbox{(i)}
 \hspace*{3ex}
 &
 \displaystyle
 |\mu_i-\overline{\mu}|
 \leq
 \displaystyle
 \frac{\sqrt{2}\big[(\mu_i-\overline{\mu})^2+\sigma_i^2\big]}
 {\sqrt{\sum_{j=1}^n
 \big\{(\mu_j-\overline{\mu})^2+\sigma_j^2\big\}}},
 \vspace{1ex}
 \\
 \mbox{(ii)} \hspace*{3ex}
 &
 \displaystyle
 \frac{(\mu_i-\overline{\mu})^2+\sigma_i^2}{\sum_{j=1}^n
 \big\{(\mu_j-\overline{\mu})^2+\sigma_j^2\big\}}
 \leq
 \frac{1}{2},
 \end{array}
 \hspace{3ex}
 i=1,\ldots,n.
 \ee
 Provided that (i) and (ii) are fulfilled,
 any extremal random vector
 ${\bXX}\in{\cal E}_n(\bmu,{\bsigma})$
 has the representation
 \be
 \label{eq.extremal}
 \mbox{$
 {\bXX}=g(X,Y):=\overline{{\mu}}\, \bone
 +{\displaystyle\frac{{\bbee}(X)-{\bbee}(Y)}{\sqrt{2}}}
 \sqrt{\sum_{j=1}^n
 \big\{(\mu_j-\overline{\mu})^2+\sigma_j^2\big\}},
 $}
 \ee
 where ${\bone}=(1,\ldots,1)\in\R^n$,
 ${\bbee}(i)=(0,\ldots,1,\ldots,0)$,
 and $(X,Y)$ is a discrete random pair
 satisfying $\Pr[X=Y]=0$, with
 marginal distributions
 \be
 \label{eq.marginals}
 \begin{array}{l}
 \displaystyle
 p_i^{+}
 =\Pr[X=i]
 =
 \frac{(\mu_i-\overline{\mu})^2
 +\sigma_i^2+\frac{1}{2}(\mu_i-\overline{\mu})AG_n}{\sum_{j=1}^n
 \big\{(\mu_j-\overline{\mu})^2+\sigma_j^2\big\}},
 \vspace{1ex}
 \\
 \displaystyle
 p_i^{-}=
 \Pr[Y=i]
 =
 \frac{(\mu_i-\overline{\mu})^2
 +\sigma_i^2-\frac{1}{2}(\mu_i-\overline{\mu})AG_n}{\sum_{j=1}^n
 \big\{(\mu_j-\overline{\mu})^2+\sigma_j^2\big\}},
 \end{array}
 \hspace{3ex} i=1,\ldots,n.
 \ee
 Moreover, if the inequalities in (\ref{eq.3.1}) are strict for all
 $i$, we can find infinitely many extremal
 random vectors; and if (\ref{eq.3.1}) is satisfied
 and for some
 $i$
 we have equality in (ii), then the extremal random vector 
 is unique.
 \end{THEO}
 \begin{REM}
 \label{rem.3.1}
 Let $(\mu_1,\mu_2,\mu_3)=(-1,0,1)$,
 $(\sigma_1^2,\sigma_2^2,\sigma_3^2)=(1,3,2)$, so that
 (\ref{eq.3.1}) holds. However,
 (\ref{eq.3.1})(ii) is satisfied with strict
 inequalities for all $i$, while this is not true for
 (\ref{eq.3.1})(i).
 We find $AG_3=4$ and
 ${\bpp}^+=\big(0,\frac{3}{8}, \frac{5}{8}\big)$,
 ${\bpp}^-=\big(\frac{1}{2},\frac{3}{8}, \frac{1}{8}\big)$.
 It is easily seen that the distribution of $(X_1,X_2,X_3)$
 (given in (\ref{eq.extremal}))
 is uniquely defined: it assigns probabilities $\frac{2}{8},
 \frac{2}{8},\frac{3}{8},\frac{1}{8}$, to the points
 $(-2,2,0)$, $(-2,0,2)$, $(0,-2,2)$, $(0,2,-2)$, respectively.
 It follows that a random vector that attains
 the $AG_n$ bound
 can be unique even if (\ref{eq.3.1})(ii) is satisfied
 with strict
 inequalities
 for all $i$.
 \end{REM}
  \begin{EXAMPLE}
 \label{ex.1} {\it The homogeneous case $\mu_i=\mu$, $\sigma_i=\sigma>0$.} Conditions
 (\ref{eq.3.1}) are obviously satisfied with strict inequalities
 (for $n\geq 3$) and the $AG_n$ bound is sharp (see also (\ref{eq.Rychlik})):
 \[
 \sup \E R_n=AG_n=\sigma\sqrt{2 n}.
 \]
 Moreover, $p_i^+=p_i^-=\frac{1}{n}$ and
 from Theorem \ref{theo.AG} we see that
 infinitely many random vectors
 attain the equality. The totality of them is characterized
 by (\ref{eq.extremal}) via the probability matrices $Q$ of $(X,Y)$.
 Recall that $X$ and $Y$ are, respectively, the positions where
 $\mu+\sigma\sqrt{n/2}$ and $\mu-\sigma\sqrt{n/2}$ appears in
 the extremal vector $(X_1,\ldots,X_n)$; the rest entries are equal to
 $\mu$. Thus,
 $Q$ has uniform marginals and vanishing principal diagonal.
 A famous theorem of Birkhoff on {\it magic matrices}
 asserts that any matrix with nonnegative elements
 having row/column sums equal to $1$
 is a convex combination of
 {\it permutation matrices},
 i.e., matrices with entries $0$ or $1$, having exactly one $1$
 in each row and in each column; see Theorem 2.54 in
 Giaquinta and Modica (2012).
 From Birkhoff's result it is evident that the probability matrix $Q$ of $(X,Y)$,
 corresponding to any extremal random vector
 ${\bXX}=\mu {\bone}+\sigma[{\bbee}(X)-{\bbee}(Y)]\sqrt{n/2}$,
 can be written as
 \[
 \mbox{$
 Q=\sum_{i=1}^k{\lambda_i}D_i, \ \ \lambda_i\geq0, \ \ \sum_{i=1}^k {\lambda_i}=\frac1n,
 $}
 \]
 where the $D_i$'s are {\it derangement matrices}, i.e.\ permutation matrices with
 vanishing diagonal entries. It is well-known that there exist
 $n!\sum_{k=0}^{n}\frac{(-1)^k}{k!}\approx
 e^{-1}n!$ different derangement matrices; they coincide with
 the extremal points of the convex polytope
 $\big\{D=(d_{ij}):\sum_i d_{ij}=\sum_j d_{ij}=1,
 \ d_{ij}\geq 0, \ d_{ii}=0
 \mbox{ for all }i,j\big\}$.
 In general, a convex polytope has a finite (often quite large) number of
 extremal points, but
 it is rather difficult to evaluate them exactly,
 since their total number depends on the marginals in
 an ambiguous way (cf.\ Example \ref{ex.2}, below).
 \end{EXAMPLE}
 \begin{EXAMPLE}
 \label{ex.2} {\it The case $\mu_i=\mu$.} Assume $0<\sigma_1\leq\cdots\leq \sigma_n$
 without loss of generality. From Theorem \ref{theo.AG} we
 see that if the larger variance does not dominate the sum of the
 other variances then the $AG_n$ bound is tight:
 \[
  \sup \E R_n=AG_n=\sqrt{2\mbox{$\sum_{i=1}^n\sigma_i^2$}}, \ \
  \mbox{ whenever } \ \ \mbox{$\sigma_n^2\leq  \sum_{i=1}^{n-1}\sigma_i^2$}.
 \]
 Moreover, if $\sigma_n^2= \sum_{i=1}^{n-1}\sigma_i^2$, the
 equality is uniquely attained  by the random vector
 ${\bXX}$
 taking values
 \begin{eqnarray*}
 \mbox{$
 {\bxx}_i=\big(\mu,\ldots,\mu,\mu+ \frac{AG_n}{2},\mu,\ldots,\mu;\mu-\frac{AG_n}{2}\big)
 $},
  & \mbox{ with probability }p_i,
 \\
 \mbox{$
 {\byy}_i=\big(\mu,\ldots,\mu,\mu- \frac{AG_n}{2},\mu,\ldots,\mu;\mu+\frac{AG_n}{2}\big)
 $},
 & \mbox{ with probability }p_i,
 \end{eqnarray*}
 where $p_i=\frac{\sigma_i^2}{\sum_{j=1}^n\sigma_j^2}$, $i=1, \ldots, n-1$.
 Of course, if
 $\sigma_n^2< \sum_{i=1}^{n-1}\sigma_i^2$ then there exist
 infinitely many extremal random vectors. They have the form $g(X,Y)$
 (see (\ref{eq.extremal})), with $\Pr[X=Y]=0$, $X\sim{\bpp}$, $Y\sim{\bpp}$,
 where ${\bpp}=(p_1,\ldots,p_{n-1}, p_n)$.

 However, if
 $\sigma_n^2>\sum_{i=1}^{n-1}\sigma_i^2$ then the $AG_n$
 is no longer tight: The infimum
 in (\ref{eq.main}) is attained at $c_0=\mu$,
 $\lambda_0=\frac{1}{2}\sqrt{\sum_{i=1}^{n-1}\sigma_i^2}<\frac14 AG_n$,
 and we get the inequality
 \[
 \mbox{$
 \E R_n \leq \phi_n(c_0,\lambda_0)
 =\sigma_n+\sqrt{\sum_{i=1}^{n-1}\sigma_i^2}
 \ \ \ \ \  \
 \big(\sigma_n^2>\sum_{i=1}^{n-1}\sigma_i^2\big).
 $}
 \]
 From $\sqrt{x}+\sqrt{y}<\sqrt{2(x+y)}$ for
 $x\neq y$
 we conclude that this bound
 is strictly better than $AG_n$. Moreover,
 the new bound is tight; one can verify
 that the equality is (uniquely) attained by the random vector
 ${\bXX}$
 taking values
 \begin{eqnarray*}
  \mbox{$
 {\bxx}_i=\big(\mu,\ldots,\mu,\mu+ \sqrt{\sum_{i=1}^{n-1}\sigma_i^2},\mu,\ldots,\mu;
 \mu-\sigma_n\big)
 $},
  & \mbox{ with probability }p_i,
 \\
 \mbox{$
 {\byy}_i=\big(\mu,\ldots,\mu,\mu-
 \sqrt{\sum_{i=1}^{n-1}\sigma_i^2},\mu,\ldots,\mu;
 \mu+\sigma_n\big)
 $},
 & \mbox{ with probability }p_i,
 \end{eqnarray*}
 where $p_i=\frac{\sigma_i^2}{2\sum_{j=1}^{n-1}\sigma_j^2}$, $i=1, \ldots, n-1$.
 Thus, the tight upper bound on the expected
 range from dependent observations with equal means
 admits a simple closed form:
 \be
 \label{eq.tight}
 \sup \E R_n=\left\{
 \begin{array}{lll}
 \vspace{.5ex}
 \sqrt{2\sum_{i=1}^n \sigma_i^2}, & \mbox{if} & 2\max_i\{\sigma_i^2\}
 \leq \sum_{i=1}^n\sigma_i^2,
 \\
 \max_i\{\sigma_i\} +\sqrt{\sum_{i=1}^{n}\sigma_i^2-\max_i \{\sigma_i^2\}},
 & \mbox{if} & 2\max_i\{\sigma_i^2\}
 \geq \sum_{i=1}^n\sigma_i^2.
 \end{array}
 \right.
 \ee
 Assuming that one variance tends to infinity (and keeping all
 other variances bounded),
 the limit $\lim_{\sigma_i\to\infty}\frac{\sup \E R_n}{AG_n}
 =\frac{1}{\sqrt{2}}\approx .707$ says that we
 can gain of an up to $30\%$ improvement over the $AG_n$ bound.
 \end{EXAMPLE}



 The following lemma will play an important role
 in verifying existence of extremal random vectors.
 \begin{LEM}
 \label{lem.3}
 Let ${\bpp}=(p_1,\ldots,p_n)$ and ${\bqq}=(q_1,\ldots,q_n)$
 be two probability vectors.
 A necessary and sufficient condition for the existence of
 a random pair $(X,Y)$ with
 \be
 \label{eq.bivariate}
 \Pr[X=Y]=0, \  X\sim{\bpp}, \ Y\sim{\bqq}
 \ee
 is the following:
 \be
 \label{eq.condition}
  \max_{1\leq i\leq n}\big\{p_i+q_i\big\}\leq 1.
 \ee
 If the equality holds in (\ref{eq.condition}), the random pair
 $(X,Y)$ is uniquely defined. If strict inequality holds in
 (\ref{eq.condition}) and, furthermore,
 $\min_{i}\{p_i\}>0$, $\min_i\{q_i\}>0$, then
 there exist infinitely many random pairs
 satisfying (\ref{eq.bivariate}).
 \end{LEM}

 \section{Convexity}
 \label{sec.4}
 \setcounter{equation}{0}
 The purpose of the present section
 is to verify that for any given values of
 $\bmu$, ${\bsigma}$,
 the function $\phi_n(c,\lambda)$ of
 (\ref{eq.phi}) is convex.
 For convenience we set $T:=\R\times(0,\infty)$
 for the domain of both functions $U$ (of (\ref{eq.u}))
 and $\phi_n$.

 We begin with a simple lemma.

 \begin{LEM}
 \label{lem.4.1}
 The function $U(x,y):T\to(2,\infty)$ of
 (\ref{eq.u})
 has continuous partial derivatives, that is,
 $U\in C^1(T)$.
 \end{LEM}

 \noindent
 We also need another simple lemma;
 see, e.g., Giaquinta and Modica (2012).
 \begin{LEM}
 \label{lem.4.2}
 Let $K$ be a convex subset of $\R^n$ and $f:K\to \R$.
 For ${\bxx}$ and ${\byy}$ in $K$
 consider the function $g:[0,1]\to \R$
 given by
 \[
 g(t):=f\big({\bxx}+t({\byy}-{\bxx})\big), \ \ 0\leq t\leq 1.
 \]
 Then, $f$ is convex if and only if
 $g$ is convex for any choice of
 ${\bxx}$ and ${\byy}$ in $K$.
 \end{LEM}
 \noindent
 Also, we shall  make use of the following lemma.
 \begin{LEM}
 \label{lem.4.3}
 Consider a finite interval $[\alpha,\beta]$, a partition
 \[
 \alpha=t_0<t_1<\cdots<t_k<t_{k+1}=\beta
 \]
 and the convex functions
 $g_i:[\alpha,\beta]\to\R$ ($i=1,\ldots,k+1$).
 Assume that
 \be
 \label{eq.4.3}
 g_i(t_i)=g_{i+1}(t_i) \  \mbox{ and }  \
 g_i'(t_i-)\leq g'_{i+1}(t_i+),  \ \ i=1,\ldots,k,
 \ee
 where $g'(t-)$ and $g'(t+)$ denote, respectively,
 the left and right hand side derivatives of $g$ at $t$.
 Then, the function
 \be
 \label{eq.4.4}
 g(t):=\left\{
 \begin{array}{cl}
 g_1(t),
 &
 \alpha\leq t\leq t_1,
 \\
 g_2(t),
 &
 t_1\leq t\leq t_2,
 \\
 \vdots
 \\
 g_k(t),
 &
 t_{k-1}\leq t\leq t_k,
 \\
 g_{k+1}(t),
 &
 t_{k}\leq t\leq \beta,
 \end{array}
 \right.
 \ee
 is convex.
 \end{LEM}
 \noindent
 \begin{Proof} Since all $g_i$ have non-decreasing left and right hand side derivatives,
 it is easily seen that
 the same is true for $g$.
 $\Box$
 \medskip
 \end{Proof}

 \noindent
 Now we can verify the following result.
 \begin{PROP}
 \label{prop.4.1}
 The function $U:T\to(2,\infty)$ in (\ref{eq.u}) is convex.
 \end{PROP}

 \noindent
 Finally, we shall make use of the following property,
 which seems to be of some independent interest.
 \begin{LEM}
 \label{lem.4.4}
 Let $f(x,y):T\to\R$ and for fixed $x_0\in\R$, $y_0>0$, consider the function
 $h(c,\lambda):T\to\R$ with
 \[
 \mbox{$
 h(c,\lambda):=\lambda\, f\big(\frac{x_0-c}{\lambda},\frac{y_0}{\lambda}\big),
 \ \ \ (c,\lambda)\in \R\times(0,\infty).
 $}
 \]
 (i) If $f$ is convex then $h$ is convex for all choices of $x_0\in\R$, $y_0>0$.
 \\
 (ii) If $h$ is convex for a particular choice of $x_0\in\R$, $y_0>0$, then $f$ is
 convex.
 \end{LEM}

 \noindent
 We can now state and prove the final conclusion of the present section:
 \begin{THEO}
 \label{th.4.1}
 For any given $\bmu$ and ${\bsigma}$,
 the function $\phi_n(c,\lambda)$ in (\ref{eq.phi}) is convex
 and belongs to $C^1(T)$, $T=\R\times (0,\infty)$.
 \end{THEO}
  \noindent
 \begin{Proof} The fact that $\phi_n\in
 C^1(T)$ follows by an obvious application of Lemma \ref{lem.4.1}.
 Also, the function $U(x,y)$ in (\ref{eq.u})
 is convex by
 Proposition \ref{prop.4.1}. Hence, by Lemma \ref{lem.4.4},
 the same is true for the
 function
 $h_i(c,\lambda)=\frac{1}{2}\lambda\, U\big(\frac{\mu_i-c}{\lambda},
 \frac{\sigma_i}{\lambda}\big)$ ($i=1,\ldots,n)$. Since
 $h(c,\lambda)=-(n-2)\lambda$ is trivially convex,
 $\phi_n(c,\lambda)$ is a sum of convex functions.
 $\Box$
 \end{Proof}

 \section{Attainability of the infimum in (\ref{eq.main}) at a unique point}
 \label{sec.5}
 \setcounter{equation}{0}
 From now on we assume that $n\geq 3$. The simple
 (but interesting)
 case $n=2$
 is deferred to the last section, noting that
 the optimal upper bound for
 $\E R_2$ is closely related to
 the bound $BNT_2$ of (\ref{eq.BNT}).

 In the present section we shall prove that
 the minimum value of
 $\phi_n(c,\lambda)$ is achieved
 at a unique
 point $(c_0,\lambda_0)\in T$.
 Of course, since $\phi_n$ is differentiable,
 a minimizing point
 (if exists) has to satisfy the system
 of equations
 \be
 \label{eq.5.1}
 \frac{\partial}{\partial c}\phi_n(c,\lambda)=0,
 \ \ \ \
 \frac{\partial}{\partial \lambda}\phi_n(c,\lambda)=0.
 \ee
 However, due to the complicated form of the derivatives
 (see (\ref{eq.4.1}), (\ref{eq.4.2})),
 it is not a trivial fact to solve $(\ref{eq.5.1})$, or even
 to verify its consistency analytically. On the other hand,
 as we shall see in the sequel,
 it is important to know the existence (and uniqueness) of a
 minimizing point; it will be used
 in an essential way in the construction of
 extremal random vectors, concluding
 tightness of the bound (\ref{eq.main}).

 The attainability of the infimum can be seen as follows:

 \noindent
 Set $\epsilon_0:=\frac{1}{4}\min_i\{\sigma_i\}>0$.
 For $c\in\R$ and $\lambda\in (0,\epsilon_0]$,
 $(\mu_i-c)^2+\sigma_i^2\geq 4\lambda^2$
 ($i=1,\ldots,n$).
 Thus,
 $\lambda\, U(\frac{\mu_i-c}{\lambda},\frac{\sigma_i}{\lambda})
 =2\sqrt{(\mu_i-c)^2+\sigma_i^2}$
 for all $i$, and
 \begin{eqnarray*}
 &&
 \hspace{-15ex}
 \mbox{$
 \phi_n(c,\lambda)=-(n-2)\lambda
 +\sum_{i=1}^n \sqrt{(\mu_i-c)^2+\sigma_i^2}
 $}
 \\
 &&
 \mbox{$
 \geq -(n-2)\epsilon_0+\sum_{i=1}^n \sqrt{(\mu_i-c)^2+\sigma_i^2}
 =\phi_n(c,\epsilon_0).
 $}
 \end{eqnarray*}
 The function $c\mapsto \sum_{i=1}^n \sqrt{(\mu_i-c)^2+\sigma_i^2}$
 is strictly convex, tending to $\infty$ as $|c|\to\infty$;
 thus, its minimum is attained at a unique $c=c_1$.
 From $\phi_n(c,\epsilon_0)\geq \phi_n(c_1,\epsilon_0)$ we get
 \[
 \phi_n(c,\lambda)
 \geq
 \phi_n(c_1,\epsilon_0)
 =
 -(n-2)\epsilon_0 +\sum_{i=1}^n \sqrt{(\mu_i-c_1)^2+\sigma_i^2},
 \ \
 c\in\R, \ 0<\lambda\leq \epsilon_0.
 \]
 We now chose $\lambda_1:=\frac{1}{2}\min_i\{\sigma_i\}$,
 so that
 $\lambda_1>\epsilon_0$ and
 $(\mu_i-c_1)^2+\sigma_i^2\geq 4\lambda_1^2$
 for all $i$.
 Therefore,
 $\lambda_1\, U(\frac{\mu_i-c_1}{\lambda_1},\frac{\sigma_i}{\lambda_1})
 =2\sqrt{(\mu_i-c_1)^2+\sigma_i^2}$
 ($i=1,\ldots,n$), and  it follows that
 $\phi_n(c_1,\lambda_1)=-(n-2)\lambda_1+
 \sum_{i=1}^n \sqrt{(\mu_i-c_1)^2+\sigma_i^2}$.
 Since $\lambda_1>\epsilon_0$ and $n\geq 3$,
 the inequality $-(n-2)\epsilon_0>-(n-2)\lambda_1$
 leads to $\phi_n(c_1,\epsilon_0)>\phi_n(c_1,\lambda_1)$.
 Moreover,
 $U(x,y)\geq U(0,y)=2+\int_{0}^y \min\{t,2\} dt >2$
 for all $x\in\R$ and $y>0$.
 We thus obtain
 $
 \phi_n(c,\lambda)=-(n-2)\lambda+\frac{\lambda}{2}\sum_{i=1}^n
 U(\frac{\mu_i-c}{\lambda}
 ,\frac{\sigma_i}{\lambda})>-(n-2)\lambda+n\lambda=2\lambda
 $
 for all $c$ and $\lambda>0$.
 Setting $M_0:=\frac{1}{2}\phi_n(c_1,\epsilon_0)>\epsilon_0$
 we see that
 \[
 \phi_n (c,\lambda)\geq \phi_n(c_1,\epsilon_0)>\phi_n(c_1,\lambda_1)
 \
 \mbox{ for all } c\in\R, \ \lambda\in(0,\epsilon_0]\cup[M_0,\infty).
 \]

 Assume now that $\lambda\in(\epsilon_0,M_0)$ with $\epsilon_0$, $M_0$
 as above. From the obvious inequality $U(x,y)\geq 2\max\{|x|,1\}\geq 2|x|$
 we get
 \[
 \mbox{$
 \phi_n(c,\lambda)
 \geq
 -(n-2)\lambda
 +\sum_{i=1}^n |\mu_i-c|
 \geq
 -(n-2) M_0  +\sum_{i=1}^n |\mu_i-c|.
 $}
 \]
 The last inequality shows that $\phi_n(c,\lambda)\to \infty$ as $|c|\to\infty$,
 uniformly in $\lambda\in(\epsilon_0,M_0)$;
 thus, we can find a constant $C_0$ such that
 \[
 \phi_n(c,\lambda)\geq \phi_n(c_1,\epsilon_0)
 \
 \mbox{ for all } |c|\geq C_0, \ \lambda\in(\epsilon_0,M_0).
 \]
 Since $\phi_n(c_1,\epsilon_0)>\phi_n(c_1,\lambda_1)$,
 we arrived at the conclusion
 \[
 \phi_n(c,\lambda)
 >\phi_n(c_1,\lambda_1)
 \
 \mbox{ for all } (c,\lambda)
 \mbox { with }
 |c|\geq C_0 \mbox{ or } \lambda\leq \epsilon_0
 \mbox{ or } \lambda\geq M_0.
 \]
 This inequality shows that any minimizing point $(c_0,\lambda_0)$
 of (the continuous function)
 $\phi_n(c,\lambda)$
 over the compact rectangle
 $R:=[-C_0,C_0]\times[\epsilon_0,M_0]$
 must lie in the interior of $R$.
 The convexity of $\phi_n$
 implies that
 its global
 minimum
 is attained at $(c_0,\lambda_0)$.
 On the other hand, the differentiability of $\phi_n$
 shows that
 $(c,\lambda)=(c_0,\lambda_0)$ is a solution to (\ref{eq.5.1});
 and the convexity of $\phi_n$ implies that any such solution
 is a minimizing point.

 Let us now define
 \be
 \label{eq.5.2}
 T_0:=\big\{(c,\lambda)\in T: (c,\lambda) \mbox{ is a solution to }
 \mbox{(\ref{eq.5.1})}
 \big\},
 \ee
 so that $T_0\neq \emptyset$.
 The minimizing points of the convex function $\phi_n$
 are exactly the points
 of $T_0$; thus, $T_0$ is a convex compact
 subset of $T$, and we have shown the following
 \begin{PROP}
 \label{prop.5.1}
 If $n\geq 3$ then for any given values of $\bmu$ and ${\bsigma}$,
 the system (\ref{eq.5.1}) is consistent,
 and the set of solutions, $T_0$,
 is a convex compact subset of $T$.
 Moreover, for any $(c_0,\lambda_0)\in T_0$,
 \[
 \phi_n(c,\lambda)\geq \phi_n(c_0,\lambda_0)
  \ \
 \mbox{ for all } (c,\lambda)\in T=\R\times(0,\infty),
 \]
 with equality if and only if
 $(c,\lambda)\in T_0$.
 \end{PROP}

 We now proceed to show that $T_0$ is a singleton.
 Let as fix $c=c_1\in\R$.
 For this particular value $c_1$ we consider the function
 \[
 \mbox{$
 \psi_n(\lambda):=\phi_n(c_1,\lambda)=-(n-2)\lambda
 +\sum_{i=1}^{n} u_i(\lambda), \ \ \lambda>0,
 $}
 \]
 where
 \[
 \mbox{$
 u_i(\lambda):=\frac{1}{2}\lambda\,
 U(\frac{\mu_i-c_1}{\lambda},\frac{\sigma_i}{\lambda}), \ \
 \lambda >0  \ \ (i=1,\ldots,n).
 $}
 \]
 The function $u_i$ can be written more precisely as follows:
 \[
 u_i(\lambda)=\left\{
 \begin{array}{cl}
 \sqrt{(\mu_i-c_1)^2+\sigma_i^2}, & 0<\lambda\leq t_i,
 \\
 \lambda+\frac{1}{4\lambda}
 \big[(\mu_i-c_1)^2+\sigma_i^2\big], & t_i\leq \lambda< \gamma_i,
 \\
 \frac{1}{2}\Big\{|\mu_i-c_1|+\lambda+
 \sqrt{(|\mu_i-c_1|-\lambda)^2+\sigma_i^2}\Big\}, & \lambda\geq
 \gamma_i,
 \end{array}
 \right.
 \]
 where
 $t_i=t_i(c_1)$ and $\gamma_i=\gamma_i(c_1)$ are given by
 \be
 \label{eq.5.3}
 \mbox{$
 t_i:=\frac{1}{2}\sqrt{(\mu_i-c_1)^2+\sigma_i^2},
 \
 \gamma_i:=
 \frac{(\mu_i-c_1)^2+\sigma_i^2}{2|\mu_i-c_1|},
 \
 \
 0<t_i<\gamma_i \leq \infty
 \ \ (i=1,\ldots,n).
 $}
 \ee
 Each function $u_i$ is continuously
 differentiable with derivative
 \be
 \label{eq.5.4}
 u'_i(\lambda)=\left\{
 \begin{array}{cl}
 0, & 0<\lambda\leq t_i,
 \\
 1-\frac{(\mu_i-c_1)^2+\sigma_i^2}{4\lambda^2},
 & t_i\leq \lambda< \gamma_i,
 \\
 \frac{1}{2}\Big(1+\frac{\lambda-|\mu_i-c_1|}
 {\sqrt{(\lambda-|\mu_i-c_1|)^2+\sigma_i^2}}\Big),
  & \lambda\geq
 \gamma_i,
 \end{array}
 \right.
 \ \
 i=1,\ldots,n.
 \ee
 Obviously, $u_i(\lambda)$ is constant (equal to $2 t_i$)
 in the interval $(0,t_i]$ and then it is strictly increasing;
 its non-decreasing continuous derivative $u_i'(\lambda)$
 satisfies $0\leq u_i'(\lambda)<1$ for all $\lambda$, and
 $\lim_{\lambda\to \infty} u_i'(\lambda)=1$.
 It follows that
 \[
 \mbox{$
 \psi_n'(\lambda)=-(n-2)+\sum_{i=1}^n u_i'(\lambda)
 $}
 \]
 is non-decreasing and, thus, $\psi_n$ is convex.
 Let $t_{1:n}, \ldots, t_{n:n}$ be the ordered
 values of $t_1,\ldots,t_n$. Noting that $n\geq 3$ and
 $0<t_{1:n}\leq \cdots \leq t_{n:n}<\infty$,
 we see that $\psi_n'(\lambda)=-(n-2)<0$ for
 $\lambda\leq t_{1:n}$, and the function $\psi_n$
 is strictly decreasing in the interval $(0,t_{1:n}]$.
 Also, $\psi_n(\lambda)$ is strictly convex in the interval
 $(t_{1:n},\infty)$,
 because
 $\psi_n'(\lambda)$
 is strictly increasing in
 that interval.  Observe that $\psi_n$ is eventually
 strictly increasing:
 $\lim_{\lambda\to\infty}
 \psi_n'(\lambda)=-(n-2)+\sum_{i=1}^n \lim_{\lambda\to\infty}u_i'(\lambda)=2$.
 It follows that
 $\psi_n(\lambda)$ attains its minimum value at a unique point
 $\lambda=\lambda_1
 >t_{1:n}$;
 clearly, $\lambda_1=\lambda_1(c_1)$ is the unique
 solution to the equation $\psi_n'(\lambda)=0$, $0<\lambda<\infty$.
 \begin{LEM}
 \label{lem.5.1}
 Let $n\geq 3$ and fix an arbitrary $c_1\in\R$.
 The function
 $\psi_n:(0,\infty)\to(0,\infty)$, with
 $\psi_n(\lambda):=\phi_n(c_1,\lambda)$,
 attains its minimum value at a
 unique point $\lambda_1=\lambda_1(c_1)$.
 The minimizing point $\lambda_1$
 is the unique solution of the equation
 \be
 \label{eq.5.5}
 \mbox{$
 \sum_{i=1}^n u_i'(\lambda)=n-2,
 \ \ \ \ \
 t_{n-1:n}<\lambda<\sum_{i=1}^n t_i,
 $}
 \ee
 where $t_i=t_i(c_1)$ are as in (\ref{eq.5.3}),
 $0<t_{1:n}\leq \cdots\leq t_{n:n}$ are the ordered values
 of $t_i$ in (\ref{eq.5.3}), and the functions
 $u_i'(\lambda)$ are given by (\ref{eq.5.4}).
 \end{LEM}
 \noindent
 \begin{Proof}
 It remains to verify that the unique solution,
 $\lambda=\lambda_1$, of
 $\sum_{i=1}^n u_i'(\lambda)=n-2$
 lies in the interval $(t_{n-1:n},\sum_{i=1}^n t_i)$.
 First observe that if $\lambda\leq t_{n-1:n}$,
 then we can find
 two indices $s\neq r$ with $\lambda\leq t_s$ and
 $\lambda\leq t_r$. Since $u_r'(\lambda)=u_s'(\lambda)=0$,
 the sum $\sum_{i=1}^n u_i'(\lambda)$ contains
 at most $n-2$ strictly positive
 terms $u_i'(\lambda)$; from $u_i'(\lambda)<1$
 it follows that $\sum_{i=1}^n u_i'(\lambda)<n-2$.
 This shows that $\lambda_1>t_{n-1:n}$.
 On the other hand, we observe that
 $\lim_{\lambda\searrow 0} \psi_n(\lambda)=2\sum_{i=1}^n t_i$.
 Thus, $\psi_n(\lambda_1)\leq 2\sum_{i=1}^n t_i$
 (because $\lambda=\lambda_1$ minimizes $\psi_n(\lambda)$).
 However, we know that
 $\phi_n(c,\lambda)>2\lambda$ for all $(c,\lambda)\in T$;
 thus,
 $2\sum_{i=1}^n t_i\geq \psi_n(\lambda_1)=\phi_n(c_1,\lambda_1)>2\lambda_1$.
 $\Box$
 \medskip
 \end{Proof}

 \begin{REM}
 \label{rem.5.1} Fix a point $(c_1,\lambda_1)\in T_0$ and
 define the following (possibly empty) sets of indices:
 \be
 \label{eq.6.2}
 \begin{array}{l}
 I_1:=\big\{ i\in\{1,\ldots,n\}: (\mu_i-c_1)^2+\sigma_i^2\geq
 4\lambda_1^2\big\}=\{i: \lambda_1\leq t_i\},
 \\
 I_2:=\big\{ i\in\{1,\ldots,n\}:
 2\lambda_1|\mu_i-c_1|<(\mu_i-c_1)^2+\sigma_i^2<4\lambda_1^2\big\}
 =\{i: t_i<\lambda_1<\gamma_i\},
 \\
 I_3:=\big\{ i\in\{1,\ldots,n\}: (\mu_i-c_1)^2+\sigma_i^2\leq
 2\lambda_1 (\mu_i-c_1)\big\}=\{i: \lambda_1\geq \gamma_i \mbox{ and } \mu_i>c_1\},
 \\
 I_4:=
 \big\{ i\in\{1,\ldots,n\}: (\mu_i-c_1)^2+\sigma_i^2\leq
 2\lambda_1 (c_1-\mu_i)\big\}=\{i: \lambda_1\geq \gamma_i \mbox{ and } \mu_i<c_1\}.
 \end{array}
 \ee
 By definition, $I_i\cap I_j=\emptyset$ for $i\neq j$ and
 $I_1\cup I_2\cup I_3\cup I_4=\{1,\ldots,n\}$. Since $(c_1,\lambda_1)\in T_0$
 it follows that $\lambda_1$ must solves (\ref{eq.5.5}) (for this
 particular value of $c_1$), that is,
 \[
 \sum_{i\in I_2} \Big\{1-\frac{(\mu_i-c_1)^2+\sigma_i^2}{4\lambda_1^2}\Big\}
 +
 \sum_{i\in I_3\cup I_4}
 \frac{1}{2}\Big\{1+\frac{\lambda_1-|\mu_i-c_1|}
 {\sqrt{(\lambda_1-|\mu_i-c_1|)^2+\sigma_i^2}}\Big\}
 =n-2,
 \]
 where an empty sum should be treated as zero. Observe that
 all summands are (strictly positive and) strictly less than $1$; thus,
 $N(I_2)+N(I_3)+N(I_4)\geq n-1$, and it follows that
 $N(I_1)\leq 1$, where $N(I)$ denotes the cardinality of $I$.
 Furthermore, $(c,\lambda)=(c_1,\lambda_1)$ is a
 solution to $\frac{\partial}{\partial c}\phi_n(c,\lambda)=0$.
 Using $\frac{\partial}{\partial c} \phi_n(c,\lambda)=
 -\frac{1}{2}\sum_{i=1}^n U_1\big(\frac{\mu_i-c}{\lambda},
 \frac{\sigma_i}{\lambda}\big)$
 and the explicit form of $U_1$, given by (\ref{eq.4.1}), we obtain
 \be
 \label{eq.5.6}
 \mbox{$
 \sum_{i\in I_1} \frac{\mu_i-c_1}{\sqrt{(\mu_i-c_1)^2+\sigma_i^2}}
  +
 \sum_{i\in I_2} \frac{\mu_i-c_1}{2\lambda_1}
 +
 \sum_{i\in I_3\cup I_4}
 \frac{
 \mbox{\footnotesize sign}(\mu_i-c_1)}{2}\Big\{1+\frac{|\mu_i-c_1|-\lambda_1}
 {\sqrt{(|\mu_i-c_1|-\lambda_1)^2+\sigma_i^2}}\Big\}
 =0.
 $}
 \ee
 This equality shows that $N(I_3)\leq n-1$ and $N(I_4)\leq n-1$;
 for if, e.g., $N(I_3)=n$ then we would have $I_1=I_2=I_4=\emptyset$ and,
 since $\mu_i>c_1$ whenever $i\in I_3$, the above equation leads
 to the (obviously impossible) relation
 \[
 \sum_{i=1}^n
 \frac{1}{2}\Big\{1+\frac{(\mu_i-c_1)-\lambda_1}
 {\sqrt{((\mu_i-c_1)-\lambda_1)^2+\sigma_i^2}}\Big\}
 =0.
 \]
 We have thus concluded the following key-property
 of a minimizing point:
 \be
 \label{eq.5.7}
 \mbox{If $(c_1,\lambda_1)\in T_0$ then
 $\max\big\{N(I_3),N(I_4)\big\}\leq n-1$ and
 $N(I_1)\leq 1$.}
 \ee
 Most cases
 suggested by (\ref{eq.5.7}) may
 appear for some values
 of $\bmu,{\bsigma}$
 (one of the rare exceptions is $N(I_1)=N(I_2)=0$,
 $\max\{N(I_3),N(I_4)\}=n-1$).
 Note that Theorem \ref{theo.AG} is, in fact, concerned
 with the particular situation
 where $N(I_2)=n$ (thus, $N(I_1)=N(I_3)=N(I_4)=0$).
 It is, essentially, the unique
 situation in which the
 $AG_n$ bound is tight (plus boundary subcases).
 Due to (\ref{eq.5.7}),
 it seems that this particular (but plausible) case is
 quite restricted.

 Behind the tedious calculations, the rough meaning of
 the argument the led to (\ref{eq.5.7}),
 is the following: For a particular $(c,\lambda)$
 to be optimal (i.e., to minimize $\phi_n$) it is
 necessary that $c$ is not ``too far away'' from the
 $\mu_i$'s and $\lambda$ is not ``too small'' or ``too large''
 compared to $\frac{1}{2}\sum_{i=1}^n {\sigma_i}$.
 In particular, (\ref{eq.5.6}) shows that
 an optimal $c$ can never lie outside the interval
 $\big[\min_i\{\mu_i\}, \max_i\{\mu_i\}\big]$, and it is located
 in an interior point when the $\mu_i$'s are not all equal;
 of course this fact is intuitively obvious.
 \end{REM}

 \begin{LEM}
 \label{lem.5.2}
 If the set $T_0$ of (\ref{eq.5.2})
 contains two different elements, then
 it must be a compact line segment
 which is not parallel to the $\lambda$-axis.
 That is, $T_0$ has to be of the form
 $T_0=[{\bxx},{\byy}]=\{{\bxx}+t({\byy}-{\bxx}), 0\leq t\leq 1\}$,
 for some ${\bxx}=(c_1,\lambda_1)\in T$ and ${\byy}=(c_2,\lambda_2)\in T$
 with $c_1\neq c_2$.
 \end{LEM}

 \begin{LEM}
 \label{lem.5.3}
 Let $A_0=(c_0,\lambda_0)\neq A_1=(c_1,\lambda_1)$ be two points in $T$.
 Fix $\mu\in\R$, $\sigma>0$
 and consider the points
 $B_0=\big(\frac{\mu-c_0}{\lambda_0},\frac{\sigma}{\lambda_0}\big)\in T$,
 $B_1=\big(\frac{\mu-c_1}{\lambda_1},\frac{\sigma}{\lambda_1}\big)\in T$.
 Let $A=(c,\lambda)$ and $B=\big(\frac{\mu-c}{\lambda},\frac{\sigma}{\lambda}\big)$.
 As the point $A$ is moving linearly in the line segment $[A_0,A_1]$ (from $A_0$ to $A_1$),
 the point $B=B(A)$ is moving continuously in the line segment
 $[B_0,B_1]$ (from $B_0$ to $B_1$).
 \end{LEM}

 We are now ready to state
 the conclusion of the present section.
 \begin{THEO}
 \label{theo.5.1}
 If $n\geq 3$ then for any given values of $\bmu$ and ${\bsigma}$,
 there exists a unique solution $(c,\lambda)=(c_0,\lambda_0)$ of (\ref{eq.5.1}), and
 \be
 \label{eq.5.8}
 \phi_n(c,\lambda)\geq \phi_n(c_0,\lambda_0) \ \ \
 \mbox{ for all }
 (c,\lambda)\in \R\times(0,\infty),
 \ee
 with equality if and only if  $(c,\lambda)=(c_0,\lambda_0)$.
 \end{THEO}

 \begin{REM}
 \label{rem.5.2} For $n=2$, Theorem \ref{theo.5.1}
 (as well as several conclusions of the present section)
 is no longer true. It is again true that the convex function
 $\phi_2(c,\lambda)$ attains its minimum value,
 $\rho_2=\sqrt{(\mu_2-\mu_1)^2+(\sigma_1+\sigma_2)^2}$,
 at the solutions of the system
 \eqref{eq.5.1}, but now $T_0$ is not a singleton:
 it contains points arbitrarily close to the boundary
 of the domain of $\phi_2$.
 More precisely, one can verify that for $n=2$,
 the exact set of minimizing points
 is the line segment
 $T_0=\{(c_0,\lambda); 0<\lambda\leq \lambda_0\}$,
 where
 \[
 \mbox{
 $c_0=\frac{\sigma_1}{\sigma_1+\sigma_2}\mu_2+\frac{\sigma_2}{\sigma_1+\sigma_2}\mu_1$,
 \ \ \
 $\lambda_0=\frac{\rho_2}{2(\sigma_1+\sigma_2)}\min\{\sigma_1,\sigma_2\}$.
 }
 \]
 However, the set
 ${\cal E}_2(\mu_1,\mu_2,\sigma_1,\sigma_2)$ is a singleton, and
 this fact can be seen directly (see
 Section \ref{sec.last});
 thus, the above calculation is completely unnecessary.
 Also, it is worth pointing out that, for $n=2$, $N(I_1)=2=n$
 (compare with \eqref{eq.5.7}).
 \end{REM}

 \section{Tightness and characterization of extremal random vectors}
 \label{sec.6}
 \setcounter{equation}{0}
 Let $n\geq 3$, $\bmu$, ${\bsigma}$ be fixed (with $0<\sigma_i<\infty$
 for all $i$). Let $(c,\lambda)$ be the unique solution of (\ref{eq.5.1}).
 With the help of $(c,\lambda)$ we shall
 give a complete description of the set  ${\cal E}_n(\bmu,{\bsigma})$
 of extremal random vectors in
 ${\cal F}_n(\bmu,{\bsigma})$.
 These are the random vectors ${\bXX}$ satisfying
 $\E {\bXX}=\bmu$, $\Var {\bXX}={\bsigma}^2$ and
 $\E R_n({\bXX})=\rho_n=\rho_n(\bmu,{\bsigma})$,
 where
 \be
 \label{eq.6.1}
 \mbox{$
 \rho_n:=\phi_n(c,\lambda)=-(n-2)\lambda+\frac{\lambda}
 {2}\sum_{i=1}^n
 U\big(\frac{\mu_i-c}{\lambda},\frac{\sigma_i}{\lambda}\big);
 $}
 \ee
 recall that $U(\cdot,\cdot)$ is given by (\ref{eq.u}).
 The construction, though more complicated, follows parallel
 arguments as for the attainability of the $AG_n$ bound
 (Theorem \ref{theo.AG}).

 We start by considering the partition $I_1,\ldots,I_4$
 of
 $\{1,\ldots,n\}$ as in \eqref{eq.6.2},
 and the
 corresponding cardinalities $n_1,\ldots,n_4$.
 The main difference from Remark \ref{rem.5.1} is that, now,
 each $I_j$ has been stabilized, because $(c,\lambda)$ is unique; thus,
 one has to substitute $c_1=c$ and $\lambda_1=\lambda$
 in \eqref{eq.6.2}.
 Clearly some of the sets $I_j$ may be empty; then $n_j=0$. The situation
 with all $I_j$ being nonempty may also appear; this is the case, e.g.,
 for $\bmu=(4,0,4,0)$, ${\bsigma}=(10,5,1,1)$.
 From Remark \ref{rem.5.1} (see (\ref{eq.5.7})) we know that
 $n_1$ $n_2$, $n_3$, $n_4$ (with $n_j\geq 0$, $\sum n_j=n$)
 cannot be completely arbitrary; they have to satisfy the restrictions:
 \be
 \label{eq.6.8old}
 n_3=N(I_3)\leq n-1, \ \  n_4=N(I_4)\leq n-1, \ \ n_1=N(I_1)\leq 1.
 \ee
 Other impossible cases are given by $n_3=1,n_4=n-1$ and $n_3=n-1,n_4=1$;
 this is a by-product of Lemma \ref{lem.6.1}, below.

 For notational simplicity it is helpful
 to consider the following numbers $\xi_i$,
 $\theta_i$:
 \be
 \label{eq.6.3}
 \hspace{-.2ex}
 \xi_i:=\left\{
 \begin{array}{ll}
 \mu_i-c,
 &
 i\in I_1\cup I_2;
 \\
 \mu_i-c-\lambda,
 &
 i\in I_3;
 \\
 c-\mu_i-\lambda,
 &
 i\in I_4;
 \end{array}
 \right.
 \theta_i:=\left\{
 \begin{array}{ll}
 \sqrt{(\mu_i-c)^2+\sigma_i^2},
 &
 i\in I_1\cup I_2;
 \\
 \sqrt{(\mu_i-c-\lambda)^2+\sigma_i^2},
 &
 i\in I_3;
 \\
 \sqrt{(c-\mu_i-\lambda)^2+\sigma_i^2},
 &
 i\in I_4.
 \end{array}
 \right.
 \ee
 We note that  $|\xi_i|<\theta_i$ for all $i$
 and $2\lambda|\xi_i|<\theta_i^2<4\lambda^2$ for all $i\in I_2$ (if any).
 Following Corollary \ref{cor.1}
 we define
  the
  probabilities
 \be
 \label{eq.6.4}
 \begin{array}{llll}
 \vspace{.5ex}
 p_i^-:=\frac{1}{2}\big(1-\frac{\xi_i}{\theta_i}\big),
 &
 p_i^{o}:=0,
 &
 p_i^+:=\frac{1}{2}\big(1+\frac{\xi_i}{\theta_i}\big),
 &
 i\in I_1;
 \\
 \vspace{.5ex}
 p_i^-:=\frac{1}{8\lambda^2}\left[\theta_i^2-2\lambda \xi_i\right],
 &
 p_i^{o}:=1-\frac{\theta_i^2}{4\lambda^2},
 &
 p_i^+:=\frac{1}{8\lambda^2}\left[\theta_i^2+2\lambda \xi_i\right],
 &
 i\in I_2;
 \\
 \vspace{.5ex}
 p_i^-:=0,
 &
 p_i^{o}:=\frac{1}{2}\big(1-\frac{\xi_i}
 {\theta_i}\big),
 &
 p_i^+:=\frac{1}{2}\big(1+\frac{\xi_i}
 {\theta_i}\big),
 &
 i\in I_3;
 \\
 p_i^-:=\frac{1}{2}\big(1+\frac{\xi_i}
 {\theta_i}\big),
 &
 p_i^{o}:=\frac{1}{2}\Big(1-\frac{\xi_i}
 {\theta_i}\big),
 &
 p_i^+:=0,
 &
 i\in I_4,
 \end{array}
 \ee
 and the corresponding (univariate) supporting points
 \be
 \label{eq.6.5}
 \begin{array}{llll}
 x_i^-:=c-\theta_i,
 &&
 x_i^+:=c+\theta_i,
 & i\in I_1;
 \\
 x_i^-:=c-2\lambda,
 &
 x_i^{o}:=c,
 &
 x_i^+:=c+2\lambda,
 &
 i\in I_2;
 \\
 &
 x_i^{o}:=c+\lambda-\theta_i,
 &
 x_i^+:=c+\lambda+\theta_i,
 &
 i\in I_3;
 \\
 x_i^-:=c-\lambda-\theta_i,
 &
 x_i^{o}:=c-\lambda+\theta_i,
 &
 &
 i\in I_4.
 \end{array}
 \ee
 By definition, each
 ${\bpp}_i:=\big(p_i^-,p_i^{o},p_i^+\big)$ is a
 probability vector.
 Clearly, one could assign an arbitrary value to a missing point,
 since its corresponding probability
 is $0$. The most convenient choice is to assign the respective
 values $c-2\lambda$, $c$, $c+2\lambda$, whenever $x_i^{-}$, $x_i^{o}$, $x_i^{+}$
 is not specified from (\ref{eq.6.5}). With this convention,
 \be
 \label{eq.6.6}
 x_i^{-}<c-\lambda<x_i^{o}<c+\lambda<x_i^{+}, \ \ \  i=1,\ldots,n.
 \ee


 \noindent
 Let $X_i$ be a random variable
 which assumes values $x_i^{-},x_i^{o},x_i^{+}$
 with respective probabilities $p_i^{-},p_i^{o},p_i^{+}$.
 Corollary \ref{cor.1} asserts that
 (the distribution of) $X_i$
 is
 characterized be the
 fact that maximizes the expectation of
 $|(X-c)-\lambda|+|(X-c)+\lambda|$ as
 $X$ varies in ${\cal F}_1(\mu_i,\sigma_i)$.

 The following lemma provides the most
 fundamental tool
 for the main result.
 \begin{LEM}
 \label{lem.6.1}
 The probabilities $p_i^+$, $p_i^-$ in (\ref{eq.6.4}) satisfy the relation
 \be
 \label{eq.6.7}
 \sum_{i=1}^n {p_i^+}=\sum_{i=1}^n p_i^{-}=1.
 \ee
 \medskip
 \end{LEM}

 Lemma \ref{lem.6.1} enables us to define the $n$-variate
 probability vectors
 \be
 \label{eq.6.9}
 {\bpp}^{+}=(p_1^{+},\ldots,p_n^+), \ \
 {\bpp}^{-}=(p_1^{-},\ldots,p_n^-).
 \ee
 By definition,
 ${\bpp}^+$ has its zero elements at exactly the positions $i$ where
 $i\in I_4$
 (if $I_4= \emptyset$, all $p_i^+$'s are positive), and
 ${\bpp}^-$ has its zero elements at exactly the positions $i$ where
 $i\in I_3$
 (if any).

 \begin{PROP}
 \label{prop.6.1}
 Assume we are given $n\geq 3$, $\bmu$, ${\bsigma}$.
 Then, (i) and (ii) are
 equivalent:
 \vspace{-1ex}
 \begin{itemize}
 \item[(i)]  We can find a random vector
 ${\bXX}\in{\cal F}_n(\bmu,{\bsigma})$ such that
 \vspace{-1ex}
 $\E R_n=\rho_n$.
 \item[(ii)] There exists a $n\times n$ probability matrix
 $Q\in {\cal M}({\bpp}^{+},{\bpp}^{-})$
 such that $q_{ii}=0$ for all
 $i\in \{1,\ldots,n\}$.
 \end{itemize}
 Moreover,
 with ${\cal L}({\bXX})$ denoting the probability law of the random vector
 ${\bXX}$, the correspondence ${\cal L}({\bXX})
 \rightleftarrows
 Q$ is a bijection; the
 explicit formula for the
 transformation $Q=(q_{ij})\mapsto {\cal L}({\bXX})$
 is given by
 \be
 \label{eq.6.10}
 \begin{array}{c}
 \Pr[{\bXX}={\bxx}_{ij}]=q_{ij}, \mbox{ where }
 {\bxx}_{ij}:=\big(x_1^{o},\ldots,x_{i-1}^{o},x_i^+,x_{i+1}^{o},\ldots,
 x_{j-1}^{o},x_j^-,x_{j+1}^{o},\ldots,x_n^{o}\big),
 \\
 \mbox{~}\hspace{50ex} \ \
 i\neq j, \ \ i,j=1,\ldots,n.
 \end{array}
 \ee
 \end{PROP}

 The main result of the present work reads as follows:
 \begin{THEO}
 \label{theo.6.1}
 Let $n\geq 3$, $\mu_i\in\R$, $\sigma_i>0$ ($i=1,\ldots,n$). Then,

 \noindent
 (a)
 \be
 \label{eq.6.17}
 \sup \E R_n =\rho_n,
 \ee
 where the supremum is taken over
 ${\bXX}\in{\cal F}_{n}(\bmu,{\bsigma})$
 and $\rho_n=\rho_n(\bmu,{\bsigma})$ is given
 by (\ref{eq.6.1}), with $(c,\lambda)=(c(\bmu,{\bsigma}),
 \lambda(\bmu,{\bsigma}))$ being the
 unique solution to the system of equations
 \medskip
 (\ref{eq.5.1}).

 \noindent
 (b)
 The set ${\cal E}_n(\bmu,{\bsigma})$
 is nonempty.
 Any
 extremal ${\bXX}\in{\cal E}_n(\bmu,{\bsigma})$
 is produced by (\ref{eq.6.10}),
 with $x_i^-$, $x_i^o$, $x_i^+$ as in (\ref{eq.6.5}),
 and corresponds uniquely to a $n\times n$
 probability matrix $Q\in {\cal M}({\bpp}^+, {\bpp}^-)$
 with zero diagonal entries, where ${\bpp}^+$,
 ${\bpp}^-$ are given by (\ref{eq.6.9}).
 \end{THEO}
 \noindent
 \begin{Proof} From Theorem \ref{th.main}
 we know that $\E R_n\leq \rho_n$ and it suffices to
 prove (b). In view of Proposition
 \ref{prop.6.1}, it remains to verify that the class
 of $n\times n$ probability matrices with zero diagonal entries
 and marginals ${\bpp}^+$, ${\bpp}^-$
 is nonempty.
 However,
 this fact follows immediately from Lemma \ref{lem.3},
 because $\max_i\{p_i^++p_i^-\}\leq 1$ (see (\ref{eq.6.4})),
 and the proof is complete.
 $\Box$
 \end{Proof}

 \begin{REM}
 \label{rem.6.1}
 Since $\E R_n({\bXX})=\rho_n$ for any
 ${\bXX}\in{\cal E}_n(\bmu,\bsigma)$,
 \[
 \mbox{$
 \rho_n=\sum_{i=1}^n \{ (x_i^+-c)p_i^+ +(c-x_i^-)p_i^-\}.
 $}
 \]
 \end{REM}
 \begin{COR}
 \label{cor.6.1}
 If $I_1\neq \emptyset$ (see (\ref{eq.6.2})) then $I_1=\{k\}$
 for some $k\in\{1,\ldots,n\}$, and 
 the equality in
 (\ref{eq.6.17}) characterizes the random vector ${\bXX}$
 with probability law
 \be
 \label{eq.6.19}
 \Pr[{\bXX}={\bxx}_{ik}]=p_i^+, \ \
 \Pr[{\bXX}={\bxx}_{ki}]=p_i^-, \ \ i\neq k, \ i=1,\ldots,n.
 \ee
 \end{COR}
 \noindent
 \begin{Proof} From (\ref{eq.6.8old})
 we know that $N(I_1)\leq 1$, and thus,
 $I_1=\{k\}$ for some $k$. Since
 $k\in I_1$, (\ref{eq.6.4}) shows that
 $\max_i\{p_i^++p_i^-\}=p_k^++p_k^{-}=1$.
 [Note that, by Lemma \ref{lem.6.1},
 $\sum_{i\neq k}p_i^{-}+\sum_{i\neq k}p_i^{+}=(1-p_k^-)+(1-p_k^+)=1$
 and, hence, (\ref{eq.6.19}) defines a probability law.]
 Lemma \ref{lem.3} implies uniqueness
 of $Q$, hence of ${\cal L}({\bXX})$
 (see (\ref{eq.6.10})).
 It is easily seen that the matrix $Q$, obtained
 by (\ref{eq.6.19}) through (\ref{eq.6.16}), is
 indeed the unique probability matrix
 with vanishing diagonal entries and marginals ${\bpp}^+$,
 ${\bpp}^-$.
 \medskip
 $\Box$
 \end{Proof}

 Corollary \ref{cor.6.1}
 implies uniqueness (denoted by (U)) for the second counterpart
 of the bound (\ref{eq.tight}) in Example \ref{ex.2}.
 It should be noted that the converse of Corollary \ref{cor.6.1}
 does not hold; that is, the condition $I_1\neq \emptyset$
 is not necessary for concluding uniqueness
 of the extremal
 random vector ${\bXX}$.
 A particular example was given by Remark \ref{rem.3.1}.

 Clearly, the most interesting situations in practice
 arise when $I_1=\emptyset$. In such
 cases it is fairly expected that there will be infinitely
 many extremal vectors, as in Theorem \ref{theo.AG}.
 This is, indeed, true in general, but not always.
 Lemma \ref{lem.3} guarantees infiniteness
 (denoted by (I)) only
 if all $p_{i}^+$, $p_i^-$ are nonzero, and this corresponds
 to the quite restricted case where $I_2=\{1,\ldots,n\}$.
 Of course, given the existence of two extremal vectors,
 one can deduce (I) by considering convex combinations of the
 corresponding matrices;  cf.\ Example \ref{ex.1}.
 If $I_1=\emptyset$,
 the complete distinction between (U) and
 (I) depends upon the values of $n$, $n_3=N(I_3)$ and
 $n_4=N(I_4)$  (see (\ref{eq.6.2}) and (\ref{eq.6.8old}));
 and if $n_3=n_4=0$ we already know that (I) results.

 We briefly discuss all remaining situations where $I_1=\emptyset$:
 If $n_2=N(I_2)=0$ and $n_3\geq 2$, $n_4\geq 2$,
 it is obvious that (I)
 holds; note that $n_3=1,n_4=n-1$ and
 $n_3=n-1,n_4=1$ are impossible
 by  
 Lemma \ref{lem.6.1}.
 If $n_2=n_3=n_4=1$ or $n_2=2,n_3=1, n_4=0$ or $n_2=2,n_3=0,n_4=1$
 then we are in (U), while
 (I) results if $n_2=n_3=1, n_4\geq 2$ or $n_2=n_4=1, n_3\geq 2$.
  If $n_2=1, n_3\geq 2, n_4\geq 2$
 then we get (I), as well as in
 all remaining cases where
 $n_2\geq 2, n_3\geq 0, n_4\geq 0$.

 The final conclusion is as follows: If $I_1=\emptyset$,
 the situations where the extremal distribution is uniquely
 defined are described by $n_2=n_3=n_4=1$ or $n_2=2,n_3=1,n_4=0$ or
 $n_2=2,n_3=0,n_4=1$
 (and thus, $n=3$); this provides an explanation
 to Remark \ref{rem.3.1}. However, we note that
 knowledge of the values $n_j$ actually requires knowledge
 of the region where the optimal $(c,\lambda)$
 appears,
 and this may be, or may not be, an easy task
 for particular $\bmu$, ${\bsigma}$.
 \begin{REM}
 \label{rem.6.2}
 The range
 $R_n({\bXX})$ of an extremal vector ${\bXX}$ need not be
 a degenerate random variable. An example is provided
 by $\bmu=(-2,0,2)$, ${\bsigma}=(1,3,1)$. Then,
 $n_1=0$, $n_2=n_3=n_4=1$ and it can be shown that
 \[
 \mbox{$
 \lambda\approx 1.737, \ \ \
 \rho_3=\frac{64 \lambda^3-72\lambda-81}{4\lambda(4\lambda^2-9)}\approx 6.066
 $}
 \]
 ($\lambda$ is the unique solution of
 $4\lambda^2(2-\lambda)=(4\lambda^2-9)\sqrt{\lambda^2-4\lambda+5}$, and this
 reduces to a four-degree polynomial equation).
 The range $R_3$ of the unique extremal vector assumes
 values $2\lambda+\frac{\lambda(8\lambda-9)}{4\lambda^2-9}\approx5.542$
 and $\frac{2\lambda(8\lambda-9)}{4\lambda^2-9}\approx6.245$ with respective
 probabilities $\frac{9}{4\lambda^2}\approx .254$ and
 $1-\frac{9}{4\lambda^2}\approx .746$.
 However, the improvement over the bound $AG_3=\sqrt{38}\approx 6.164$
 is negligible. As a general observation, even for small $n$,
 the value of $\rho_n$ is difficult to evaluate when more
 than two index sets
 $I_j$ are nonempty.
 \end{REM}

 \begin{EXAMPLE}
 \label{ex.6.1} {\it Homoscedastic observations from two balanced groups.}
 Let $n=2k$, $\sigma_i^2=\sigma^2$
 and $\mu_i=-\mu$ or $\mu$ according to $i\leq k$
 or $i>k$, respectively ($\mu\geq 0$).
 The Arnold-Groeneveld bound
 (\ref{eq.AG}) takes here the form
 \[
 \E R_{2k}\leq AG_{2k}=2\sqrt{k(\mu^2+\sigma^2)},
 \]
 and it is tight if $\mu\leq \frac{\sigma}{\sqrt{k-1}}$
 (in particular, if $n=2$ or $\mu=0$). Also, we
 know from Theorem \ref{theo.AG} the nature of the random vectors
 that attain the equality.
 However, for $\mu\geq \frac{\sigma}{\sqrt{k-1}}$ one finds
 $N(I_3)=N(I_4)=k$, and the tight bound
 of Theorem \ref{theo.6.1} becomes
 \[
 \mbox{$
 \E R_{2k}\leq \rho_{2k}=2\mu+2\sigma\sqrt{k-1} \ \ \ \big(\mu\geq \frac{\sigma}{\sqrt{k-1}}\big);
 $}
 \]
 note that $\rho_{2k}$ is equal to $AG_{2k}$ only in the boundary case
 $\sigma=\mu\sqrt{k-1}$. For $\mu\geq \frac{\sigma}{\sqrt{k-1}}$ the
 nature
 of extremal random vectors is different:
 They assume values
 \[
 {\byy}_{ij}=\big(-x,\ldots,-x,-y,-x,\ldots,-x \ ; \ x,\ldots,x,y,x,\ldots,x\big), \ \ i,j=1,\ldots,k,
 \]
 where $-y$ is located at the $i$-th place and
 $y$ is located at the $(k+j)$-th place of the vector.
 Here, $0\leq x=\mu-\frac{\sigma}{\sqrt{k-1}}<y=\mu
 +\sigma\sqrt{k-1}$. The respective probabilities $p_{ij}=\Pr[{\bXX}={\byy}_{ij}]$, $i,j=1,\ldots,k$, correspond to a probability matrix $P_{k\times k}$
 with uniform marginals. Both limits
 \[
 \mbox{
  $\lim_{\mu\to\infty}\frac{\rho_{2k}}{AG_{2k}}=\frac{1}{\sqrt{k}}$
 \ ($k$, $\sigma$ fixed),  \ \ \
 $\lim_{k\to\infty}\frac{\rho_{2k}}{AG_{2k}}
 =\frac{\sigma}{\sqrt{\mu^2+\sigma^2}}$
 \
 ($\mu$, $\sigma$ fixed)
 }
 \]
 show that, under some circumstances,
 the improvement that is achieved
 by using $\rho_n$ instead of $AG_n$
 can become arbitrarily large.
 \end{EXAMPLE}
 \begin{EXAMPLE}
 \label{ex.6.2} {\it Homoscedastic data with a single outlier.}
 Let $\sigma_i^2=\sigma^2$ for all $i$, $\mu_i=0$
 ($i=1,\ldots,n-1$) and $\mu_n=\mu\geq 0$. Theorem \ref{theo.AG}
 asserts that the bound
 \[
 \mbox{$
 \E R_n\leq AG_n=\sqrt{2\frac{n-1}{n}\mu^2+2n\sigma^2}
 $}
 \]
 is not tight for $n\geq 3$ and $\mu>\frac{n}{\sqrt{n-1}}\sigma$.
 The tight bound has the form
 \[
 \E R_n\leq \rho_n=\sqrt{(n-1)(c^2+\sigma^2)}+\sqrt{(\mu-c)^2+\sigma^2},
 \]
 where $c$ is the unique root of the equation
 \[
 \mbox{
 $\frac{c\sqrt{n-1}}{\sqrt{c^2+\sigma^2}}=\frac{\mu-c}{\sqrt{(\mu-c)^2+\sigma^2}}$, \ \ \
 $0<c<\min\big\{\frac{\mu}{n},\frac{\sigma}{\sqrt{n-2}}\big\}$.
 }
 \]
 Although $\rho_n<AG_n$ (for $\mu\sqrt{n-1}>n\sigma$),
 it is not easy to make direct comparisons. However,
 $c^2<\frac{\sigma^2}{n-2}$ and $(\mu-c)^2<\mu^2$, so that
 $\rho_n<\rho_n'=\frac{n-1}{\sqrt{n-2}}\sigma+\sqrt{\mu^2+\sigma^2}$.
 Hence, for the (non-tight) upper bound $\rho_n'$,
 \[
 \mbox{$
 \lim_{\mu\to\infty}\frac{\rho_n'}{AG_n}=\frac{\sqrt{n}}{\sqrt{2n-2}}
 \ \ \ \ (n\geq3, \ n,\sigma \mbox{ fixed}).
 $}
 \]
 \end{EXAMPLE}

 \begin{REM}
 \label{rem.6.3}
 Example \ref{ex.6.2} and Remark \ref{rem.6.2}
 entail that $\rho_n$ may have a rather
 complicated form when the $\mu_i$'s
 are not all equal. On the other hand,
 $\rho_n$ becomes quite plausible in the case of
 equal $\mu_i$'s; see Example \ref{ex.2}.
 This particular case is useful in concluding
 some facts about the behavior of $\rho_n$ in
 general. Indeed,
 taking into account the obvious relation
 $U(x,y)\geq U(0,y)$, we see that
 for any given $\bmu$ and ${\bsigma}$,
 \[
 \mbox{$
 \rho_n=\phi_n(c,\lambda)\geq
 -(n-2)\lambda+\frac{\lambda}{2}\sum_{i=1}^n
 U\big(0,\frac{\sigma_i}{\lambda}\big)
 =\widehat{\phi}_n(0,\lambda)\geq\widehat{\rho}_n:=\inf_{x\in \R, y>0}\widehat{\phi}_n(x,y),
 $
 }
 \]
 where $\widehat{\rho}_n$ is the upper bound of Theorem
 \ref{th.main}, calculated under $\mu_i=\mu$ for all $i$,
 and for the given ${\bsigma}$.
 Since $\widehat{\rho}_n =\min_{y>0}\widehat{\phi}_n(0,y)$
 admits a simple closed form, see (\ref{eq.tight}),
 we get the following lower bound:
 \[
 \rho_n\geq
 \widehat{\rho}_n=
 \left\{
 \begin{array}{lll}
 \sqrt{2\sum_{i=1}^n \sigma_i^2}, & \mbox{if} & 2\max_i\{\sigma_i^2\}
 \leq \sum_{i=1}^n\sigma_i^2,
 \\
 \max_i\{\sigma_i\} +\sqrt{\sum_{i=1}^{n}\sigma_i^2-\max_i \{\sigma_i^2\}},
 & \mbox{if} & 2\max_i\{\sigma_i^2\}
 \geq \sum_{i=1}^n\sigma_i^2,
 \end{array}
 \right.
 \mbox{any }\bmu,{\bsigma}.
 \]
 Since $U(x,y)>U(0,y)$ for $x\neq 0$,
 the equality holds only if all the $\mu_i$'s are equal.
 Despite its weakness, this lower bound provides an idea of
 what can  be expected
 for the actual size of $\rho_n$.
 It is also helpful in giving some light to
 the observation that, provided
 the means are small compared to the variances,
 the $AG_n$ bound tends to be tight.
 More precisely,
 assume that $\min_i \{\sigma_i^2\} \to \infty$ and
 $(\sum_{i=1}^n (\mu_i-\overline{\mu})^2)/(\sum_{i=1}^n \sigma_i^2)\to 0$
 (in particular, $\max_i |\mu_i-\overline{\mu}|\leq C<\infty$ suffices for this).
 Then, the homogeneity assumption
 $\max_i\{\sigma_i^2\}\leq (n-1)\min_i\{\sigma_i^2\}$
 is sufficient for the asymptotic tightness of the $AG_n$
 bound (for fixed $n\geq 3$).
 Indeed, from this assumption we get
 $\sum_{i=1}^n \sigma_i^2\geq \max_{i}\{\sigma_i^2\}+(n-1)\min_i \{\sigma_i^2\}
 \geq 2\max\{\sigma_i^2\}$, and thus,
 $\widehat{\rho}_n=\sqrt{2\sum_{i=1}^n\sigma_i^2}$.
 Hence,
 \[
 1\geq
 \Big(\frac{\rho_n}{AG_n}\Big)^2
 \geq
 \Big(\frac{\widehat{\rho}_n}{AG_n}\Big)^2
 =
 \frac{\sum_{i=1}^n\sigma_i^2}
 {\sum_{i=1}^n\{(\mu_i-\overline{\mu})^2+\sigma_i^2\}}
 =
 \frac{1}
 {1+(\sum_{i=1}^n (\mu_i-\overline{\mu})^2)/(\sum_{i=1}^n\sigma_i^2)}\to 1.
 \]
 Therefore, under the above circumstances,
 the improvement achieved by using
 $\rho_n$ instead of $AG_n$ becomes negligible.
 \end{REM}

 \section{The case $\bbb{n}\, \bbb{=}\, \bbb{2}$ and
 further remarks}
 \setcounter{equation}{0}
 \label{sec.last}
 For $n=2$ the bound $\rho_2$ admits a closed form. More precisely,
 from Theorem \ref{th.main},
 \be
 \label{eq.7.1}
 \E R_2\leq \rho_2, \ \
 \mbox{ where }
 \rho_2:=\inf_{c\in\R,\lambda>0}\phi_2(c,\lambda)
 =\sqrt{(\mu_1-\mu_2)^2+(\sigma_1+\sigma_2)^2};
 \ee
 see Remark \ref{rem.5.2}. The inequality \eqref{eq.7.1}
 is tight, since the equality
 is attained
 by
 (and characterizes)
 the random pair $(X_1,X_2)$ with distribution
 given by
 \be
 \label{eq.7.2}
 \begin{array}{c}
 \vspace{1ex}\mbox{$
 \Pr\big[
 X_1=\frac{\sigma_2\mu_1+\sigma_1\mu_2}{\sigma_1+\sigma_2}
 +\frac{\sigma_1}
 {\sigma_1+\sigma_2}\rho_2,
 \
 X_2=\frac{\sigma_2\mu_1+\sigma_1\mu_2}{\sigma_1+\sigma_2}
 -\frac{\sigma_2}
 {\sigma_1+\sigma_2}\rho_2
 \big]
 =\frac{1}{2}\big(1+\frac{\mu_1-\mu_2}{\rho_2}\big),
 $}
 \\
 \mbox{$
 \Pr\big[
 X_1=\frac{\sigma_2\mu_1+\sigma_1\mu_2}{\sigma_1+\sigma_2}
 -\frac{\sigma_1}
 {\sigma_1+\sigma_2}\rho_2,
 \
 X_2=\frac{\sigma_2\mu_1+\sigma_1\mu_2}{\sigma_1+\sigma_2}
 +\frac{\sigma_2}
 {\sigma_1+\sigma_2}\rho_2
 \big]
 =\frac{1}{2}\big(1+\frac{\mu_2-\mu_1}{\rho_2}\big).
 $}
 \end{array}
 \ee
 Therefore, ${\cal E}_2(\bmu,{\bsigma})$ is a singleton.
 Also, $AG_2=\sqrt{(\mu_1-\mu_2)^2+2\sigma_1^2+2\sigma_2^2}$, and it is
 worth pointing out that the bound $AG_2$ is tight if and
 only if $\sigma_1=\sigma_2$. Another observation is that
 the extremal random vector for the expected range
 coincides with the (unique) extremal random vector for the expected
 maximum (see \eqref{eq.BNT}). However,
 this is not a coincidence. In view
 of the obvious relationship
 \be
 \label{eq.max12}
 \mbox{$
 R_2=|X_1-X_2|=2\max\{X_1,X_2\}-X_1-X_2
 =2 X_{2:2}-X_1-X_2,
 $}
 \ee
 a bound for the maximum
 can be translated to a bound for the range, and vice-versa
 (provided that the expectations, $\mu_1,\mu_2$, of $X_1,X_2$,
 are known). In this sense, the bound
 $\rho_2$ turns to be a particular case of the results given by
 Bertsimas, Natarajan and Teo (2004, 2006), namely
 \[
 \hspace{-1ex}
 \mbox{$
 \rho_2=\sup \E R_2
 =\sup \E\{ 2 X_{2:2}-X_1-X_2\}
 =2\sup \E X_{2:2} -\mu_1-\mu_2=2BNT_2-\mu_1-\mu_2,
 $}
 \]
 and the equality characterizes the same extremal distribution
 as for the maximum. Consequently, it is of some interest to observe that
 the bound $BNT_2$ admits a closed form, namely,
 \[
 \mbox{$
 BNT_2=\frac12(\mu_1+\mu_2)
 +\frac12\sqrt{(\mu_1-\mu_2)^2+(\sigma_1+\sigma_2)^2}.
 $}
 \]
 Note also that
 the $BNT_2$--bound improves
 the corresponding Arnold-Groeneveld bound (\ref{eq.AG.general})
 for the expected maximum only in the case where $\sigma_1\neq \sigma_2$.

 It is also worth pointing out that a particular application
 of the main result in Papadatos (2001a) yields an even better
 (than $BNT_2$, $AG_2$ and $\rho_2$) bound. Indeed, setting
 $\rho:=\Corr(X_1,X_2)$, it follows from Papadatos' results
 that for any $(X_1,X_2)\in{\cal F}_2(\bmu,{\bsigma})$,
 \be
 \label{eq.7.3}
 \E R_2\leq \gamma_2:=\sqrt{(\mu_1-\mu_2)^2
 +(\sigma_1+\sigma_2)^2-2(1+\rho)\sigma_1\sigma_2}.
 \ee
 Obviously, $\gamma_2\leq \rho_2$ with equality if and only if
 $\rho=-1$. This inequality explains the fact that the extremal
 random pair $(X_1,X_2)$
 (that attains the bounds $\rho_2$ and $BNT_2$)
 has correlation $\rho=-1$; see (\ref{eq.7.2}).

 The preceding inequalities have some interest because they provide a basis
 for the investigation of the dependence structure of an ordered pair.
 This kind of investigation is particularly useful for its application to
 reliability systems; see Navarro and Balakrishnan (2010). On the other hand,
 in view of the obvious facts $X_{1:2}+X_{2:2}=X_1+X_2$ and
 $X_{1:2}X_{2:2}=X_1X_2$,
 we get the relation
 \be
 \label{eq.cov12}
 \mbox{$
 \Cov[X_{1:2},X_{2:2}]=\rho\sigma_1\sigma_2-\frac14 (\mu_2-\mu_1)^2+\frac14 (\E R_2)^2,
 \ \ (X_1,X_2)\in {\cal F}_2(\bmu,{\bsigma}),
 $}
 \ee
 where $\rho=\Corr(X_1,X_2)$.
 Thus, any bound (upper or lower) for $\E R_2$ can be translated
 to a bound for $\Cov(X_{1:2},X_{2:2})$ as well as for $\E X_{2:2}$; see
 Papathanasiou (1990), Balakrishnan and Balasubramanian (1993).
 Therefore,
 it is of some interest to know whether the bound in \eqref{eq.7.3} is tight
 for given $\rho$. This is indeed the case but,
 to the best of our knowledge, this elementary fact does not
 seem to be well-known, and we shall provide a simple proof here.
 To this end, let $\bmu=(\mu_1,\mu_2)$, ${\bsigma}=(\sigma_1,\sigma_2)$
 (with $\sigma_1>0$, $\sigma_2>0$), $-1\leq \rho\leq 1$, and define the section
 \[
  {\cal F}_2(\bmu,{\bsigma};\rho):=
  \big\{(X_1,X_2)\in {\cal F}_2(\bmu,{\bsigma})
  :\Corr(X_1,X_2)=\rho\big\}.
 \]
 Then we have the following.
 \begin{THEO}
 As $(X_1,X_2)$ varies in ${\cal F}_2(\bmu,{\bsigma};\rho)$,
 \label{theo.7.1}
  \be
  \label{eq.7.4}
   \inf \E R_2=|\mu_2-\mu_1|,  \ \ \
   \sup  \E R_2=\sqrt{(\mu_2-\mu_1)^2+(\sigma_1+\sigma_2)^2-2(1+\rho)\sigma_1\sigma_2}.
  \ee
 \end{THEO}

 \begin{REM}
 \label{rem.7.1} From the proof it follows that (the probability law of)
 the extremal vector $(X_1,X_2)$
  $\in{\cal F}_2(\bmu,{\bsigma};\rho)$
 that attains the equality in \eqref{eq.7.3}
 is unique if and only if
 either (i) $\rho=-1$ or (ii) $\sigma_1\neq \sigma_2$ and $\rho=1$.
 With this in mind, let us keep $\mu_1,\mu_2,\sigma_1,\sigma_2$ constant,
 and write
 $\gamma_2=\gamma_2(\rho)$ for the quantity defined by \eqref{eq.7.3}.
 Then, $\gamma_2(\rho)$ is strictly decreasing in $\rho$
 (recall that $\sigma_1>0$, $\sigma_2>0$), attaining its maximum value
 at $\rho=-1$. By definition, $\gamma_2(-1)=\rho_2$ (see \eqref{eq.7.1}),
 and thus,
 for the equality $\E R_2=\rho_2$ it is necessary that $\rho=-1$.
 This observation verifies that the unique distribution that attains
 the equality
 in \eqref{eq.7.1} is the $BNT_2$--distribution, given by (\ref{eq.7.2}).
 \end{REM}

 In view of \eqref{eq.max12}, \eqref{eq.cov12},
 the following result is straightforward
 from Theorem \ref{theo.7.1}.
 \begin{COR}
 \label{cor.7.1}
 Let $(X_1,X_2)\in{\cal F}_2(\mu_1,\mu_2,\sigma_1,\sigma_2;\rho)$ with
 $\sigma_1>0,\sigma_2>0$. Then,
 \[
 \begin{array}{l}
 \vspace{1.3ex}
 \max\{\mu_1,\mu_2\}\leq \E\big\{\hspace{-.5ex}\max\{X_1,X_2\}\hspace{-.2ex}\big\}\leq
 \frac{1}{2}(\mu_1+\mu_2)+\frac{1}{2}
 \sqrt{(\mu_2-\mu_1)^2+\sigma_1^2+\sigma_2^2-2\rho\sigma_1\sigma_2},
 \\
 \rho\sigma_1\sigma_2\leq \Cov\hspace{-.4ex}\big[\min\{X_1,X_2\},\max\{X_1,X_2\}\big]
 \leq
 \frac{1}{4}
 \big(\sigma_1^2+\sigma_2^2+2\rho\sigma_1\sigma_2\big).
 \end{array}
 \]
 All bounds are best possible.
 \end{COR}

 \noindent
 It is worth pointing out that, as Corollary \ref{eq.7.1} shows,
 the covariance of an ordered pair can never be
 smaller than the covariance of the observations and, in particular,
 an ordered pair formed from non-negatively correlated observations
 is non-negatively correlated. While these facts,
 as well as the lower covariance bound
 of an ordered pair,
 are well-known (see eq.'s (2.9), (2.11)
 in Navarro and Balakrishnan (2010)),
 the upper bound seems to be of some interest.

 There are some propositions and questions for further
 research. An obvious one is in extending the main result
 of Theorem \ref{theo.6.1} and of (\ref{eq.BNT}) to more general
 $L$-statistics.
 Recall that the tight bound for any $L$-statistic under the i.d.\
 assumption is known from the work of Rychlik (1993b). However,
 Rychlik's result is not applicable
 if arbitrary multivariate distributions are
 allowed for the data.

 A second one concerns extension to other $L$-statistics
 of the bounds given in
 Corollary \ref{cor.7.1} and Theorem \ref{theo.7.1}
 for $n\geq 3$, noting that these bounds have a different nature, because
 they use covariance information from the data.
 It is particularly interest to know the tight bounds for the
 the expected range and the expected maximum under mean-variance-covariance
 information on the observations.
 Non-tight bounds of this form are
 given, e.g., in Aven (1985), Papadatos (2001a).
 It is worth pointing out that
 some sophisticated optimization techniques
 (semidefinite programming)
 have been fruitfully applied to this kind of problems, especially
 for the maximum and the range.
 The interested reader is referred to Natarajan and Teo (2014),
 where some financial applications of the range bounds are also included.
 However, note that one would hardly discover the simple formula (\ref{eq.7.3})
 from the (reduced) semidefinite program in Natarajan and Teo's Section 4.

 A lot of research has been devoted in deriving
 distribution and expectation bounds for
 $L$-statistics based on random vectors
 with given marginals; see Arnold (1980, 1985, 1988), Caraux and Gascuel (1992),
 Gascuel and Caraux (1992),
 Meilijson and Nadas (1979), Papadatos (2001b), Rychlik (1992b,
 1993a, 1994, 1995, 1998, 2007), Gajek and Rychlik (1996, 1998).
 The results by Lai and Robbins (1976), Nagaraja (1981)
 and Arnold and Balakrishnan (1989)
 show that some deterministic
 inequalities play an important role in
 the derivation of tight bounds for $L$-statistics;
 see Rychlik (1992a).
 On the other hand, the deterministic inequality (\ref{eq.deterministic})
 can be viewed as a range analogue of the inequality
 from Lai and Robbins (1976). Noting that
 the Lai-Robbins
 inequality yields the tight bound for the expected
 maximum under completely known marginal distributions
 (see Bertsimas, Natarajan and Teo (2006),
 Meilijson and Nadas (1979)), it would not be
 surprising if (\ref{eq.deterministic}) could
 produce
 the best possible bound
 for the expected range. Thus, a natural question is
 whether it is true that for all multivariate vectors
 with given marginal distributions $F_1,\ldots,F_n$
 and finite
 first moment,
 \[
 \mbox{$
 \sup \E R_n =\inf_{c\in \R, \lambda>0}\big\{ \hspace{-.5ex}
 -(n-2)\lambda+\frac{1}{2}\sum_{i=1}^n
 \E\big[ \big|(X_i-c)-\lambda\big|+\big|(X_i-c)+\lambda\big|\big]\big\}.
 $}
 \]
 Note that the RHS is an upper bound for the LHS, and
 depends only on $F_1,\ldots,F_n$.



 {
 \begin{appendix}
 \section{\normalsize Appendix: Proofs}
 \label{app}
 \small
 \noindent
 \noindent
 \begin{PR}{\small\bf Proof of Lemma \ref{lem.1}:}
 Fix $c\in\R$ and $\lambda>0$ and set
 $y_1=c-\lambda$, $y_2=c+\lambda$, so that $y_1<y_2$.
 Observe that $R_n=X_{n:n}-X_{1:n}$ and
 \[
 \mbox{$
 \sum_{i=1}^n \big\{|X_i-y_1|+|X_i-y_2|\big\}
 =
 \sum_{i=1}^n \big\{|X_{i:n}-y_1|+|X_{i:n}-y_2|\big\}.
 $}
 \]
 Hence,
 \begin{eqnarray*}
 &&
 \hspace{-15ex}
 \mbox{$
 \sum_{i=1}^n \big\{|X_i-y_1|+|X_i-y_2|\big\}
 -(n-2)(y_2-y_1)-2R_n
 $}
 \\
 &&
 \mbox{$
 =
 \sum_{i=2}^{n-1}
 \big\{|X_{i:n}-y_1|+|X_{i:n}-y_2|-(y_2-y_1)\big\}
 $}
 \\
 &&
 \hspace{5ex}
 \mbox{$
  +\big\{|X_{1:n}-y_1|+|X_{n:n}-y_1|-(X_{n:n}-X_{1:n})\big\}
 $}
 \\
 &&
 \hspace{5ex}
 \mbox{$
  +\big\{|X_{1:n}-y_2|+|X_{n:n}-y_2|-(X_{n:n}-X_{1:n})\big\}.
 $}
 \end{eqnarray*}
 For each $i\in\{2,\ldots,n-1\}$ we have
 \[
 y_2-y_1=|y_2-y_1|=\big|(X_{i:n}-y_1)-(X_{i:n}-y_2)\big|\leq
 \big|X_{i:n}-y_1\big|+\big|X_{i:n}-y_2\big|,
 \]
 with equality if and only if $y_1\leq X_{i:n}\leq y_2$.
 Since the sum $\sum_{i=2}^{n-1}
 \big\{|X_{i:n}-y_1|+|X_{i:n}-y_2|-(y_2-y_1)\big\}$
 contains only non-negative terms,
 it follows that
 \[
 \mbox{$
 \sum_{i=2}^{n-1}
 \big\{|X_{i:n}-y_1|+|X_{i:n}-y_2|-(y_2-y_1)\big\}\geq 0,
 $}
 \]
 with equality if and only if $y_1\leq X_{2:n}\leq\cdots\leq X_{n-1:n}\leq y_2$.
 Also, for $y=y_1$ or $y_2$,
 \[
  X_{n:n}-X_{1:n}=\big| (X_{n:n}-y)-(X_{1:n}-y)\big|\leq
  \big| X_{1:n}-y\big|+
  \big| X_{n:n}-y\big|
 \]
 with equality if and only if $X_{1:n}\leq y\leq X_{n:n}$.
 Therefore,
 \[
 \mbox{$
 -2R_n-(n-2)(y_2-y_1)+
 \sum_{i=1}^n \big\{|X_i-y_1|+|X_i-y_2|\big\}\geq 0
 $}
 \]
 with equality if and only if
 $
 X_{1:n}\leq y_1
 \leq X_{2:n}\leq \cdots\leq X_{n-1:n}\leq y_2\leq X_{n:n}.
 \ \ \
 \Box
 $
 \medskip
 \end{PR}

 \noindent
 \begin{PR}{\small\bf Proof of Lemma \ref{lem.2}:}
 In case $\mu^2+\sigma^2\geq 4$
 it suffices to use the inequality
 \[
 \mbox{$
 |X-1|+|X+1|\leq \sqrt{\mu^2+\sigma^2}+\frac{X^2}{\sqrt{\mu^2+\sigma^2}},
 $}
 \]
 where the equality holds if and only if
 $X\in\{-\sqrt{\mu^2+\sigma^2},\sqrt{\mu^2+\sigma^2}\}$.
 Taking expectations we get
 \[
 \mbox{$
 \E\big\{|X-1|+|X+1|\big\}\leq
 \sqrt{\mu^2+\sigma^2}+\frac{\E X^2}{\sqrt{\mu^2+\sigma^2}}
 =2\sqrt{\mu^2+\sigma^2}.
 $}
 \]
 For equality $X$ has to assume
 the values $\pm\sqrt{\mu^2+\sigma^2}$.
 Set $p=\Pr[X=\sqrt{\mu^2+\sigma^2}]$ so that  $1-p=\Pr[X=-\sqrt{\mu^2+\sigma^2}]$.
 The relation $\E X^2=\mu^2+\sigma^2$
 is satisfied for any value of
 $p\in[0,1]$, while the condition $\E X=\mu$ specifies $p$ to be as in (a).

 Next, we assume that $2|\mu|<\mu^2+\sigma^2<4$ and use the inequality
 \[
 \mbox{$
 |X-1|+|X+1|\leq 2+\frac{1}{2}X^2,
 $}
 \]
 in which the equality holds if and only if $X\in\{-2,0,2\}$.
 Taking expectations we again conclude (\ref{eq.in}) with
 $U(\mu,\sigma)$ given by the second line of (\ref{eq.u}).
 It is easy to see that the unique random variable in ${\cal F}_1(\mu,\sigma)$
 that assumes values in the set $\{-2,0,2\}$ is the one given by (b).

 Next, suppose that $\mu^2+\sigma^2\leq 2\mu$, and hence, $0<\mu<2$.
 Working as before, it suffices to take expectations in the inequality
 \[
 \mbox{$
 |X-1|+|X+1|\leq 2
 +\frac{(X-1+\sqrt{(\mu-1)^2+\sigma^2})^2}{2\sqrt{(\mu-1)^2+\sigma^2}},
 $}
 \]
 in which the equality holds if and only if $X\in\{x_1,x_2\}$,
 where $x_1=1-\sqrt{(\mu-1)^2+\sigma^2}$, $x_2=1+\sqrt{(\mu-1)^2+\sigma^2}$.
 Note that $0<(\mu-1)^2+\sigma^2=1-[2\mu-(\mu^2+\sigma^2)]\leq 1$; thus,
 $0\leq x_1<1<x_2\leq 2$. Now it is easily seen that the unique
 random variable in ${\cal F}_1(\mu,\sigma)$
 that assumes values in the set $\{x_1,x_2\}$ is the one given by (c).
 Observing that $|X-1|+|X+1|$ is even,
 the case $\mu^2+\sigma^2\leq -2\mu$ is reduced
 to the previous one by considering $-X\in{\cal F}_1(-\mu,\sigma)$.
 $\Box$
 \medskip
 \end{PR}

 \noindent
 \begin{PR}{\small\bf Proof of Lemma \ref{lem.3}:}
 For $n=1$ both (\ref{eq.bivariate}) and
 (\ref{eq.condition}) are invalid, so we have nothing to prove.
 For $n=2$ the result is trivial (we have uniqueness
 if  (\ref{eq.bivariate}) is satisfied; we have equality
 in (\ref{eq.condition}) whenever it is fulfilled).
 Assume $n\geq 3$
 and consider the set of all probability matrices with the given marginals,
 \[
 \mbox{$
 {\cal M}({\bpp},{\bqq})=\big\{Q=(q_{ij})\in\R^{n\times n}:q_{ij}\geq 0, \,
 \sum_{i=1}^n q_{ij}=q_j, \, \sum_{j=1}^n q_{ij}=p_i \mbox{ for all } i,j\big\}.
 $}
 \]
 The set ${\cal M}({\bpp},{\bqq})$ is nonempty since, e.g.,
 it contains the matrix $Q=(p_i q_j)$. Also, the function
 $f(Q):=\mbox{trace}(Q)=\sum_{i=1}^n q_{ii}$ is continuous with respect
 to the total variation distance,
 $d(Q,\widetilde{Q})=\sum_{i,j}|q_{ij}-\widetilde{q}_{ij}|$
 (or any other equivalent metric on $\R^{n\times n}$).
 Moreover, ${\cal M}({\bpp},{\bqq})$ is a compact subset of
 $\R^{n\times n}$, since it is obviously closed, and it is
 contained in a ball with center the null matrix $O_{n\times n}$
 and (total variation) radius $1$.
 It follows that $f(Q)$ attains its minimum value for some
 $Q^*\in {\cal M}({\bpp},{\bqq})$.

 Let $(X,Y)\sim Q^*=(q_{ij}^*)$ where $Q^*\in{\cal M}({\bpp},{\bqq})$ is
 a minimizing matrix.
 Then, $X\sim {\bpp}$, $Y\sim{\bqq}$ and $f(Q^*)=\Pr[X=Y]$.
 A simple argument shows that
 the principal diagonal of any minimizing matrix $Q^*$ can contain
 at most one nonzero entry.
 Indeed, if $q^*_{ii}>0$ and $q^*_{jj}>0$ with $i\neq j$,
 set
 $\gamma=\min\{q^*_{ii},q^*_{jj}\}>0$,
 and consider the matrix $\widetilde{Q}=(\widetilde{q}_{ij})$
 which differs from $Q^*$ only in the following four
 entries:
 $\widetilde{q}_{ii}=q^*_{ii}-\gamma$,
 $\widetilde{q}_{jj}=q^*_{jj}-\gamma$,
 $\widetilde{q}_{ij}=q^*_{ij}+\gamma$,
 $\widetilde{q}_{ji}=q^*_{ji}+\gamma$.
 Since the row/column sums are unaffected and the elements of
 $\widetilde{Q}$ are nonnegative,
 it is clear that $\widetilde{Q}\in{\cal M}({\bpp},{\bqq})$ and
 we arrived at the contradiction
 $f(\widetilde{Q})=f(Q^*)-2\gamma<f(Q^*)$.
 Therefore, all diagonal entries of a minimizing matrix
 $Q^*$ have to be zero,
 with the possible exception of at most one of them.

 {\it Sufficiency}:
 Assume that (\ref{eq.condition}) is satisfied, and suppose that
 $\min_{Q} f(Q)=f(Q^*)=\theta>0$.
 Let $q^*_{kk}=\theta$ and thus, $q^*_{ii}=0$ for all $i\neq k$.
 Then,
 \[
 \Pr[\{X=k\} \cup \{Y=k\}]=\Pr[X=k]+\Pr[Y=k]-\Pr[X=k,Y=k]=p_k+q_k-\theta.
 \]
 Since $1-p_k-q_k\geq 0$ (from (\ref{eq.condition})) we thus obtain
 \[
 \Pr[X\neq k, Y\neq k]=1-\Pr[\{X=k\} \cup \{Y=k\}]=\theta+(1-p_k-q_k)\geq
 \theta>0.
 \]
 On the other hand, since $q^*_{ii}=0$ for all $i\neq k$, we have
 \[
 \Pr[X\neq k, Y\neq k]=\sum_{(i,j):\ i\neq k, j\neq k, i\neq j} q^*_{ij}.
 \]
 The above probability is at least $\theta$, and thus,
 strictly positive.
 It follows that the sum contains at least
 one positive term. Hence, we can find
 two indices $r,s$ with $r\neq k$,
 $s\neq k$, $r\neq s$, such that $q^*_{rs}>0$.
 Set
 $\delta=\min\{\theta,q^*_{rs}\}>0$ and
 consider the matrix $\widetilde{Q}=(\widetilde{q}_{ij})$
 which differs from $Q^*$ only in the elements
 $\widetilde{q}_{kk}=q^*_{kk}-\delta=\theta-\delta$,
 $\widetilde{q}_{rs}=q^*_{rs}-\delta$,
 $\widetilde{q}_{rk}=q^*_{rk}+\delta$,
 $\widetilde{q}_{ks}=q^*_{ks}+\delta$.
 Since the row/column sums are unaffected and the elements of
 $\widetilde{Q}$ are nonnegative,
 it is clear that $\widetilde{Q}\in{\cal M}({\bpp},{\bqq})$, and
 this results to the contradiction
 $f(\widetilde{Q})=\theta-\delta<\theta$.
 Thus, $f(Q^*)=\Pr[X=Y]=0$; this proves the existence of random vectors
 satisfying (\ref{eq.bivariate}).

 {\it Necessity}: This is entirely obvious. For, if a random vector $(X,Y)$
 satisfies (\ref{eq.bivariate}) then $(X,Y)\sim Q$ for some
 $Q\in{\cal M}({\bpp},{\bqq})$ with $q_{ii}=0$ for all $i$.
 Thus, for any $i$,
 \[
 p_i+q_i=p_i+q_i-q_{ii}=\Pr[X=i]+\Pr[Y=i]-q_{ii}=\Pr[\{X=i\}\cup\{Y=i\}]\leq 1.
 \]

 {\it Uniqueness}: Assume that $\max_{i}\big\{p_i+q_i\}=1$ and choose
 $k$ with $p_k+q_k=1$. If $(X,Y)\sim Q$ satisfies (\ref{eq.bivariate}), we have
 $\Pr[\{X= k\}\cup \{Y=k\}]=p_k+q_k-q_{kk}\geq p_k+q_k-\Pr[X=Y]=p_k+q_k=1$.
 It follows that $Q$ can have non-zero entries only in its $k$-th row
 and in its $k$-th column. Thus, $q_{ik}=p_i$ for all $i\neq k$,
 $q_{kj}=q_j$ for all $j\neq k$ and $q_{ij}=0$ otherwise;
 hence, $Q$ is uniquely determined from
 ${\bpp}$, ${\bqq}$. Note that $k$ need not be unique, but $Q$ is always unique.
 For example, if ${\bpp}=(1-p,p,0,\ldots,0)$ and ${\bqq}=(p,1-p,0,\ldots,0)$
 with $0\leq p\leq 1$, we obtain the unique solution to
 (\ref{eq.bivariate}) as $\Pr[X=2,Y=1]=p=1-\Pr[X=1,Y=2]$. In fact,
 one can easily verify that this example describes the most general case
 (modulo the positions of $p, 1-p$) where the relation $p_k+q_k=1$ can hold for
 more than one index $k$.

 {\it Non-uniqueness}: Suppose that all $p_i$ and $q_i$ are positive
 and that (\ref{eq.condition}) holds as a strict inequality,
 that is,
 $p_i+q_i<1$ for all $i$.
 [The last assumption is possible only if
 $n\geq 3$.]
 Set $\beta=\frac{1}{n^2}\big[1-\max_i\big\{p_i+q_i\big\}\big]>0$,
 $\delta=\min_{i,j}\{p_iq_j\}>0$ and
 $\epsilon=\min\{\beta,\delta\}>0$. Define
 \[
 {\cal M}_{\epsilon}({\bpp},{\bqq}):=\big\{ Q\in {\cal M}({\bpp},{\bqq})
 : q_{ij}\geq \epsilon \mbox{ for all } i,j \mbox{ with } i\neq j\big\}.
 \]
 Observe that
 ${\cal M}_{\epsilon}({\bpp},{\bqq})$ is a nonempty (since it contains $Q=(p_i q_j)$)
 compact subject of $\R^{n\times n}$.
 Applying the same arguments as in the beginning of the proof
 we see that the continuous function $f(Q)=\mbox{trace}(Q)$
 attains its minimum value at
 a matrix $Q^*_{\epsilon}=(q_{ij}^*)\in{\cal M}_{\epsilon}({\bpp},{\bqq})$;
 $Q^*_{\epsilon}$ has at most one nonzero diagonal entry while,
 by the definition of ${\cal M}_{\epsilon}({\bpp},{\bqq})$, all off-diagonal
 entries are at least $\epsilon$. Let $(X,Y)\sim Q^*_{\epsilon}$.
 Assuming $\Pr[X=Y]=\theta>0$ we can find a unique index $k$ such that
 $q^*_{kk}=\theta$;
 then, $\Pr[\{X=k\}\cup \{Y=k\}]=p_k+q_k-\theta$. Since
 $q_{ii}^*=0$ for $i\neq k$, we have
 \begin{eqnarray*}
 &&
 \sum_{(i,j):\ i\neq k,j\neq k,i\neq j}q_{ij}^*=\Pr[X\neq k,Y\neq k]=\theta+(1-p_k-q_k)
 \geq \theta+\big[1-\max_{i}\big\{p_i+q_i\big\}\big] \\
 &&
 \hspace{16ex}
 =\theta+n^2 \beta \geq \theta +n^2\epsilon>n^2\epsilon.
 \end{eqnarray*}
 This sum contains $(n-1)(n-2)<n^2$ terms and
 the inequality shows that at least one of them
 is greater than $\epsilon$.
 Thus, we can find
 two indices $r,s$ with $r\neq k$,
 $s\neq k$, $r\neq s$, such that $q^*_{rs}>\epsilon$;
 say $q^*_{rs}=\epsilon+\gamma$ with $\gamma>0$.
 Set
 $\lambda=\min\{\theta,\gamma\}>0$ and
 consider the matrix $\widetilde{Q}_{\epsilon}=(\widetilde{q}_{ij})$,
 which differs from $Q^*_{\epsilon}$ at exactly
 the four elements
 $\widetilde{q}_{kk}=q^*_{kk}-\lambda=\theta-\lambda\geq 0$,
 $\widetilde{q}_{rs}=q^*_{rs}-\lambda=\epsilon+(\gamma-\lambda)\geq \epsilon$,
 $\widetilde{q}_{rk}=q^*_{rk}+\lambda$,
 $\widetilde{q}_{ks}=q^*_{ks}+\lambda$.
 It is clear that $\widetilde{Q}_{\epsilon}
 \in{\cal M}_{\epsilon}({\bpp},{\bqq})$ and, once again,
 it contradicts the definition of $Q_{\epsilon}^*$:
 $f(\widetilde{Q}_{\epsilon})=\theta-\lambda<\theta=f(Q_{\epsilon}^*)$.
 Thus, $f(Q^*_{\epsilon})=\Pr[X=Y]=0$. This shows the existence
 of random vectors $(X,Y)$
 satisfying (\ref{eq.bivariate}) with the additional
 property $\Pr[X=i,Y=j]\geq \epsilon>0$ for all $i\neq j$, provided that
 $\epsilon>0$ is sufficiently small. Given a probability matrix
 $Q^*_{\epsilon}=(q_{ij}^*)$ of this form, it is easy to construct
 a second solution, $Q=(q_{ij})$, to (\ref{eq.bivariate}); e.g., set
 $q_{12}=q_{12}^*-\epsilon/2$, $q_{13}=q_{13}^*+\epsilon/2$,
 $q_{21}=q_{21}^*+\epsilon/2$, $q_{23}=q_{23}^*-\epsilon/2$,
 $q_{31}=q_{31}^*-\epsilon/2$, $q_{32}=q_{32}^*+\epsilon/2$,
 and leave the rest entries unchanged.
 Finally, it is easy to see that if $Q_0$, $Q_1$ both solve
 (\ref{eq.bivariate}), the same is true for
 $Q_t=tQ_1+(1-t)Q_0$, $0\leq t\leq 1$, and the proof is complete.
 $\Box$
 \medskip
 \end{PR}

 \noindent
 \begin{PR}{\small\bf Proof of Theorem \ref{theo.AG}:}
 Assume that $\E R_n=AG_n$ for some random vector
 ${\bXX}$  with
 $\E{\bXX}=\bmu$ and  $\Var{\bXX}=\bsigma^2$.
 Set $c=\overline{\mu}$,
 $\lambda=\frac{1}{4}AG_n>0$
 and take expectations in (\ref{eq.deterministic})
 to get (cf.\ Remark \ref{rem.2})
 \begin{eqnarray*}
 \mbox{$
 AG_n=\E R_n \leq
 \frac{-(n-2)AG_n}{4}+
 \frac{AG_n}{8}\sum_{i=1}^{n}
 \E
 \Big\{\Big|\frac{X_i-\overline{\mu}}{AG_n/4}-1\Big|
 + \Big|\frac{X_i-\overline{\mu}}{AG_n/4}+1\Big|
 \Big\}.
 $}
 \end{eqnarray*}
 Next, from $|y-1|+|y+1|\leq 2 +\frac{1}{2}y^2$ with equality if and only
 if $y\in\{-2,0,2\}$ we get
 \[
 \mbox{$
 \sum_{i=1}^{n}
 \E
 \Big\{\Big|\frac{X_i-\overline{\mu}}{AG_n/4}-1\Big|
 + \Big|\frac{X_i-\overline{\mu}}{AG_n/4}+1\Big|
 \Big\}\leq 2n+\frac{1}{2}\sum_{i=1}^n \E
 \Big\{ \Big(\frac{X_i-\overline{\mu}}{AG_n/4}\Big)^2
 \Big\}=2n+4.
 $}
 \]
 Since $\frac{-(n-2)AG_n}{4}+\frac{AG_n}{8} (2n+4)=AG_n$,
 it follows that the preceding inequalities are, in fact, equalities.
 Therefore, $\E R_n=AG_n$ is equivalent to
 (\ref{eq.equality})
 (with $c=\overline{\mu}$, $\lambda=\frac{1}{4}AG_n$) and
 $\frac{X_i-\overline{\mu}}{AG_n/4} \in\{-2,0,2\}$,
 $i=1,\ldots,n$
 (of course, it suffices to hold with probability $1$).
 Hence, $\E R_n=AG_n$ if and only if
 \be
 \label{3.7}
 \begin{array}{ll}
 \mbox{(a)} &X_{1:n}\leq \overline{\mu}-\frac{1}{4}AG_n\leq X_{2:n}\leq
 \cdots\leq X_{n-1:n}\leq \overline{\mu}+\frac{1}{4}AG_n\leq X_{n:n}
 \vspace{-1ex}
 \\
 \hspace{-4ex}\mbox{and}
 \vspace{-1ex}
 &
 \\
 \mbox{(b)} &
 X_i\in\big\{\overline{\mu}-\frac{AG_n}{2},\overline{\mu},
 \overline{\mu}+\frac{AG_n}{2}\big\},
 \ \
 i=1,\ldots,n,
 \end{array}
 \ee
 with probability $1$.
 Therefore, the (essential) support of any extremal random vector is a subset of
 \[
 \mbox{$
 S:=\big\{\big(\overline{\mu},\ldots,\overline{\mu},\overline{\mu}+\frac{AG_n}{2},
 \overline{\mu},\ldots, \overline{\mu},
 \overline{\mu}-\frac{AG_n}{2},\overline{\mu},\ldots,\overline{\mu}\big)\big\},
  $}
 \]
 where the plus and minus signs can appear at any two (different) places.
 Clearly,
 $S$ has $n(n-1)$ elements and can be written as
 \[
 \mbox{$
 S= \big\{\overline{\mu}\, {\bone} +\frac{{\bee}(i)-{\bee}(j)}{2}
 AG_n:(i,j)\in\{1,\ldots,n\}^2, i\neq j\big\}.
 $}
 \]
 Let $S':=\big\{(i,j) \in \{1,\ldots,n\}^2:i\neq j\big\}$.
 The function $g:S'\to S$,
 that sends $(i,j)$ to
 $g(i,j)=\overline{{\mu}}\, {\bone} +\frac{{\bee}(i)-{\bee}(j)}{2}
 AG_n$, is a bijection. It follows that $(X,Y):=g^{-1}({\bXX})$
 is a random pair with values in a subset of $S'$, and
 ${\bXX}=g(X,Y)$; this verifies the representation (\ref{eq.extremal}).
 For $i\in\{1,\ldots,n\}$ we set
 \[
 \mbox{$
 p_i^+:=\Pr[X_i=\overline{\mu}+\frac{AG_n}{2}]=\Pr[X=i], \ \
 p_i^-=\Pr[X_i=\overline{\mu}-\frac{AG_n}{2}]=\Pr[Y=i],
 $}
 \]
 so that $\Pr[X_i=\overline{\mu}]=1-p_i^+-p_i^-$. From
 $\E X_i=\mu_i$ we get $p_i^+-p_i^-=\frac{2(\mu_i-\overline{\mu})}{AG_n}$ and
 from $\E\big\{ (X_i-\overline{\mu})^2\big\}=(\mu_i-\overline{\mu})^2+\sigma_i^2$
 we obtain $p_i^++p_i^-=\frac{4[(\mu_i-\overline{\mu})^2+\sigma_i^2]}{AG_n^2}$.
 Hence,
 \[
 \mbox{$
 p_i^+=\frac{2[(\mu_i-\overline{\mu})^2+\sigma_i^2]+(\mu_i-\overline{\mu})AG_n}{AG_n^2},
 \ \ \ \
 p_i^-=\frac{2[(\mu_i-\overline{\mu})^2+\sigma_i^2]-(\mu_i-\overline{\mu})AG_n}{AG_n^2},
 $}
 \]
 and (\ref{eq.marginals}) follows.
 Therefore, we can find a random vector ${\bXX}$ with $\E {\bXX}=\bmu$,
 $\Var{\bXX}={\bsigma}^2$ and $\E R_n=AG_n$ if and only if
 the above construction of a random pair $(X,Y)$, with $\Pr[X=Y]=0$, is possible.
 According to Lemma \ref{lem.3}, this is equivalent to
 $\max_{i}\big\{p_i^++p_i^-\big\}\leq 1$, which gives
 (\ref{eq.3.1})(ii) (it also guarantees
 that $\Pr[X_i=\overline{\mu}]=1-p_i^+-p_i^-\geq 0$),
 while (\ref{eq.3.1})(i) follows from
  $p_i^+\geq 0$ and  $p_i^-\geq 0$.

 Finally, the inequalities
 (\ref{eq.3.1}) are strict for all $i$ if and only if
 $p_i^++p_i^-<1$, $p_i^+>0$ and $p_i^->0$ for all $i$. Lemma
 \ref{lem.3} shows that there exist infinitely many vectors
 $(X,Y)$
 in this case.
 Also, if (\ref{eq.3.1}) is satisfied and we have equality in
 (\ref{eq.3.1})(ii) for some $i$, uniqueness follows again
 from  Lemma \ref{lem.3}.
 $\Box$
 \medskip
 \end{PR}

 \noindent
 \begin{PR}{\small\bf Proof of Lemma \ref{lem.4.1}:}
 The functions $f_i:T\to(0,\infty)$ ($i=1,2,3,4$) given by
 $f_1(x,y):=2\sqrt{x^2+y^2}$,
 $f_2(x,y):=2+\frac{1}{2}(x^2+y^2)$,
 $f_3(x,y):=x+1+\sqrt{(x-1)^2+y^2}$ and
 $f_4(x,y):=1-x+\sqrt{(x+1)^2+y^2}$ are
 obviously $C^{\infty}(T)$. The function
 $U$ can be defined as the restriction of
 $f_1$ in $A_1:=\{(x,y)\in T: x^2+y^2\geq 4\}$,
 of $f_2$ in $A_2:=\{(x,y)\in T:2|x|\leq x^2+y^2\leq 4\}$,
 of
 $f_3$ in $A_3:=\{(x,y)\in T: x^2+y^2\leq 2x\}$
 and of $f_4$ in $A_4:=\{(x,y)\in T: x^2+y^2\leq -2x\}$.
 Observe that $A_3$ and $A_4$ are the closed (with respect to $T$)
 semidisks
 $T\cap D((1,0),1)$, $T\cap D((-1,0),1)$;
 also, $A_2=T\cap [D((0,0),2)\smallsetminus A_3^{o} \cup A_4^{o}]$,
 and $A_1=T\smallsetminus A_2^{o}\cup A_3\cup A_4$.
 Therefore, $A_1\cap A_3=\emptyset$,
 $A_1\cap A_4=\emptyset$, $A_3\cap A_4=\emptyset$,
 $\partial A_1=A_1\cap A_2=\{(x,y)\in T:x^2+y^2=4\}$,
 $\partial A_3=A_2\cap A_3=\{(x,y)\in T:(x-1)^2+y^2=1\}$,
 $\partial A_4=A_2\cap A_4=\{(x,y)\in T:(x+1)^2+y^2=1\}$ and
 $\partial A_2=\partial A_1 \cup \partial A_3\cup \partial A_4$.
 It is easy to check that both partial derivatives of
 $f_1$ and $f_2$ coincide at $\partial A_1$, that both partial
 derivatives of $f_2$ and $f_3$ coincide at $\partial A_3$
 and that both partial derivatives
 of $f_2$ and $f_4$ coincide at $\partial A_4$.
 We conclude that for $(x,y)\in T$,
 \be
 \label{eq.4.1}
 U_1(x,y):=\frac{\partial}{\partial x} U(x,y)=\left\{
 \begin{array}{cll}
 \frac{x}{\sqrt{x^2+y^2}},
 &
 \mbox{ if }
 & x^2+y^2\geq 4,
 \\
 x,
 &
   \mbox{ if }
 & 2|x|\leq x^2+y^2\leq 4,
 \\
 \frac{x-1}{\sqrt{(x-1)^2+y^2}}+1,
 &
  \mbox{ if }
 &
 (x-1)^2+y^2\leq 1,
  \\
 \frac{x+1}{\sqrt{(x+1)^2+y^2}}-1,
 &
  \mbox{ if }
 & (x+1)^2+y^2\leq 1,
  \end{array}
 \right.
 \ee
 and
 \be
 \label{eq.4.2}
 U_2(x,y):=\frac{\partial}{\partial y} U(x,y)=\left\{
 \begin{array}{cll}
 \frac{y}{\sqrt{x^2+y^2}},
 &
 \mbox{ if }
 & x^2+y^2\geq 4,
 \\
 y,
 &
   \mbox{ if }
 &
 2|x|\leq x^2+y^2\leq 4,
 \\
 \frac{y}{\sqrt{(x-1)^2+y^2}},
 &
 \mbox{ if }
 &
 (x-1)^2+y^2\leq 1,
 \\
 \frac{y}{\sqrt{(x+1)^2+y^2}},
 &
 \mbox{ if }
 &
 (x+1)^2+y^2\leq 1,
 \end{array}
 \right.
 \ee
 and the above functions are obviously continuous.
 $\Box$
 \medskip
 \end{PR}

 \noindent
 \begin{PR}{\small\bf Proof of Proposition \ref{prop.4.1}:}
 Fix ${\bxx}$ and ${\byy}$ in $T$.
 The set $\partial A_2$ (where $U$ changes type) is a union of three disjoint
 semicircles, and the line segment $[{\bxx},{\byy}]=
 \{{\bxx}+t({\byy}-{\bxx})$, $0\leq t\leq 1\}$
 can have at most six common points with
 $\partial A_2=\{(x,y)\in T:x^2+y^2=4 \mbox { or } (x-1)^2+y^2=1 \mbox
 { or } (x+1)^2+y^2=1\}$;
 for the definition of $A_2$ see the proof of Lemma
 \ref{lem.4.1}.
 Consider now the function $g:[0,1]\to\R$ with
 $g(t):=U({\bxx}+t({\byy}-{\bxx}))$, $0\leq t\leq 1$,
 which is continuously differentiable from Lemma \ref{lem.4.1}.
 Also, $g$ is of the form (\ref{eq.4.4}) with $k\in\{0,\ldots,6\}$,
 where $g_i(t)=f_j({\bxx}+t({\byy}-{\bxx}))$, $0\leq t\leq 1$,
 for some
 $j=j(i)\in\{1,2,3,4\}$ (the functions $f_j:T\to (0,\infty)$
 are defined in the proof of Lemma \ref{lem.4.1}).
 It is easy to verify that each $f_j$ has
 nonnegative definite Hessian matrix and, thus, is convex.
 Lemma \ref{lem.4.2} asserts that
 $g_i(t):[0,1]\to (0,\infty)$ ($i=1,\ldots,k+1$)
 is convex.
 Since $g$ is continuously differentiable, (\ref{eq.4.3}) is automatically satisfied,
 and
 we conclude from Lemma \ref{lem.4.3} that $g$ is convex.
 Therefore, $g$ is convex for any choice of ${\bxx}$ and ${\byy}$ in $T$,
 and a final application of  Lemma \ref{lem.4.2} completes the proof.
 $\Box$
 \medskip
 \end{PR}

 \noindent
 \begin{PR}{\small\bf Proof of Lemma \ref{lem.4.4}:}
 (i) Fix $x_0\in\R$, $y_0>0$ and let
 $\alpha\in(0,1)$, $c_1,c_2\in\R$, $\lambda_1,\lambda_2>0$.
 Write $\beta_1=\frac{\alpha \lambda_1}
 {\alpha \lambda_1+(1-\alpha)\lambda_2}>0$,
 $\beta_2=\frac{(1-\alpha) \lambda_2}
 {\alpha \lambda_1+(1-\alpha)\lambda_2}>0$,
 so that $\beta_1+\beta_2=1$. We have
 \begin{eqnarray*}
 &&
 \hspace{-3ex}
 \mbox{$
 \frac{h\big(\alpha c_1+(1-\alpha) c_2,\alpha \lambda_1+(1-\alpha) \lambda_2\big)}
 {\alpha \lambda_1+(1-\alpha)\lambda_2}
 =f\big(\frac{x_0-[\alpha c_1+(1-\alpha) c_2]}
 {\alpha \lambda_1+(1-\alpha)\lambda_2},\frac{y_0}{\alpha \lambda_1+(1-\alpha)\lambda_2}\big)
 $}
 \\
 &&
 \hspace{5ex}
 \mbox{$
 =f\big(\beta_1\big(\frac{x_0- c_1}
 {\lambda_1}\big)
 +\beta_2 \big(\frac{x_0- c_2}
 {\lambda_2}\big)
 ,
 \beta_1\big(\frac{y_0}
 {\lambda_1}\big)
 +\beta_2 \big(\frac{y_0}
 {\lambda_2}\big)\big)
 $}
 \\
 &&
 \mbox{$
 \hspace{5ex}
 \leq \beta_1 f\big(\frac{x_0- c_1}{\lambda_1},\frac{y_0}{\lambda_1}\big)
 +\beta_2 f\big(\frac{x_0- c_2}{\lambda_2},\frac{y_0}
 {\lambda_2}\big)
 =\frac{\alpha h(c_1,\lambda_1)+(1-\alpha) h(c_2,\lambda_2)}
 {\alpha \lambda_1+(1-\alpha)\lambda_2},
 $}
 \end{eqnarray*}
 showing that $h$ is convex.

 \noindent
 (ii) Suppose that for a particular $(x_0,y_0)\in T$, the function
 $h_0(c,\lambda)=\lambda f\big(\frac{x_0-c}{\lambda},\frac{y_0}{\lambda}\big)$
 is convex. Set $x=\frac{x_0-c}{\lambda}$, $y=\frac{y_0}{\lambda}>0$, so that
 \[
 c=x_0-y_0\frac{x}{y}, \ \ \lambda=\frac{y_0}{y}, \
  \
  \
  y_0f(x,y)=y\, h_0\Big(x-y_0\frac{x}{y}, \frac{y_0}{y}\Big),
  \   (x,y)\in T.
 \]
 Let $\alpha\in(0,1)$, $x_1,x_2\in\R$ and $y_1,y_2>0$.
 Let us now write
 $\beta_1=\frac{\alpha y_1}{\alpha y_1+(1-\alpha) y_2}>0$,
 $\beta_2=\frac{(1-\alpha) y_2}{\alpha y_1+(1-\alpha) y_2}>0$,
 so that $\beta_1+\beta_2=1$.
 It follows that
 \begin{eqnarray*}
 &&
 \hspace{-5ex}
 y_0 f\big(\alpha x_1+(1-\alpha)x_2,\alpha y_1+(1-\alpha)y_2\big)
 \\
 &&
 =
 \big[\alpha y_1+(1-\alpha)y_2\big] \,
 h_0\Big(\beta_1 \Big(x_1-y_0\frac{x_1}{y_1}\Big)+\beta_2 \Big(x_2-y_0\frac{x_2}{y_2}\Big),
 \beta_1 \Big(\frac{y_0}{y_1}\Big)+\beta_2 \Big(\frac{y_0}{y_2}\Big)
 \Big)
 \\
 &&
 \leq
 \big[\alpha y_1+(1-\alpha)y_2\big] \,
 \Big\{
 \beta_1 h_0\Big(x_1-y_0\frac{x_1}{y_1},\frac{y_0}{y_1}\Big)
 +
 \beta_2 h_0\Big(x_2-y_0\frac{x_2}{y_2},\frac{y_0}{y_2}\Big)
 \Big\}
 \\
 &&
 =\alpha y_1 h_0\Big(x_1-y_0\frac{x_1}{y_1},\frac{y_0}{y_1}\Big)
 +(1-\alpha)y_2 h_0\Big(x_2-y_0\frac{x_2}{y_2},\frac{y_0}{y_2}\Big)
 \\
 &&
 =y_0 \, \big[\alpha f(x_1,y_1)+(1-\alpha)f(x_2,y_2)\big],
 \end{eqnarray*}
 and the proof is complete.
 $\Box$
 \medskip
 \end{PR}

 \noindent
 \begin{PR}{\small\bf Proof of Lemma \ref{lem.5.2}:}
 If $(c_0,\lambda_0)\in T_0$ then,
 by Proposition \ref{prop.5.1}, $\phi_n(c,\lambda)\geq \phi_n(c_0,\lambda_0)$
 for all $(c,\lambda)\in T$. On the other hand, for this $c_0$
 we can define the function $\psi_n(\lambda)=\phi_n(c_0,\lambda)$;
 by Lemma \ref{lem.5.1}, the function $\psi_n(\lambda)$
 is minimized at a unique $\lambda=\lambda_1=\lambda_1(c_0)$.
 Thus,
 \[
 \psi_n(\lambda_0)=\phi_n(c_0,\lambda_0)\leq \phi_n(c_0,\lambda_1)
 =\psi_n(\lambda_1)\leq \psi_n(\lambda_0);
 \]
 the first inequality follows from $(c_0,\lambda_0)\in T_0$
 and the second from the definition of $\lambda_1$. Therefore,
 $\psi_n(\lambda_0)=\psi_n(\lambda_1)$, so that
 $\lambda=\lambda_0$ is a minimizing point for $\psi_n(\lambda)$.
 By uniqueness, $\lambda_1=\lambda_0$.
 Thus, $\lambda_0=\lambda_1(c_0)$, where
 $\lambda_1(\cdot):\R\to(0,\infty)$ is a well-defined function;
 it is described (implicitly) in Lemma \ref{lem.5.1}.
 Hence, if $(c_0,\lambda_0)\neq (c_2,\lambda_2)$
 are any two points in $T_0$
 then $c_0\neq c_2$; indeed, $c_0=c_2$ implies
 $\lambda_0=\lambda_1(c_0)=\lambda_1(c_2)=\lambda_2$,
 contradicting the assumption $(c_0,\lambda_0)\neq (c_2,\lambda_2)$.

 Let $L$ be the straight line that passes through the points
 $(c_0,\lambda_0)$ and $(c_2,\lambda_2)$.
 We now verify that if $(c_3,\lambda_3)\in T_0$ then
 $(c_3,\lambda_3)\in L$. Indeed, if $(c_3,\lambda_3)\in
 T_0\smallsetminus L$ then the convex hull $B$ of the
 triangle
 $\{(c_0,\lambda_0)$, $(c_2,\lambda_2)$, $(c_3,\lambda_3)\}$
 must be a subject of $T_0$, because $T_0$ is convex.
 Since, however, $(c_3,\lambda_3)\notin L$, the set
 $B$ contains a line segment of positive length, parallel to the
 $\lambda$-axis and, by the previous argument,
 this is impossible.
 It follows that $T_0\subseteq L\cap T$, and since $T_0$
 is compact and convex, it must be a compact line segment.
 $\Box$
 \medskip
 \end{PR}

 \noindent
 \begin{PR}{\small\bf Proof of Lemma \ref{lem.5.3}:}
 By assumption, $A$ is moving linearly in the line
 segment $[A_0,A_1]$ from $A_0$ to $A_1$, thus we may write
 $A=A(t):=(c(t), \lambda(t))$ where
 $c(t)=c_0+t(c_1-c_0)$, $\lambda(t)=\lambda_0+t(\lambda_1-\lambda_0)$,
 $0\leq t\leq 1$.
 Then $B=B(t)=\big(\frac{\mu-c(t)}{\lambda(t)},\frac{\sigma}{\lambda(t)}\big)$,
 so that $B(0)=B_0$, $B(1)=B_1$ and $B(t)$ is continuous in $t$.
 It follows that for all $t\in[0,1]$,
 \[
 \det[B_0, B(t), B_1]:=
 \left|
 \begin{array}{ccc}
 \frac{\mu-c_0}{\lambda_0} & \frac{\sigma}{\lambda_0} &  1 \\
 \frac{\mu-c(t)}{\lambda(t)} & \frac{\sigma}{\lambda(t)} &  1 \\
 \frac{\mu-c_1}{\lambda_1} & \frac{\sigma}{\lambda_1} &  1
 \end{array}
 \right|=
 \frac{\sigma}{\lambda_0\lambda(t)\lambda_1}
 \left|
 \begin{array}{ccc}
 c_0 & \lambda_0 &  1 \\
 c(t) & \lambda(t) &  1 \\
 c_1 & \lambda_1 &  1
 \end{array}
 \right|=0.
 \ \ \Box
 \]
 \end{PR}

 \noindent
 \begin{PR}{\small\bf Proof of Theorem \ref{theo.5.1}:}
 According to Proposition \ref{prop.5.1}, it remains to
 verify that $T_0$ in (\ref{eq.5.2}) is a singleton.
 Assume, in contrary, that $T_0$ contains two points
 $(c_0,\lambda_0)\neq (c_1,\lambda_1)$.
 From Lemma \ref{lem.5.2} we know that $c_0\neq c_1$,
 and that all points
 $(c,\lambda)\in T_0$ can be written
 as $(c,\lambda)=(c,\alpha c+\beta)$, $c_2\leq c\leq c_3$,
 for some $\alpha,\beta, c_2,c_3\in\R$ with $c_2<c_3$. Therefore,
 we can write
 $\lambda(c)=\alpha c+\beta$, $c_2\leq c\leq c_3$, and
 \[
 T_0=\{(c,\alpha c+\beta), \ c_2\leq c\leq c_3\},
 \ \ \alpha, \beta, c_2, c_3\in\R, \
 c_2<c_3.
 \]
 Note that the parameters $\alpha$, $\beta$, $c_2$, $c_3$ have to fulfill
 additional restrictions so that $\lambda(c)>0$ for all $c\in[c_2,c_3]$;
 namely, $\alpha c_2+\beta>0$  and $\alpha c_3+\beta>0$.

 Consider now the points $A(c):=(c,\lambda(c))$ and
 $B_i(c):=\big(\frac{\mu_i-c}{\lambda(c)},\frac{\sigma_i}{\lambda(c)}\big)$,
 $i=1,\ldots,n$, $c_2\leq c\leq c_3$. As $c$ varies in $[c_2,c_3]$,
 the point $A=A(c)$ is moved from $A(c_2)$ to $A(c_3)$,
 generating the
 line
 segment $[A(c_2),A(c_3)]=T_0\subset T$.
 It follows from Lemma \ref{lem.5.3}
 that each point $B_i=B_i(c)$, $i=1,\ldots,n$, produces
 a line segment too; that is, $B_i$ generates its corresponding
 segment $L_i:=[B_i(c_2), B_i(c_3)]\subset T$.
 Consider now the region
 $A_2=\{(x,y)\in T : 2|x|\leq x^2+y^2\leq 4\}\subset T$.
 The function $U(x,y)$ (see (\ref{eq.u})) changes types (and it is
 not even $C^{2}$)
 only at the boundary points of $A_2$,
 i.e., at those
 $(x,y)\in T$ that belong to the set
 \[
 C :=\{x^2+y^2=4\}\cup \{(x-1)^2+y^2=1\}\cup \{(x+1)^2+y^2=1\}\subset \R^2.
 \]
 The set $\partial A_2=C\cap T$ is a union of
 three (disjoint) semicircles, and thus, any line segment
 can have
 at most six common points with it.
 It follows that only of finite number of points
 of the set $\cup_{i=1}^n L_i=\cup_{i=1}^n\cup_{c_2\leq c \leq c_3} B_i(c)$
 can intersect $\partial A_2$. Let $\Gamma_1,\ldots, \Gamma_k$
 be all these points.
 Each $\Gamma_j$ belongs to some $L_i$;
 that is, for any $j\in\{1,\ldots,k\}$
 we can find an index $i=i(j)\in\{1,\ldots,n\}$,
 and then a unique number $t=t_{ij}\in[c_2,c_3]$ such that
 $B_i(t)=\Gamma_j$.
 Clearly, for a particular index $j$,
 the maximal number of different $t$'s that can be
 found (satisfying $B_i(t)=\Gamma_j$ for some $i$)
 is $n$, because $B_i(t_1)\neq B_i(t_2)$ if $t_1\neq t_2$.
 Therefore, the set
 \[
 N:
 =
 \{t\in[c_2,c_3]: B_i(t)=\Gamma_j \mbox{ for some } i \mbox{ and } j\}
 \]
 is finite, say $N=\{t_1,\ldots,t_m\}$ with $c_2\leq t_1<\cdots<t_m\leq c_3$.
 Fix now an interval $[t,s]\subseteq (c_2,c_3)$, of positive length,
 such that $[t,s]\cap N=\emptyset$.
 Since $[t,s]$ has no common points with $N$,
 it is clear that the line segment $J_i:=[B_i(t),B_i(s)]\subseteq L_i$ does not
 intersect $\partial A_2$, and this is true for all $i\in\{1,\ldots, n\}$.
 In this way we obtain a subset $T_1$ of $T_0$, namely
 \[
 T_1:=\{(c,\alpha c+\beta), \ t\leq c\leq s\},
 \ \
 \mbox{ with } c_2<t<s<c_3.
 \]
 The boundary of $A_2$
 divides $T$ into four disjoint open regions, namely
 \[
 \begin{array}{ll}
 G_1:=\{(x,y)\in T: x^2+y^2>4\}, & G_2:=\{(x,y)\in T: 2|x|<x^2+y^2<4\},
 \\
 G_3:=\{(x,y)\in T: (x-1)^2+y^2<1\}, & G_4:=\{(x,y)\in T: (x+1)^2+y^2<1\}.
 \end{array}
 \]
 Compared to $T_0$, the set $T_1$ has the additional
 property that, as $c$ varies, every line segment
 $\{B_i(c), t\leq c\leq s\}$ stays in the same open region.
 This means that the sets of indices $I_1$, $I_2$, $I_3$, $I_4$,
 defined in Remark \ref{rem.5.1}, do not depend on
 $c$.
 Recall that
 \begin{eqnarray*}
 &&
 B_i(c)\in G_1
 \Leftrightarrow
 (\mu_i-c)^2+\sigma_i^2 >4\lambda^2
 \Rightarrow i\in I_1,
 \\
 &&
 B_i(c)\in G_2
 \Leftrightarrow
 2\lambda|\mu_i-c|<(\mu_i-c)^2+\sigma_i^2 <4\lambda^2
 \Rightarrow i\in I_2,
 \\
 &&
 B_i(c)\in G_3
 \Leftrightarrow
 (\mu_i-c)^2+\sigma_i^2 <2\lambda (\mu_i-c)
 \Rightarrow i\in I_3,
 \\
 &&
 B_i(c)\in G_4
 \Leftrightarrow
 (\mu_i-c)^2+\sigma_i^2 <-2\lambda (\mu_i-c)
 \Rightarrow i\in I_4,
 \end{eqnarray*}
 where $\lambda=\lambda(c)=\alpha c+\beta$.

 Consider now the function $g_n:(t,s)\to\R$ with
 \[
 g_n(c):=\phi_n(c,\lambda(c))=\phi_n(c,\alpha c+\beta), \ \ t<c<s.
 \]
 The explicit form of $g_n$ is quite complicated:
 \begin{eqnarray*}
 g_n(c)&=&
 \mbox{$
 -(n-2)(\alpha c+\beta) + \sum_{i\in I_1}\sqrt{(\mu_i-c)^2+\sigma_i^2}
 $}
 \\
 &&
 \mbox{$
 +
 \sum_{i\in I_2} \Big\{(\alpha c+\beta)
 +\frac{1}{4(\alpha c+\beta)}
 \big[(\mu_i-c)^2+\sigma_i^2\big]
 \Big\}
 $}
 \\
 &&
 \mbox{$
 +
 \sum_{i\in I_3}
 \frac{1}{2}\Big\{
 \mu_i-c+(\alpha c+\beta)+
 \sqrt{\big[\mu_i-c-(\alpha c+\beta)\big]^2+\sigma_i^2}
 \Big\}
 $}
 \\
 &&
 \mbox{$
 +
 \sum_{i\in I_4}
 \frac{1}{2}\Big\{
 c-\mu_i+(\alpha c+\beta)+
 \sqrt{\big[c-\mu_i-(\alpha c+\beta)\big]^2+\sigma_i^2}
 \Big\}.
 $}
 \end{eqnarray*}
 Since, however, the sets $I_j$ do not depend on $c$, it is obvious that
 $g_n\in C^{\infty}(t,s)$. By assumption,
 $(c,\lambda(c))$ minimizes $\phi_n(c,\lambda)$ for all $c\in(t,s)$, and this
 means that $g_n(c)$ is constant, implying that
 $g_n''(c)=0$, $t<c<s$.
 A straightforward computation
  shows that for all $c\in(t,s)$,
 \begin{eqnarray*}
 g''_n(c)&=&
 \mbox{$
 \sum_{i\in I_1}\frac{\sigma_i^2}{\left[(\mu_i-c)^2+\sigma_i^2\right]^{3/2}}
 +\frac{1}{2\left[\lambda(c)\right]^3}
 \sum_{i\in I_2}
 \Big\{
 \alpha^2 \sigma_i^2 +(\alpha \mu_i+\beta)^2
 \Big\}
 $}
 \\
 &&
 \mbox{$
 +
 \frac{(\alpha+1)^2}{2}
 \sum_{i\in I_3}
 \frac{\sigma_i^2}{\left[\left(\beta+(\alpha+1)c-\mu_i\right)^2
 +\sigma_i^2\right]^{3/2}}
 +
 \frac{(\alpha-1)^2}{2}
 \sum_{i\in I_4}
 \frac{\sigma_i^2}{\left[\left(\beta+(\alpha-1)c+\mu_i\right)^2
 +\sigma_i^2\right]^{3/2}}.
 $}
 \end{eqnarray*}
 Obviously, all summands are nonnegative.
 If $\alpha\neq 0$, the only two possibilities
 which are compatible with $g_n''(c)=0$ are the following:
 (i) either $I_1=I_2=I_4=\emptyset$ (and thus, $N(I_3)=n$)
 and $\alpha=-1$ or
 (ii) $I_1=I_2=I_3=\emptyset$ (and $N(I_4)=n$) and $\alpha=1$.
 However, because of (\ref{eq.5.7}), neither (i) nor (ii)
 is allowed for a minimizing point $(c,\lambda)$, and in particular
 for $(c,\lambda(c))$. Finally, if $\alpha=0$ then we must have
 $I_1=I_3=I_4=\emptyset$ and, therefore, $N(I_2)=n$. The
 condition
 $\lambda(c)>0$ now yields $\beta>0$; thus,
 $g_n''(c)=\frac{n}{2\beta}>0$
 and $g_n(c)$ could not be a
 constant function in the interval $t<c<s$.

 The resulting contradiction implies that the set $T_0$ cannot contain
 two
 distinct elements, and the proof is complete.
 $\Box$
 \medskip
 \end{PR}

\noindent
 \begin{PR}{\small\bf Proof of Lemma \ref{lem.6.1}:}
 From (\ref{eq.phi}), (\ref{eq.4.1}),
 (\ref{eq.4.2}), and in view of (\ref{eq.6.3}), (\ref{eq.6.4}),
 \begin{eqnarray*}
 \mbox{$
 \frac{\partial}{\partial \lambda} \phi_n(c,\lambda)
 $}
 \hspace{-1ex}
 &=&
 \hspace{-1ex}
 \mbox{$
 -(n-2)
 +
 \frac12
 \sum_{i\in I_2} \Big\{
 2-\frac{1}{2\lambda^2}
 \big[(\mu_i-c)^2+\sigma_i^2\big]
 \Big\}
 $}
 \\
 &&
 \mbox{$
 +
 \frac12
 \sum_{i\in I_3}
 \Big\{
 1-\frac{\mu_i-c-\lambda}{\sqrt{\left(\mu_i-c-\lambda\right)^2+\sigma_i^2}}
 \Big\}
 +
 \frac12
 \sum_{i\in I_4}
 \Big\{
 1-\frac{c-\mu_i-\lambda}{\sqrt{\left(c-\mu_i-\lambda\right)^2+\sigma_i^2}}
 \Big\}
 $}
 \\
 \hspace{-1ex}
 &=&
 \hspace{-1ex}
 -(n-2)
 +
 \sum_{i\in I_2\cup I_3\cup I_4} p_i^{o}.
 \end{eqnarray*}
 Since $\frac{\partial}{\partial\lambda}\phi_n(c,\lambda)=0$ and
 $p_i^{o}=0$ for $i\in I_1$,  it follows that
 $\sum_{i=1}^n p_i^{o} =n-2$.
 Taking into account the fact that
 $p_i^{o}=1-p_i^+-p_i^-$, we obtain
 $\sum_{i=1}^n p_i^- + \sum_{i=1}^n p_i^+ =2.$

 Similarly, we have
 \begin{eqnarray*}
 \mbox{$
 \frac{\partial}{\partial c} \phi_n(c,\lambda)
 $}
 \hspace{-1ex}
 &=&
 \hspace{-1ex}
 \mbox{$
 -\sum_{i\in I_1} \frac{\mu_i-c}{\sqrt{(\mu_i-c)^2+\sigma_i^2}}
 -
 \sum_{i\in I_2} \frac{\mu_i-c}{2\lambda}
 $}
 \\
 &&
 \mbox{$
 -
 \sum_{i\in I_3}
 \frac12
 \Big\{
 1+\frac{\mu_i-c-\lambda}{\sqrt{\left(\mu_i-c-\lambda\right)^2+\sigma_i^2}}
 \Big\}
 +
 \sum_{i\in I_4}
 \frac12
 \Big\{
 1+\frac{c-\mu_i-\lambda}{\sqrt{\left(c-\mu_i-\lambda\right)^2+\sigma_i^2}}
 \Big\}
 $}
 \\
 \hspace{-1ex}
 &=&
 \hspace{-1ex}
 -\sum_{i\in I_1} (p_i^+-p_i^-)
 -\sum_{i\in I_2} (p_i^+-p_i^-)
 -\sum_{i\in I_3} p_i^+
 + \sum_{i\in I_4} p_i^-,
 \end{eqnarray*}
 that is, $\frac{\partial}{\partial c}\phi_n(c,\lambda)
 =-\sum_{i\in I_1\cup I_2\cup I_3} p_i^+ + \sum_{i\in I_1\cup I_2\cup I_4}p_i^-$.
 From the fact
 that $p_i^+=0$ for $i\in I_4$ and $p_i^-=0$ for $i\in I_3$
 (see (\ref{eq.6.4})),
 the relation $\frac{\partial}{\partial c}\phi_n(c,\lambda)=0$
 implies the equality $\sum_{i=1}^n p_i^{+}=\sum_{i=1}^n p_i^{-}$,
 and (\ref{eq.6.7}) follows.
 $\Box$
 \medskip
 \end{PR}

\noindent
 \begin{PR}{\small\bf Proof of Proposition \ref{prop.6.1}:}
 [$\mbox{(ii)}\Rightarrow\mbox{(i)}$].
 Suppose we are
 given a probability matrix $Q$ satisfying
 (ii).
 By assumption,
 $Q$ has vanishing principal diagonal.
 Define
 ${\bXX}=(X_1,\ldots,X_n)$
 as in (\ref{eq.6.10}).
 Since $\sum_{(i,j):\ i\neq j} q_{ij}=\sum_{i,j}q_{ij}=1$,
 this procedure maps $Q$ to a well-defined probability
 law ${\cal L}({\bXX})$ on $\R^n$,
 and the
 map $Q\mapsto {\cal L}({\bXX})$ is, obviously, one to one.
 Due to (\ref{eq.6.6}), the order
 statistics of ${\bXX}$
 satisfy
 \be
 \label{eq.6.11}
 X_{1:n}<c-\lambda<X_{2:n}\leq \cdots \leq X_{n-1:n}<c+\lambda<X_{n:n}
 \ \  \mbox{ with probability }  1.
 \ee
 Thus, from Lemma \ref{lem.1} it follows
 that, with probability $1$,
 \be
 \label{eq.6.12}
 \mbox{$
 R_n=-(n-2)\lambda+\frac{1}{2}
 \sum_{i=1}^n \big\{ \big|(X_i-c)-\lambda\big|
 +\big|(X_i-c)+\lambda\big|\big\}.
 $}
 \ee
 The assumptions $Q\in{\cal M}({\bpp}^{+},{\bpp}^{-})$
 and $q_{ii}=0$
 for all $i$ now
 show that for any fixed $j$,
 $\Pr[X_j=x_j^{-}]=\sum_{i\neq j} q_{ij}=\sum_{i=1}^n q_{ij}=p_j^{-}$.
 Similarly
 we conclude that for any fixed $i$,
 $\Pr[X_i=x_i^{+}]=\sum_{j\neq i} q_{ij}=\sum_{j=1}^n q_{ij}=p_i^{+}$.
 Thus, $\Pr[X_i=x_i^{o}]=1-p_i^{-}-p_i^{+}=p_i^{o}$, and the marginal
 $X_i$ of ${\bXX}$ is the extremal random variable in
 ${\cal F}_1(\mu_i,\sigma_i)$.
 That is, it has mean $\mu_i$, variance $\sigma_i^2$, and maximizes
 $\E\big\{ |(X-c)-\lambda\big|
 +\big|(X-c)-\lambda|\big\}$ as $X$ varies in ${\cal F}_1(\mu_i,\sigma_i)$.
 Since this holds for all $i$, taking expectations in (\ref{eq.6.12})
 we see that
 \be
 \label{eq.6.13}
 \mbox{$
 \E R_n=-(n-2)\lambda+\frac{\lambda}{2}
 \sum_{i=1}^n U\big(\frac{\mu_i-c}{\lambda},\frac{\sigma_i}{\lambda}\big)
 =\rho_n,
 $}
 \ee
 completing the proof.

 \noindent
 [$\mbox{(i)}\Rightarrow\mbox{(ii)}$].
 Assumptions ${\bXX}\in{\cal F}_n(\bmu,{\bsigma})$ and $\E R_n=\rho_n$
 imply that (repeat the proof of Theorem \ref{th.main})
 \begin{eqnarray*}
 \rho_n
 \hspace{-1ex}
 &=&
 \hspace{-1ex}
 \mbox{$
 \E R_n
 \leq
 \E
 \big\{-(n-2)\lambda
 +\frac{1}{2}\sum_{i=1}^{n}\big[
 \big| (X_i-c)-\lambda\big|
 +
 \big| (X_i-c)+\lambda\big|\big]
 \big\}
 $}
 \\
 \hspace{-1ex}
 &=&
 \hspace{-1ex}
 \mbox{$
 -(n-2)\lambda
 +\frac{1}{2}\sum_{i=1}^{n}\E\big\{
 \big| (X_i-c)-\lambda\big|
 +
 \big| (X_i-c)+\lambda\big|\big\}
 $}
 \\
 \hspace{-1ex}
 &&
 \hspace{-1ex}
 \mbox{$
 \leq
 -(n-2)\lambda
 +\frac{\lambda}{2}\sum_{i=1}^{n}
 U\big(\frac{\mu_i-c}{\lambda},\frac{\sigma_i}{\lambda}\big)
 =\rho_n.
 $}
 \end{eqnarray*}
 Thus, all displayed inequalities are attained as equalities.
 In view of
 Lemma \ref{lem.1} and
 Corollary \ref{cor.1}, this
 can happen only if the law ${\cal L}({\bXX})$ of the
 given random vector ${\bXX}=(X_1,\ldots,X_n)$ satisfies
 \be
 \label{eq.6.14}
 \begin{array}{ll}
 \hspace{-1.5ex}
 \mbox{(a)} &
 \ \Pr\big[X_{1:n}\leq c-\lambda \leq X_{2:n}\leq
 \cdots\leq X_{n-1:n}\leq c+\lambda\leq X_{n:n}\big]=1
 \\
 \hspace{-1.5ex}\mbox{and}
 &
 \\
 \hspace{-1.5ex}
 \mbox{(b)}
 &
 \begin{array}{l}
 X_i
 \mbox{ is extremal in } {\cal F}_1(\mu_i,\sigma_i)
 \mbox{ for all } i,
 \mbox{ or, equivalently, }
 \\
 \Pr[X_i=x_i^-]=p_i^{-},
 \
 \Pr[X_i=x_i^{o}]=p_i^{o},
 \
 \Pr[X_i=x_i^+]=p_i^{+},
 \
 i=1,\ldots,n.
 \end{array}
 \end{array}
 \ee
 Taking into account (\ref{eq.6.6}) we conclude that
 (\ref{eq.6.14}) can happen only if
 the (essential) support of ${\bXX}$ is contained in the set
 \be
 \label{eq.6.15}
 S:=\big\{{\bxx}_{ij},  \ i\neq j, \  i,j=1,\ldots,n\big\},
 \ee
 with ${\bxx}_{ij}$ as in (\ref{eq.6.10}).
 We can thus define the $n\times n$ matrix $Q$ as follows:
 \be
 \label{eq.6.16}
 Q:=(q_{ij}), \mbox{ with } q_{ii}:=0, \ q_{ij}:=\Pr[{\bXX}={\bxx}_{ij}],
 \ i\neq j, \  i,j=1,\ldots,n.
 \ee
 By definition, $Q$ has vanishing principal diagonal
 and nonnegative entries,
 and the relation
 $\Pr[{\bXX}\in S]=1$ implies that $Q$ is a probability matrix.
 By the assumption ${\bXX}\in{\cal F}_n(\bmu,{\bsigma})$
 and $\E R_n=\rho_n$,
 the marginal $X_i$ of ${\bXX}$ has to fulfill (\ref{eq.6.14})(b),
 that is,
 $\sum_{j=1}^n q_{ij}=\sum_{j\neq i} q_{ij}
 =\sum_{j} \Pr[{\bXX}={\bxx}_{ij}] =\Pr[X_i=x_{i}^{+}]=
 p_i^+$; similarly,
 $\sum_{i=1}^n q_{ij}=p_j^-$. Therefore, we have
 constructed a matrix $Q\in{\cal M}({\bpp}^{+},{\bpp}^-)$
 with $q_{ii}=0$ for all $i$. Clearly, if
 two random vectors ${\bXX}$, $\bYY$, with
 ${\cal L}(\bYY)\neq {\cal L}({\bXX})$,
 satisfy the assumptions in (i),
 the corresponding matrices (obtained through
 (\ref{eq.6.16})) will be distinct. Consequently,
 the above procedure determines a one to one mapping
 ${\cal L}({\bXX})\mapsto Q$,
 completing the proof.
 $\Box$
 \medskip
 \end{PR}

 \noindent
 \begin{PR}{\small\bf Proof of Theorem \ref{theo.7.1}:}
 For the infimum, a proof (for any $n\geq 2$) is
 given in the beginning of Section \ref{sec.main.inequality},
 following the arguments of Bertsimas, Doan, Natarajan and Teo
 (2010).
 Regarding the supremum: The key-observation is that
 \eqref{eq.7.3} is a special application
 of the Cauchy-Schwarz inequality,
 \[
 \mbox{$
 \E R_2=\E |X_1-X_2|\leq \sqrt{\E\big[(X_1-X_2)^2\big]}
 =\sqrt{(\mu_1-\mu_2)^2+\sigma_1^2+\sigma_2^2-2\rho\sigma_1\sigma_2}=\gamma_2.
 $}
 \]
 This means that, in order to justify the equality, we have to construct
 a vector $(X_1,X_2)\in{\cal F}_2(\bmu,{\bsigma};\rho)$
 such that the random variable $|X_1-X_2|$ is degenerate.
 Let $\delta:=\Var[X_1-X_2]=\sigma_1^2+\sigma_2^2-
 2\rho\sigma_1\sigma_2\geq 0$. We
 distinguish cases $\delta>0$, $\delta=0$.

 Assume $\delta>0$, so that $\gamma_2>0$.
 First, we consider
 a $0$--$1$ Bernoulli random variable $I_p$
 with probability of success
 $p:=\frac{1}{2}\big(1+\frac{\mu_1-\mu_2}{\gamma_2}\big)$.
 Next, we consider another random variable
 $T$ with mean
 $\mu_T:=\mu_1\sigma_2^2+\mu_2\sigma_1^2-\rho\sigma_1\sigma_2(\mu_1+\mu_2)$
 and variance
 $\sigma_T^2:=\delta\sigma_1^2\sigma_2^2(1-\rho^2)\geq 0$,
 stochastically independent
 of $I_p$.
 Finally, we define
 \[
 \mbox{$
 (X_1,X_2):=\frac{1}{\delta}\
 \big[\gamma_2(\sigma_1^2-\rho\sigma_1\sigma_2)(2I_p-1)+T,
 \
 \gamma_2(\rho\sigma_1\sigma_2-\sigma_2^2)(2I_p-1)+T\big].
 $}
 \]
 It is easily seen that $(X_1,X_2)\in{\cal F}_2(\bmu,{\bsigma};\rho)$
 and $|X_1-X_2|=\gamma_2$ with probability 1.

 Let us now assume $\delta=0$.
 This implies that $X_1-X_2=\mu_1-\mu_2$ with probability $1$, and hence,
 $\sigma_1=\sigma_2$ and $\rho=1$.
 Let $\sigma^2>0$ be the common variance and consider
 the pair $(X_1,X_2):=(\mu_1+T,\mu_2+T)$, where $T$
 is any random variable with mean zero and variance $\sigma^2$.
 It follows that $(X_1,X_2)$ satisfies the moment requirements and
 $|X_1-X_2|=|\mu_1-\mu_2|=\gamma_2$ with probability 1. This completes the proof.
 $\Box$
 \end{PR}
 \end{appendix}
 }



\begin{thebibliography}{99}
 {\small


 \bibitem{01}  Arnold, B.C.\  (1980). Distribution-free bounds on
 the mean of the maximum of a dependent sample.
 {\it SIAM J.\ Appl.\ Math.} {\bf 38}, 163--167.
 \vspace{-1.4ex}

 \bibitem{02}   Arnold, B.C.\  (1985). $p$-Norm bounds on the
 expectation of the maximum of possibly dependent sample.
  {\it  J.\ Multivariate Anal.} {\bf 17}, 316--332.
 \vspace{-1.4ex}

 \bibitem{03}   Arnold, B.C.\  (1988). Bounds on the
  expected maximum.
  {\it  Commun.\ Statist.--Theory Meth.}
  {\bf 17}, 2135--2150.
 \vspace{-1.4ex}

 \bibitem{04}
 Arnold, B.C.; Balakrishnan, N.\
 (1989).
 {\it Relations,
 Bounds and Approximations
 for Order Statistics}.
 Lecture Notes in Statistics, Vol. 53, Springer, New York.
 \vspace{-1.4ex}

 \bibitem{05}   Arnold, B.C.; Groeneveld, R.A.\ (1979). Bounds
 on expectations of linear systematic statistics based on
 dependent samples. {\it Ann.\ Statist.} {\bf 7}, 220--223.
 Correction: {\bf 8}, 1401.
 \vspace{-1.4ex}

 \bibitem{06}
 Aven, T.\ (1985). Upper (lower) bounds on the mean of the maximum (minimum)
 of a number of random variables.
 {\it J. Appl.\ Probab.} {\bf 22}, 723--728.
 \vspace{-1.4ex}

 \bibitem{07}
 Balakrishnan, N.; Balasubramanian, K., (1993). Equivalence of
 Hartley-David-Gumbel and Papathanasiou bounds and some further remarks.
 {\it Statist.\ Probab.\ Lett.} {\bf 16}, 39--41.
 \vspace{-1.4ex}

 \bibitem{08}
 Barvinok, A.\ (2012). Matrices with prescribed row and column sums.
 {\it Linear Algebra Appl.}
 {\bf 436}, 820--844.
 \vspace{-1.4ex}

 \bibitem{09}
 Bertsimas, D.; Doan, X.V.; Natarajan, K.; Teo, C.-P.\ (2010).
 Models for minimax stochastic linear optimization problems
 with risk aversion. {\it Math.\ O.\ R.}
 {\bf 35}(3),  580--602.
 \vspace{-1.4ex}

 \bibitem{10}
 Bertsimas, D.; Natarajan, K.; Teo, C.-P.\ (2004).
 Probabilistic combinatorial optimization:
 moments,
 semidefinite programming and asymptotic bounds.
 {\it SIAM J.\ Optimiz.} {\bf 15}(1), 185--209.
  \vspace{-1.4ex}

 \bibitem{11}
 Bertsimas, D.; Natarajan, K.; Teo, C.-P.\ (2006).
 Tight bounds on expected order statistics.
 {\it Prob.\ Engineer.\ Inform.\ Sci.}
 {\bf 20},  667--686.
 \vspace{-1.4ex}


 \bibitem{12}   Caraux, G.\ and Gascuel, O.\ (1992).
 Bounds
 on distribution functions of order statistics
 for dependent variates. {\it  Statist.\ Probab.\ Lett.}
 {\bf 14}, 103--105.
 \vspace{-1.4ex}

 \bibitem{13}
 David, H.A.\ (1981).
 {\it Order Statistics}, 2nd ed. Wiley, N.Y.
 \vspace{-1.4ex}

 \bibitem{14}
 David, H.A.;  Nagaraja, H.N.\ (2003).
 {\it Order Statistics}, 3rd ed. Wiley, N.Y.
 \vspace{-1.4ex}

 \bibitem{15}
 Gajek, L.; Rychlik, T.\ (1996).
 Projection method for moment bounds on order
 statistics from restricted
 families. I. Dependent case. {\it
 J.\ Multivariate Anal.} {\bf 57}, 156--174.
 \vspace{-1.4ex}

 \bibitem{16}
 Gajek, L.; Rychlik, T.\ (1998). Projection method for moment bounds on
 order statistics from restricted
 families. II. Independent case. {\it
 J.\ Multivariate Anal.} {\bf 64}, 156--182.
 \vspace{-1.4ex}

 \bibitem{17}   Gascuel, O.; Caraux, G.\ (1992). Bounds
 on expectations  of order statistics via
 extremal dependencies.
 {\it  Statist.\ Probab.\ Lett.} {\bf 15}, 143--148.
 \vspace{-1.4ex}

 \bibitem{18}
 Giaquinta, M.; Modica, G.\ (2012). {\it Mathematical Analysis.
 Foundations and Advanced Techniques for Functions of
 Several Variables}. Springer,
 \vspace{-1.4ex}
 Birkh\"{a}user.

 \bibitem{19}
 Gumbel, E.J.\ (1954).
 The maxima of the mean largest value and of the range.
 {\it Ann.\ Math.\ Statist.} {\bf 25}, 76--84.
 \vspace{-1.4ex}

 \bibitem{20}
 Hartley, H.O.; David, H.A.\ (1954).
 Universal bounds for mean range and extreme observations.
 {\it Ann.\ Math.\ Statist.} {\bf 25}, 85--99.
 \vspace{-1.4ex}

 \bibitem{21}
 Isii, K.\ (1963).
 On the sharpness of Chebyshev-type inequalities
 {\it Ann.\ Inst.\ Statist.\ Math.} {\bf 14},
 185--197.
 \vspace{-1.4ex}

 \bibitem{22}
 Kaluszka, M.; Okolewski, A.; Szymanska, K.\ (2005).
 Sharp bounds for
 $L$-statistics from dependent samples of
 random length. {\it J.\ Statist.\ Plann.\ Inference}
 {\bf 127}, 71--89.
 \vspace{-1.4ex}

 \bibitem{23}
 Karlin, S.;  Studden, W.J.\ (1966).
 {\it Tchebycheff Systems: With
 Applications in Analysis and Statistics.}
 Wiley-Interscience, N.Y.
 \vspace{-1.4ex}

 \bibitem{24}
 Lai, T.L.; Robbins, H.\ (1976).
 Maximally dependent random variables.
 {\it Proceedings of the National
 Academy of the Sciences of the United States of America}
 {\bf 73}(2), 286--288.
 \vspace{-1.4ex}

 \bibitem{25}
 Lef\`{e}vre, C.\ (1986). Bounds on the expectations
 of linear combinations of
 order statistics with applications to
 Pert networks. {\it Stochast.\ Anal.\ Appl.}
 {\bf 4}, 351--356.
 \vspace{-1.4ex}

 \bibitem{26}
 Meilijson, I.;  Nadas, A.\ (1979).
 Convex majorization with an application to the length of critical
 path. {\it J.\ Appl.\ Probab.} {\bf 16},
 671--677.
 \vspace{-1.4ex}


 \bibitem{27}
 Nagaraja, H.N.\ (1981). Some finite sample results for the
 selection differential.
 {\it Ann.\ Inst.\ Statist.\ Math.}
 {\bf 33}, 437--448.
 \vspace{-1.4ex}

 \bibitem{28}
 Natarajan, K.; Teo, C.-P.\ (2014). Semidefinite programming
 reformulation of completely positive programs: range estimation
 and best-worst choice modeling. Available at
 {\tt http://people.}
 \vspace{-1.4ex}
 {\tt sutd.edu.sg}.

 \bibitem{29}
 Navarro, J.; Balakrishnan, N.\ (2010).
 Study of some measures of dependence between
 order statistics and systems.
 {\it J.\ Multivariate Anal.}
 {\bf 101}, 52--67.
 \vspace{-1.4ex}

 \bibitem{30}   Papadatos, N.\  (2001a). Expectation bounds on
 linear estimators from dependent samples.
 {\it J.\ Statist.\ Plann.\ Inference}
 {\bf 93}, 17--27.
 \vspace{-1.4ex}

 \bibitem{31}   Papadatos, N.\  (2001b). Distribution and expectation
 bounds on order statistics from possibly
 dependent variates.
 {\it  Statist.\ Probab.\ Lett.} {\bf 54}, 21--31.
 \vspace{-1.4ex}

 \bibitem{32}
 Papathanasiou, V.\ (1990).
 Some characterizations of distributions based on
 order statistics. {\it Statist.\ Probab.\ Lett.}
 {\bf 9}, 145--147.
 \vspace{-1.4ex}

 \bibitem{33}
 Placket, R.L.\ (1947).
 Limits of the ratio of mean range to
 standard deviation. {\it Biometrika}
 {\bf 34},
 120--122.
  \vspace{-1.4ex}



 \bibitem{34}
 Rychlik, T.\ (1992a). Sharp inequalities for
 linear combinations of elements of monotone sequences.
 {\it Bull.\ Polish Acad.\ Sci.\ Math.} {\bf 40}, 247--254.
 \vspace{-1.4ex}

 \bibitem{35}   Rychlik, T.\ (1992b). Stochastically extremal
 distributions of order statistics for dependent samples.
 {\it  Statist.\ Probab.\ Lett.} {\bf 13}, 337--341.
 \vspace{-1.4ex}

 \bibitem{36}   Rychlik, T.\ (1993a). Bounds for expectations of
  $L$-estimates for dependent samples.
  {\it  Statistics} {\bf 24}, 9--15.
  \vspace{-1.4ex}

 \bibitem{37}   Rychlik, T.\ (1993b). Sharp bounds on
 $L$-estimates and their expectations for dependent samples.
 {\it  Commun.\ Statist.--Theory \& Meth.} {\bf 22},
 1053--1068.
 Erratum {\bf 23},
 \vspace{-1.4ex}
 305--306.

 \bibitem{38}
 Rychlik, T.\ (1994). Distributions and expectations of order
 statistics for possibly dependent random variables.
 {\it J.\ Multivariate Anal.} {\bf 48}, 31--42.
  \vspace{-1.4ex}

 \bibitem{39}
 Rychlik, T.\ (1995). Bounds for order statistics based on
 dependent variables with given nonidentical
 distributions. {\it Statist.\ Probab.\ Lett.}
 \vspace{-1.4ex}
 {\bf 23}, 351--358.

 \bibitem{40}   Rychlik, T.\ (1998). Bounds on expectations of
 $L$-estimates. In: {\it Order Statistics: Theory and Methods}
 (N.\ Balakrishnan and C.R.\ Rao, eds.), Handbook of Statistics,
 vol.\ 16, North-Holland, Amsterdam,
 \vspace{-1.4ex}
 105--145.

 \bibitem{41}   Rychlik, T.\ (2001). {\it Projecting Statistical
 Functionals}. Lecture Notes in Statistics, {\bf 160},
 Springer-Verlag, N.Y.
 \vspace{-1.4ex}

 \bibitem{42}   Rychlik, T.\ (2007). Optimal deterministic
 bounds on $L$-statistics.
 In: {\it Recent Developments in Ordered Random Variables}
 (M.\ Ahsanullah and M.Z.\ Raqab, Eds.),
 Nova Science Publishers, N.Y., pp. 1--18.


 }
 \end{thebibliography}
 \end{document}